\documentclass[12pt]{article}
\usepackage{epsf,amsmath,amssymb,amsfonts}
\usepackage{graphicx}
\usepackage{epsfig}
\usepackage{verbatim}
\usepackage{color}
\usepackage{hyperref}
\usepackage{cite}

\definecolor{darkred}{rgb}{0.5,0,0}
\definecolor{darkgreen}{rgb}{0,0.5,0}
\definecolor{darkblue}{rgb}{0,0,0.5}

\newcommand{\be}{\begin{equation}}
\newcommand{\ee}{\end{equation}}
\newcommand{\bea}{\begin{eqnarray}}
\newcommand{\eea}{\end{eqnarray}}
\newcommand{\ba}{\begin{eqnarray}}
\newcommand{\ea}{\end{eqnarray}}

\newcommand{\beq}{\begin{equation}}
\newcommand{\eeq}{\end{equation}}
\newcommand{\beqa}{\begin{eqnarray}}
\newcommand{\eeqa}{\end{eqnarray}}
\newcommand{\beqar}{\begin{eqnarray*}}
\newcommand{\eeqar}{\end{eqnarray*}}
\numberwithin{equation}{section}

\newcommand{\sH}{\mathsf{H}}
\newcommand{\sx}{\mathsf{x}}

\newcommand{\Dp}[1]{\mathrm{D {#1}}}
\newcommand{\antiDp}[1]{\overline{\Dp{#1}}}
\def\G{\Gamma}

\begin{document}
\setlength{\unitlength}{1mm}

\thispagestyle{empty} \rightline{\hfill \small arXiv:0811.0263v2
 }
\rightline{\hfill \small
 ITFA-2008-43, UPR-T-1199}

\vspace*{2cm}

\begin{center}
{\bf \Large 
Black Holes as Effective Geometries
}\\

\vspace*{1.4cm}

\vspace*{0.3cm}

{\bf Vijay Balasubramanian$\,^\dag$, Jan de Boer$\,^\clubsuit$, Sheer
El-Showk$\,^\clubsuit$ and Ilies Messamah$\,^\clubsuit$}

\vspace{.8cm}

{\it $^\dag$ \, David Rittenhouse Laboratories, University of Pennsylvania,\\
Philadelphia, PA, 19104, USA \\ }
\vspace{.8cm}
{\it $^\clubsuit$  \, Instituut voor Theoretische Fysica,\\ Valckenierstraat 65,
1018XE Amsterdam, The Netherlands.}

\vspace*{0.6cm} {\tt vijay@physics.upenn.edu, J.deBoer@uva.nl,
S.ElShowk@uva.nl, I.Messamah@uva.nl}

\vspace{.8cm}
{\bf Abstract}

\end{center}
Gravitational entropy arises in string theory via coarse graining over an
underlying space of microstates.  In this review we would like to address the
question of how the classical black hole geometry itself arises as an effective
or approximate description of a pure state, in a closed string theory, which
semiclassical observers are unable to distinguish from the ``naive'' geometry.
In cases with enough supersymmetry it has been possible to explicitly construct
these microstates in spacetime, and understand how coarse-graining of
non-singular, horizon-free objects can lead to an effective description as an
extremal black hole.  We discuss how these results arise for examples in Type II
string theory on AdS${}_5\times$S${}^5$ and on
AdS${}_3\times$S${}^3\times$T${}^4$ that preserve 16 and 8 supercharges
respectively.  For such a picture of black holes as effective geometries to
extend to cases with finite horizon area the scale of quantum effects in
gravity would have to extend well beyond the vicinity of the singularities in
the effective theory.  By studying examples in M-theory on
AdS${}_3\times$S${}^2\times$CY that preserve 4 supersymmetries we show how this
can happen.

This review is a revised and extended version of proceedings submitted for the
Sower's Theoretical Physics Workshop 2007 ``What is String Theory?'' and of
lecture notes for the CERN RTN Winter School on Strings, Supergravity and Gauge
Theories \cite{Balasubramanian:2008cqg}.
\noindent

\vfill \setcounter{page}{0} \setcounter{footnote}{0}
\vspace{2cm}
\newpage
\tableofcontents

\vspace{2cm}

\setcounter{equation}{0}

\section{Introduction}
A spacetime geometry can carry an entropy in string theory via coarse graining
over an underlying set of  microstates.   Since the initial success of string
theory in accounting for the entropy of supersymmetric black holes by counting
states in a field theory \cite{Strominger:1996sh} there has been an ongoing
effort to understand exactly what the structure of these microstates is and how
they manifest themselves in gravity.    It has been shown that in examples with
enough supersymmetry, including some extremal black holes,  one can construct a
basis of ``coherent'' microstates whose spacetime descriptions in the $\hbar \to
0$ limit approach non-singular, horizon-free geometries which resemble a
topologically complicated ``foam''.  Conversely, in these cases the quantum
Hilbert space of states can be constructed by directly quantizing a moduli space
of smooth classical solutions.  Nevertheless, the {\it typical} states in these
Hilbert spaces respond to semiclassical probes as if the underlying geometry was
singular, or an extremal black hole.  In this sense, these black holes are
effective, coarse-grained descriptions of underlying non-singular, horizon-free
states.  We discuss how these results arise for states in Type II string theory
on AdS${}_5\times$S${}^5$ and on AdS${}_3\times$S${}^3\times$T${}^4$ that
preserve 16 and 8 supercharges respectively.   We also discuss the connection
between ensembles of microstates and coarse-grained effective geometries.  Such
results suggest the idea, first put forward by Mathur and collaborators
\cite{Lunin:2001jy, Mathur:2005zp}, that {\it all} black hole geometries in
string theory, even those with finite horizon area, can be seen as the effective
coarse-grained descriptions of complex underlying horizon-free
states\footnote{The idea here is that a single microstate does not have an
entropy, even if its coarse-grained description in gravity has a horizon.  Thus
the spacetime realization of the microstate, having no entropy, should be in
some sense horizon-free, even though the idea of a horizon, or even a geometry,
may be difficult to define precisely at a microscopic level.}   which have an
extended spacetime structure.   This idea seems initially unlikely because one
might expect that  the quantum effects that correct the classical black hole
spacetime would be largely confined to regions of high curvature near the
singularity, and would thus not modify the horizon structure.  To study this we
examine states of M-theory on AdS${}_3\times$S${}^2\times$CY with 4
supercharges, where a finite horizon area can arise.    We work with a large
class of these states whose spacetime descriptions are amenable to study using split
attractor flows and some of which give rise to ``long throats'' of the kind needed to give
effective black hole behavior.  These states are related to distributions of
D-branes in spacetime.  Surprisingly, it turns out that the quantized solution
space has large fluctuations even at macroscopic proper distances, suggesting
that the scale of quantum effects in gravity could extend beyond the vicinity of
singularities in the effective theory.  Thus, the idea that all black holes
might simply be effective descriptions of underlying horizon-free objects
tentatively survives this test.

\subsection{Background}

In string theory, black holes can often be constructed by wrapping $D$-branes on
cycles in a compact manifold $X$ so they appear as point like objects in the
spatial part of the non-compact spacetime, $\mathbb{R}^{1,d-1}$.  As the string
coupling is increased, these objects backreact on spacetime and can form
supersymmetric spacetimes with macroscopic horizons.  The entropy associated
with these objects can be determined ``microscopically'' by counting BPS states
in a field theory living on the branes and this has been shown in many cases to
match the count expected from the horizon area (see \cite{Strominger:1996sh,
Maldacena:1997de} for the prototypical calculations).  Although the field theory
description is only valid for very small values of the string coupling $g_s$ the
fact that the entropy counting in the two regimes coincides can be attributed to
the protected nature of BPS states that persist in the spectrum at any value of
the coupling unless a phase transition occurs or a wall of stability is
crossed\footnote{It is also possible that BPS states pair up and get lifted from
the spectrum.  For the systems considered in these notes, however, this does not appear to
be an important phenomena.}.  The fact that the (leading) contribution
to the entropy of the black hole could be reproduced from counting states in a
sector of the field theory suggests that the black hole microstates dominate the
entropy in this sector.

While it is very helpful that these states can be counted at weak coupling,
understanding the nature of these states in gravity at finite coupling remains
an open problem.  As $g_s$ is increased the branes couple to gravity and we
expect them to start backreacting on the geometry.  The main tools we have to
understand the spacetime or closed-string picture of the system are the AdS/CFT
correspondence and the physics of D-branes.

Within the framework of the AdS/CFT correspondence black holes with near horizon
geometries of the form AdS$_m\times{\mathcal M}$ must correspond to objects in a dual
conformal field theory that have an associated entropy\footnote{More generally
objects in AdS with horizons, microscopic or macroscopic, are expected to have an
associated entropy which should manifest itself in the dual CFT.}.  A natural
candidate is a thermal ensemble or density matrix, in the CFT, composed of
individual pure states (see e.g.  \cite{Maldacena:1998bw}).  AdS/CFT then
suggests that there must be corresponding pure states in the closed string
picture and that these would comprise the microstates of the black hole.  It is
not clear, however, that such states are accessible in the supergravity
description.  First, the dual objects should be closed string {\em states} and
may not admit a classical description.  Even if they do admit a classical
description they may involve regions of high curvature and hence be inherently
stringy.  For BPS black holes\footnote{Here ``BPS'' can mean either 1/2, 1/4 or
1/8 BPS states or black holes in the full string theory. The degree to which
states are protected depends on the amount of supersymmetry that they preserve
and our general remarks should always be taken with this caveat.}, however, we
may restrict to the BPS sector in the Hilbert space where the protected nature
of the states suggests that they might persist as we tune continuous parameters
(barring phase transitions or wall crossings). We may then hope to see a supergravity manifestation of these
states, and indeed this turns out to be the case for systems with sufficient
supersymmetry.  However, the large $N$ limit\footnote{$N$ measures the size of
the system. For black holes it is usually related to mass in the bulk and
conformal weight in the CFT.}, which must be taken for supergravity to be a
valid description, bears many similarities with the $\hbar \rightarrow 0$ limit
in quantum mechanics where we know that most states do not have a proper
classical limit.  As we will see, if a supergravity description can be obtained
at all, it will only be for appropriately ``semiclassical'' or ``coherent''
states.

Despite these
potential problems, recently, a very fruitful program has been undertaken to
explore and classify the smooth supergravity duals of coherent CFT states in the
black hole ensemble.  Smoothness here is important because if these geometries
exhibit singularities we expect these to either be resolved by string-scale
effects, making them inaccessible in supergravity, or enclosed by a horizon
implying that the geometry corresponds, not to a pure state, but rather an
ensemble with some associated entropy.

Large classes of such smooth supergravity solutions, asymptotically
indistinguishable from black hole\footnote{Throughout this paper we will be
discussing ``microstates'' of various objects in string theory but the objects
will not necessarily be holes (i.e. spherical horizon topology) nor will they
always have a macroscopic horizon.  In fact, there is no 1/2 BPS solution in
AdS$_5\times$S$^5$ with any kind of a horizon.  We will, none-the-less, somewhat
carelessly continue to refer to these as ``microstates'' of a black hole for the
sake of brevity.}  solutions, have indeed been found \cite{Lin:2004nb,
Lunin:2002iz, Lunin:2001fv,
Bena:2005va,Bena:2004de,Berglund:2005vb,Balasubramanian:2006gi,Lunin:2004uu}
(and related \cite{Gaiotto:2005gf,Cheng:2006yq} to previously known black hole
composites \cite{Behrndt:1997ny,Denef:2002ru,Bates:2003vx}).  These are complete
families of solutions preserving a certain amount of supersymmetry with fixed
asymptotic charges\footnote{The question of which asymptotic charges of the
microstates should match those of the black hole is somewhat subtle and depends
on which ensemble the black hole is in.  In principle some of the asymptotic
charges might be traded for their conjugate potentials.  Moreover, the solutions
will, in general, only have the same isometries
asymptotically.\label{footnote_charges}} and with no (or very mild)
singularities.

In constructing such solutions it has often been possible to start with a
suitable probe brane solution with the correct asymptotic charges in a flat
background and to generate a supergravity solution by backreacting the probe
\cite{Lin:2004nb, Bena:2005va,Emparan:2001ux}.    In a near-horizon limit these
back-reacted probe solutions are asymptotically AdS, and by identifying the
operator corresponding to the probe and the state it makes in the dual CFT, the
backreacted solution can often be understood as the spacetime realization of a
coherent state in the CFT.  Lin, Lunin and Maldacena  \cite{Lin:2004nb} showed
that the back-reaction of such branes (as well their transition to flux) was
identified with a complete set of asymptotically AdS$_5$ supergravity solutions
(as described above) suggesting that the latter should be related to 1/2 BPS
states of the original $D3$ probes generating the geometry.  Indeed, in
\cite{Grant:2005qc, Maoz:2005nk} it was shown that quantizing the space of such
supergravity solutions as a classical phase space reproduces the spectrum of BPS
operators in the dual $\mathcal{N}=4$ superconformal Yang-Mills (at
$N\rightarrow\infty$).

In a different setting Lunin and Mathur \cite{Lunin:2001fv} were able to
construct supergravity solutions related to configurations of a $D1$-$D5$ brane in six
dimensions (i.e. compactified on a $T^4$) by utilizing dualities that relate
this system to an $F1$-$P$ system (see also \cite{Lunin:2002iz}). The latter
system is nothing more than a BPS excitation of a fundamental string quantized
in a flat background.  The back reaction of this system can be parametrized by a
profile $F^i(z)$ in $\mathbb{R}^4$ (the transverse directions). T-duality
relates configurations of this system to that of the $D1$-$D5$ system.

Recall that the naive back-reaction of a bound state of $D1$-$D5$ branes is a
singular or ``small'' black hole in five dimensions.  The geometries arising
from the $F1$-$P$ system, on the other hand, are smooth  after dualizing back to
the D1-D5 frame, though they have the same asymptotics as the naive solution
\cite{Lunin:2002iz}. Each $F1$-$P$ curve thus defines a unique supergravity
solution with the same asymptotics as the naive $D1$-$D5$ black hole but with
different subleading structure.  Smoothness of these geometries led Lunin and
Mathur to propose that these solutions should be mapped to individual states of
the $D1$-$D5$ CFT.   The logic of this idea was that {\it individual}
microstates do not carry any entropy, and hence should be represented in
spacetime by configurations without horizons.  Lunin and Mathur also conjectured
that the naive black hole geometry is somehow a coarse graining over all these
smooth solutions, i.e. that the black hole itself is simply an effective,
coarse-grained description. This  idea sometimes goes under the name of {\it{the
fuzz ball proposal}}.

The focus of these notes will be
\cite{Balasubramanian:2005mg, Alday:2005xj, Alday:2006nd, deBoer:2008fk, deBoer:2008zn, Balasubramanian:2007zt, Balasubramanian:2005qu, Balasubramanian:2007hu, Balasubramanian:2007qv, Balasubramanian:2006jt} which use well-controlled supersymmetric examples to explore the idea that black holes might be simply ``effective geometries'', i.e. that they are effective coarse-grained descriptions of underlying horizon-free objects.  This will also involve
understanding the nature of typical black hole microstates and how
they may be resolved by probes \cite{Balasubramanian:2005mg}.
The
discussion will involve  1/2
BPS states in AdS$_5\times$S$^5$,  1/4 BPS states of the
$D1$-$D5$ system, and the least controlled 1/8 BPS case where we
will study bound multicenter configurations in four and five
dimensions.  Only in the final case will genuine macroscopic
horizons be possible but the 1/2 and 1/4 BPS cases are under more
technical control and hence important to study.  In all cases we
will try to understand how the BPS spectrum emerges in
supergravity, how it is related to the BPS spectrum of a dual CFT (or
more generally the brane theory) and how such states might
contribute to the ensembles characterizing black holes in string
theory.  Much of what will be discussed is a review of work by
other authors, including \cite{Lin:2004nb, Lunin:2001jy, Grant:2005qc, Bena:2005va, Bates:2003vx, Berglund:2005vb, Lunin:2004uu}.

As the related literature is voluminous and complicated we attempt to
provide, in section \ref{sec_lit}, a brief survey of the various works and which
branches of the field they fit into.  This survey is by no means exhaustive and no
doubt neglects many important works but we feel it may, nonetheless, serve as a
useful reference for readers attempting to orient themselves within the field.

\subsection{Some answers to potential objections}\label{sec_obj}

The idea that black holes are simply effective descriptions of underlying
horizon-free objects is confusing because it runs counter to well-established
intuitions in effective field theory, most importantly the idea that near the
horizon of a large black hole the curvatures are small and hence so are the
effects of quantum gravity.   Indeed, it is not easy to formulate a precisely
stated conjecture for black holes with finite horizon area, although for
extremal black holes with enough supersymmetry a substantial amount of evidence
has accumulated for the correctness of the picture, as reviewed in this article.
To clarify some potential misconceptions, we transcribe below a dialogue between the
authors, addressing some typical objections and representing our current point
of view.   Also see \cite{Mathur:2005zp, Mathur:2008wi, MathurFAQ}.

\begin{enumerate}
    \item How can a smooth geometry possibly correspond to a ``microstate'' of a black
        hole?\label{q_microstate}\\
        \\
		Smooth geometries do {\em not} exactly correspond to states.  Rather, as
		classical solutions they define points in the phase space of a theory
		(since a coordinate and a momenta define a history and hence a solution;
		see section \ref{phase_quant} for more details) which is isomorphic to
		the solution space.  In combination with a symplectic form, the phase
		space defines the Hilbert space of the theory upon quantization.  While
		it is not clear that direct phase space quantization is the correct way
		to quantize gravity in its entirety this procedure, when applied to the
		BPS sector of the theory, seems to yield meaningful results that are
		consistent with AdS/CFT.  
		
		As always in quantum mechanics, it is not possible to write down a state
		that corresponds to a point in phase space. The best we can do is to
		construct a state which is localized in one unit of phase space volume
		near a point. We will refer to such states as coherent states. Very
		often (but not always, as we will see later in these notes) the limit in
		which supergravity becomes a good approximation corresponds exactly to
		the classical limit of this quantum mechanical system, and in this limit
		coherent states localize at a point in phase space. It is in this sense,
		and only in this sense, that smooth geometries can correspond to
		microstates. Clearly, coherent states are very special states, and a
		generic state will {\em not} admit a description in terms of a smooth
		geometry.  

    \item How can a finite dimensional solution space provide an exponential
        number of states?\\
        \\
        The number of states obtained by quantizing a given phase
        space is roughly given by the volume of the phase space as
        measured by the symplectic form $\omega$, $N\sim\int
        \omega^k/k!$ for a $2k$-dimensional phase space. Thus, all
        we need is an exponentially growing volume which is
        relatively easy to achieve.

    \item Why do we expect to be able to account for the entropy of the black
        hole simply by studying smooth supergravity solutions?\\
        \\
        Well, actually, we do not really expect this to be true.
        In cases with enough supersymmetry, one does recover all
        BPS states of the field theory by quantizing the space of
        smooth solutions, but there is no guarantee that the same
        will remain true for large black holes, and the available
        evidence does not support this point of view. We do
        however expect that by including stringy degrees of
        freedom we should be able to accomplish this, in view of
        open/closed string duality.

    \item If black hole ``microstates'' are stringy in nature then what is the
        content of the ``fuzzball proposal''?\\
        \\
        The content of the fuzzball proposal  is
        that the closed string description of a generic microstate of a black hole,
        while possibly highly stringy and quantum in nature, has interesting structure that
        extends all the way to the horizon
        of the naive black hole solution, and is well approximated by the black hole geometry
        outside  the horizon.

        More precisely the naive black hole solution is argued to correspond to
        a thermodynamic ensemble of pure states.  The {\em generic} constituent
        state will not have a good geometrical description in classical
        supergravity; it may be plagued by regions with string-scale curvature
        and may suffer large quantum fluctuations.  These, however, are not
        restricted to the region near the singularity but extend all the way to
        the horizon of the naive geometry.  This is important as it might shed light
        on information loss via Hawking radiation from the horizon as near
        horizon processes would now encode information about this state that, in
        principle, distinguish it from the ensemble average.

    \item Why would we expect string-scale curvature or large quantum
        fluctuations away from the singularity of the naive black hole solution?
        Why would the classical equations of motion break down in this regime?\\
        \\
        As mentioned in the answer to question \ref{q_microstate}, it is not always true that a
        solution to the classical equations of motion is well described by a coherent state,
        even in the supergravity limit. In particular there may be some regions of phase
        space where the density of states is too low to localize a coherent
        state at a particular point.  Such a point, which can be mapped to a
        particular solution of the equations of motion,
        is not a good classical solution because the variance of any quantum 
        state whose expectation values match the solution will necessarily be
        large.

        Another way to understand this is to recall that the symplectic
        form effectively discretizes the phase space into
        $\hbar$-sized cells.  In general all the points in a given cell
        correspond to classical solutions that are essentially indistinguishable
        from each other at large scales.  It is possible, however, for a
        cell to contain solutions to the equation of motion that {\em do} differ from each other at very
        large scales.  Since a quantum state can be localized at most to one such
        cell it is not possible to localize any state to a particular point
        within the cell.  Only in the strict $\hbar \rightarrow 0$ limit will
        the cell size shrink to a single point suggesting there might be states
        corresponding to a given solution but this is just an artifact of the
        limit.  A specific explicit example of such a scenario is discussed
        later in these lecture notes.

        Thus, even though the black hole solution satisfies the classical
        equation of motion all the way to the singularity this does not
        necessarily imply that when quantum effects are taken into account that
        this solution will correspond to a good semi-classical state with
        very small quantum fluctuations.

    \item So is a black hole a pure state or a thermal ensemble?\\
        \\
        In a fundamental theory we expect to be able to describe a quantum system
        in terms of pure states.  This applies to a black hole as well.   At first glance, since the black hole carries an entropy, it should be associated to a thermal ensemble of microstates.  But, as we know from statistical physics, the thermal ensemble can be regarded as a technique for approximating the physics of the generic microstate in the microcanonical ensemble with the same macroscopic charges.   Thus, one should be able to speak of the black hole as a coarse grained effective description of a generic underlying microstate.
         Recall that a typical or generic state in an ensemble is
        very hard to distinguish from the ensemble average without doing
        impossibly precise microscopic measurements. The entropy
        of the black hole is then, as usual in thermodynamics, a measure of the
        ignorance of macroscopic observers about the
        nature of the microstate underlying the black hole.

    \item What does an observer falling into a black hole see?\\
        \\
        This is a difficult question which cannot be answered at present.
        The above picture of a black hole does
        suggest that the observer will gradually thermalize once
        the horizon has been passed, but the rate of
        thermalization remains to be computed. It would be
        interesting to do this and to compare it to recent
        suggestions that black holes are the most efficient
		scramblers in nature \cite{Hayden:2007cs,Kim:2007vx,Sekino:2008he}.

    \item Does the fuzzball proposal follow from AdS/CFT?\\
        \\
        As we have defined it the fuzzball proposal does not follow from
        AdS/CFT. The latter is obviously useful for many purposes.
        For example, given a state or density matrix, we can try
        to find a bulk description by first computing all
        one-point functions in the state, and by subsequently
        integrating the equations subject to the boundary
        conditions imposed by the one-point function. If this bulk
        solution is unique and has a low variance (so that it
        represents a good saddle-point of the bulk path integral)
        then it is the right geometric dual description. In
        particular, this allows us to attempt to find geometries dual to
        superpositions of smooth geometries. What it does not do
        is provide a useful criterion for which states have good
        geometric dual descriptions; it is not clear that there is
        a basis of coherent states that all have decent dual
        geometric descriptions, and it is difficult to determine
        the way in which bulk descriptions of generic states
        differ from each other. In particular, it is difficult to
        show that generic microstates have non-trivial structure
        all the way up to the location of the horizon of the
        corresponding black hole.

	\item To what degree does it make sense to consider quantizing a (sub)space
	  of supergravity solutions? \\
        \\
		In some instances a subspace of the solution space corresponds to a well
		defined symplectic manifold and is hence a phase space in its own right.
		Quantizing such a space defines a Hilbert space which sits in the larger
		Hilbert space of the full theory.  Under some favourable circumstances
		the resulting Hilbert space may be physically relevant because a
		subspace of the total Hilbert space can be mapped to this smaller
		Hilbert space.  That is, there is a one-to-one map between states in the
		Hilbert space generated by quantizing a submanifold of the phase space
		and states in the full Hilbert space whose support is localized on this
		submanifold.  
		
		For instance, in determining BPS states we can imagine
		imposing BPS constraints on the Hilbert space of the full theory,
		generated by quantizing the full solution space, and expect that the
		resulting states will be supported primarily on the locus of points that
		corresponds to the BPS phase space; that is, the subset of the solution
		space corresponding to classical BPS solutions.  It is therefore
		possible to first restrict the phase space to this subspace and then
		quantize it in order to determine the BPS sector of the Hilbert space.

\end{enumerate}

\section{States, Geometries, and Quantization}\label{sec_rel_microstates}

\begin{figure}[htp]
    \includegraphics[width=140mm,angle=-90]{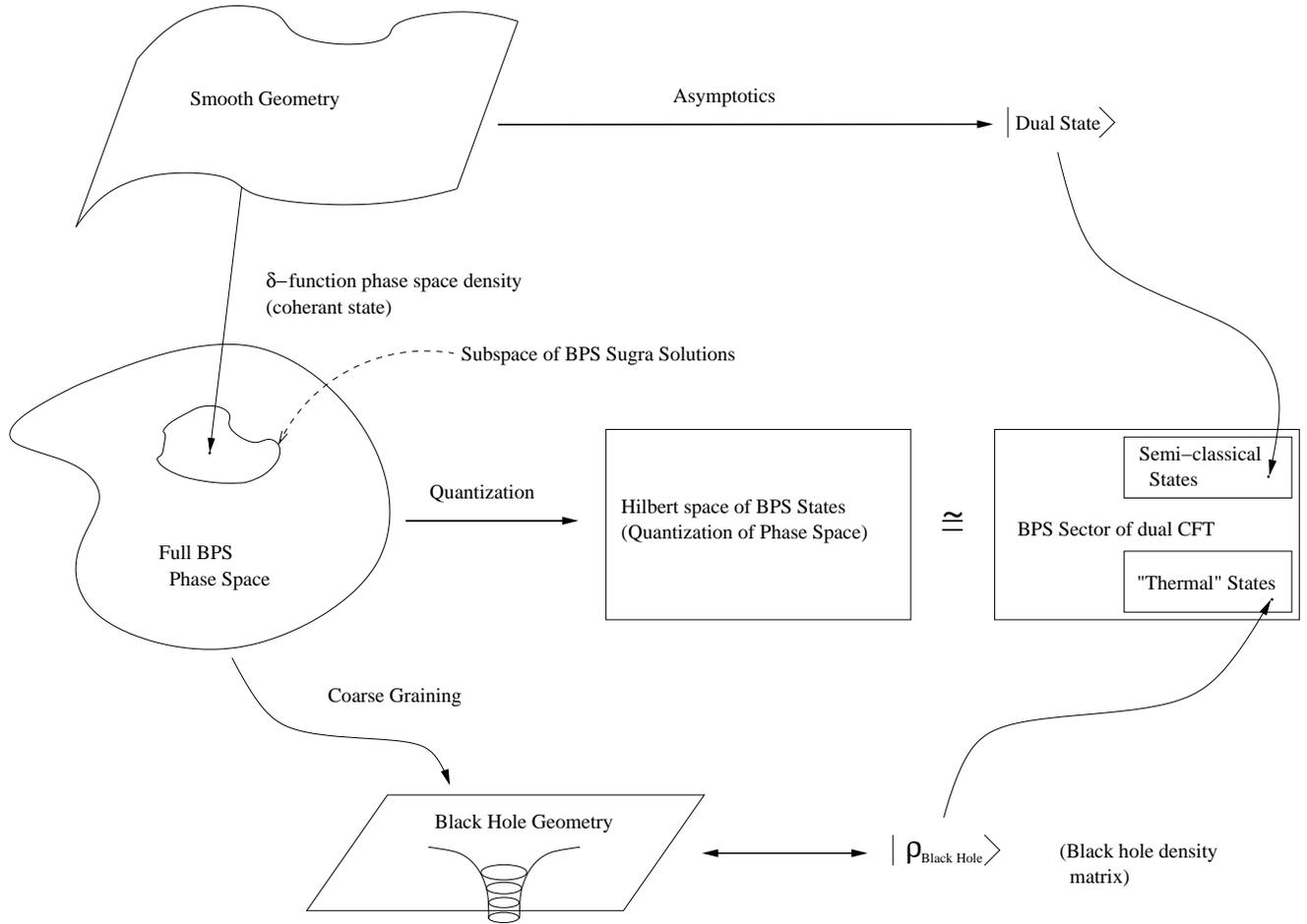}
    \caption{Relationship between various components appearing in the study of
    black hole microstates and black holes as effective geometries.  
    The smooth geometries making up the phase space can
    be thought of either as classical solutions defining a solution space
    (isomorphic to a phase space) or as highly-localized phase space densities
    corresponding to coherent states.  The black hole at the bottom of the
    figure is then to be generated by coarse graining (in some suitable
    sense) over a large number of underlying horizon-free configurations; the resultant geometry need
    not be a black hole but may, for instance, include a naked singularity.
      The details can differ significantly between examples -- the quantization, for instance, is rather different for the  1/4 and 1/8 BPS case.  The ``thermal states'' in the BPS sector box refer to ensembles with any chemical potential that couples to an operator that commutes with supersymmetry, and thus acts within the BPS Hilbert space (for example the left-moving temperature in a 2d CFT while the right-movers are kept in their ground state).
       }\label{fig_rel}
\end{figure}

Throughout these notes we will be exploring the relationship between
families of smooth supergravity solutions and coherent microstates with the charges 
of supersymmetric black holes in string theory.   The logic of this relationship
is  illustrated in Fig.~\ref{fig_rel}.

The first important component is a (complete) family of
supergravity solutions preserving the same supersymmetry and with
the same asymptotic conserved charges (see footnote
\ref{footnote_charges}) as a BPS black hole.  
 As solutions they can be related to points in a phase space and, as a family,
 they define a submanifold of the full phase space (this notion is elaborated
 upon in the next section and references therein).  Because they are BPS they
 can be argued to generate a proper decoupled phase space of their
 own\footnote{Though these arguments are for the 1/2 and 1/4 BPS case they
 should extend to 1/8 BPS bearing in mind possible discontinuities in the
 spectrum at walls of marginal stability.} \cite{Grant:2005qc,Maoz:2005nk}, at
 least for the purpose of enumerating states. Indeed, in the cases we consider,
 one can check that the restriction of the symplectic form to this space is
 non-degenerate implying that the space is actually a symplectic
 submanifold\footnote{As noted in \cite{Maoz:2005nk} this is related to the fact
 that the solutions are stationary but not static so the momenta conjugate to
 the spatial components of the metric are non-vanishing.}.  For a thorough discussion of the subtleties involved in this ``on-shell
 quantization'' the reader is referred to \cite[Sec.  2.7-2.10]{Maoz:2005nk}.
 Quantizing the space of such solutions as a phase space yields a Hilbert space
 populated by putative BPS microstates of the black hole.  In
 \cite{Grant:2005qc} this was done for 1/2 BPS geometries asymptotic to
 AdS$_5\times$S$^5$ and found to reproduce the 1/2 BPS spectrum of the dual CFT.  In general, though, we would expect that supergravity alone would not suffice and that stringy degrees of freedom would be necessary to reproduce the complete spectrum of BPS states.

The states arising from quantizing only smooth supergravity solutions, without
any stringy degrees of freedom, should thus correspond to some restricted
subspace of the full BPS Hilbert space of the theory (which corresponds to the
BPS sector of the CFT).  We would like to understand how these pure states
relate to black holes.  The latter have entropy so we expect them to be dual to
density matrices in the CFT.  The ensemble of states making up such a density
matrix should be a suitable thermodynamic average over pure states in the same
sector of the Hilbert space as the black hole.  As our ``microstates'' arise
from geometries that preserve the same supersymmetries and carry the
same asymptotic charges as the black hole, they provide suitable candidates for
states in the ensemble.   However, it is  likely that the generic state in this ensemble
will not be semi-classical and thus will only be accessible in the CFT or by directly
quantizing the BPS phase space.  It is also not clear whether a large portion
of the states in the black hole ensemble, or indeed any typical states at all, arise from quantizing only
supergravity modes.  As mentioned above this has proven to be the case when
sufficient supersymmetry is present but it is not clear if it will continue to do
so for less supersymmetric black holes.  Nevertheless, as the states arising from supergravity
are relatively accessible (compared to more stringy states) it is important to
explore their structure and also to determine how they are related to the
typical states making up the black hole ensemble.

In some cases, there is reason to believe that the Hilbert space enjoys a more
refined decomposition in subsectors than just by macroscopic quantum numbers
alone.  In \cite{Denef:2000nb, Denef:2000ar,Denef:2007vg}, for instance, a
decomposition based on split attractor trees is conjectured (this will be
discussed further in section \ref{sec_split_attr}). In these cases it is
possible that of all the states with the correct asymptotic charges the black
hole ensemble will only include microstates from a single subsector.  In
\cite{deBoer:2008fk} another (related) constraint on the constituents of the
black hole ensemble was found.  There it was argued that any geometries that do
not survive a near-horizon decoupling limit should not contribute states to the
black hole ensemble because they do not correspond to bona fide bound
configurations of the original $D$-brane system generating the black hole.

In order to study the spacetime structure of the microstates of a black hole it
is desirable to have an inverse map between the states in the CFT and classical
geometries.  Of course this map is not injective as many states in the CFT are
not semi-classical so we would also like a criteria for determining which such
states yield good classical geometries and which yield geometries with large
quantum fluctuations.  Possible criteria were discussed in
\cite{Balasubramanian:2007zt, Larjo:2007zu} and will play a role in some of the
arguments that follow.  The point of view that we would like to assume is based
on the need for a classical observer to measure the system
\cite{Balasubramanian:2005mg}.  Thus we would like to identify a set of
operators, $\mathcal{O}_\alpha$, in the CFT that are dual to ``macroscopic
observables''.  The requirement that a state yield a good classical geometry can
be translated into a constraint on the variance of the expectation values for
these observables in the semi-classical limit.

\subsection{Phase Space Quantization}\label{phase_quant}

The space of classical solutions of a theory is generally isomorphic to its
classical phase space\footnote{It is not entirely clear, however, which
solutions should be included in defining the phase space.  For instance, in
treating gravity, it is not clear if trajectories which eventually develop
singularities should also be included as points in the phase space or only
solutions which are eternally smooth.  As we will primarily be concerned with
static or stationary solutions in these notes we will largely avoid this
issue.}.  Heuristically, this is because a given point in the phase space,
comprised of a configuration and associated momenta, can be translated into an
entire history by integrating the equations of motion against this initial data;
likewise, by fixing a spatial foliation, any solution can translated into a
unique point in the phase space by extracting a configuration and momentum from
the solution evaluated on a fixed spacial slice.  This observation can be used
to quantize the theory using a symplectic form, derived from the Lagrangian, on
the space of solutions rather than on the phase space.  This is an old idea
\cite{Dedecker:1953} (see also \cite{Raju:2007uj} for an extensive list of
references and \cite{Crnkovic:1986ex} \cite{Zuckerman:1989cx} \cite{Lee:1990nz}
for more recent work) which was used in \cite{Grant:2005qc,Maoz:2005nk}
\cite{Donos:2005vs,Rychkov:2005ji} to quantize the LLM \cite{Lin:2004nb} and
Lunin-Mathur \cite{Lunin:2001fv} geometries.   An important subtlety in these
examples is that it is not the entire solution space which is being quantized
but rather a subspace of the solutions with a certain amount of supersymmetry.

In general, quantizing a subspace of the phase space will not yield the correct
physics as it is not clear that the resultant states do not couple to states
coming from other sectors.  It is not even clear that a given subspace will be a
symplectic manifold with a non-degenerate symplectic pairing.  As discussed in
\cite{Maoz:2005nk} we expect the latter to be the case only if the subspace contains
dynamics; for gravitational solutions we thus expect stationary solutions, for
which the canonical momenta are not trivial, to possibly yield a non-degenerate
phase space.  This still does not address the issue of consistency as states in
the Hilbert space derived by quantizing fluctuations along a constrained
submanifold of the phase space might mix with modes transverse to the
submanifold.  When the submanifold corresponds to the space of BPS solutions one
can argue, however, that this should not matter.  The number of BPS states is
invariant under continuous deformations that do not cross a wall of marginal
stability or induce a phase transition.  Thus if we can quantize the solutions
in a regime where the interaction with transverse fluctuations is very weak then the
energy eigenstates will be given by perturbations around the states on the BPS
phase space, and, although these will change character as parameters are varied
the resultant space should be isomorphic to the Hilbert space obtained by
quantizing the BPS sector alone.  If a wall of marginal stability is crossed states
will disappear from the spectrum but there are tools that allow us to analyze
this as it occurs (see section \ref{sec_split_attr}).

Let us emphasize that the validity of this decoupling argument depends on what
questions one is asking.  If we were interested in studying dynamics then we
would have to worry about how modes on the BPS phase space interact with
transverse modes.  For the purpose of enumerating or determining general
properties of states, however, as we have argued, it should be safe to ignore
these modes.  For an example of the relation between states obtained by
considering the BPS sector of the fully quantized Hilbert space and the states
obtained by quantizing just the BPS sector the phase space see
\cite{Denef:2002ru} and \cite{deBoer:2008zn}.

As mentioned, the LLM and Lunin-Mathur geometries have already been quantized
and the resultant states were matched with states in the dual CFTs.  We will
have occasion to mention this briefly in the sequel but we will ultimately focus
on the quantization of $\mathcal{N} = 2$ solutions in four (or $\mathcal{N}=1$
in five) dimensions.  For such solutions, although a decoupling limit has been
defined \cite{deBoer:2008zn}, the dual $\mathcal{N} = (0, 4)$ CFT is rather
poorly understood.  Thus quantization of the supergravity solutions may yield
important insights into the structure of the CFT and will be important in
studying the microstates of the corresponding extremal black objects.

\subsection{Black holes, AdS throats and dual CFTs}
\label{sec22}

One of the most powerful tools to study properties of black holes in string
theory is the AdS/CFT correspondence \cite{Maldacena:1997re}. This conjecture
relates string theory on backgrounds of the form AdS$_{p+1} \times \mathcal{M}$
to a CFT$_p$ that lives on the boundary of the AdS$_{p+1}$ space.  Such
backgrounds arise from taking a particular decoupling limit of geometries
describing black objects such as black holes, black strings, black tubes, etc.
This limit amounts to decoupling the physics in the near horizon
region\footnote{In some of the cases treated in these notes the region will not
be an actual near-horizon region as the original solutions may be horizon-free
(or, in some cases, may have multiple horizons) but the decoupling limits are
motivated by analogy with genuine black holes where the relevant region is the
near horizon one.} of the black object from that of the asymptotically flat
region by scaling the appropriate Planck length, $l_p$, to decouple the
asymptotic gravitons from the bulk. At the same time one needs to scale
appropriate spatial coordinates with powers of $l_p$ to keep the energies of
some excitations finite. This procedure should be equivalent to the field theory
limit of the brane bound states generating the geometry under consideration.

We are interested in black objects which describe
normalizable deformations in the AdS$_{p+1}$ background. These correspond to a state/density matrix on the dual CFT according to the following
dictionary

\begin{center}
   \begin{tabular}{|c|c|}
     \hline
      \bf{BULK} & \bf{BOUNDARY}\\
   \hline
      $\exp{\left( - S^{\text{on shell}}_{\text{bulk}} \right)}$ & Tr$[\rho \mathcal{O}_1 \ldots \mathcal{O}_n] = \left< \mathcal{O}_1 \ldots \mathcal{O}_n \right>_{\rho}$ \\
      classical geometries   & semiclassical states ? \\
      black hole  & $\rho \sim \exp\{- \sum_i \beta_i
      \mathcal{O}_i \}$ \\
      entropy S &  S = $-$ Tr$(\rho \log \rho)$ \\
      bulk isometry D  & $\left[ \rho, \hat{D} \right] = 0$ \\
      ADM quantum numbers of D & Tr$\left(\rho \hat{D}\right) = \left< \hat{D} \right> = D_{\text{ADM}}$ \\
    \hline
   \end{tabular}
\end{center}
In the first line $\mathcal{O}_i$ are operators dual to sources turned on in the
boundary. They are included in the bulk calculation of $\left( -
S^{\text{on shell}}_{\text{bulk}} \right)$. The second line can be
seen as the definition of the dual semiclassical state.   More specifically, a semiclassical state is the one that has an unambiguous dual bulk geometry (i.e. that in the classical limit ($N\rightarrow \infty$ and $\hbar \rightarrow 0$) macroscopic observables take on a fixed
expectation value with vanishing variance).  In some ideal
situations such semiclassical states turn out to be the analog of coherent states in the harmonic oscillator.  In the third
line, we describe a typical form of a density matrix that we
expect to describe black holes. This form is motivated by the
first law of thermodynamics: the entropy as defined in the fourth
line obeys $dS=\sum_i \beta_i d \langle {\cal O}_i \rangle$, and
by matching this to the first law as derived from the bulk
description of the black hole we can identify the relevant set of
operators ${\cal O}_i$ and potentials $\mu_i$ and guess the
corresponding density matrix. The fourth line simply states that
we expect a relation between the bulk and boundary entropies. In
the fifth and the last line, $\hat{D}$ is the current/operator
dual to the bulk isometry D.

One question that we want to shed
some light on in these notes is; ``{\it Given a density matrix $\rho$ on
the CFT side, is there a dual geometry in the bulk?}''.    On general grounds
one could have expected that a general density matrix $\rho$ should be dual to a
suitably weighted sum over geometries, each of which could be singular, have
regions with high curvature, and perhaps not have good  classical limits. As a
result the dual gravitational description of a general density matrix will not
generally be trustworthy.  However, under suitable circumstances, it can happen
that there is a dual ``{\it effective}'' geometry that describes the density
matrix $\rho$ very well. This procedure of finding the effective geometry is
what we will call ``{\it coarse graining}''. In the gravity description, this
amounts to neglecting the details that a classical observer cannot access anyway
due to limitations associated to the resolution of their apparatus. So, one can
phrase our question in the opposite direction, ``{\it What are the
characteristics of a density matrix on the CFT side, so that there is a good
dual effective geometry that describes the physics accurately?}''.

One can try to construct the dual effective geometry following the
usual AdS/CFT prescription. To do so, one should first calculate
all the non vanishing expectation values of all operators dual to
supergravity modes (assuming one knows the detailed map between
the two). On the CFT side, these VEVs are simply given by
         $$ \langle\mathcal{O}_i\rangle = \mathrm{Tr} (\rho\;\mathcal{O}_i), $$
and they determine the boundary conditions for all the
supergravity fields. The next step is to integrate the gravity
equations of motion subject to these boundary conditions to get
the dual geometry. This is in principle what has to be done
according to AdS/CFT prescription. The problem with this
straightforward approach is that it is not terribly practical, and
we will therefore revert to a different approach\footnote{Though
it would be interesting to study in some detail the connection
between the two.}. Before describing various examples in more
detail, we first describe the main idea in general terms. We will
first start by describing the connection between quantum physics
and the classical phase space. After that, we are going to briefly
describe the philosophy behind constructing effective geometries.

\subsection{Phase Space Distributions}
\label{sec:phasespace}

To have an idea about what it means to average over ensembles of geometries, or
``coarse grain'' as we will refer to it, we need to understand some general
features of the bulk theory.  In general, we will assume that we are dealing
with a supergravity theory in the bulk\footnote{Although we would
like to extend this discussion to include stringy degrees of freedom (when
relevant) we do not, at present, have any control over the latter.}.  Recall that solutions to the supergravity equation of motion can be
associated with points in a phase space (see for example \cite{Lee:1990nz}). The
boundary theory, on the other hand, is generally studied as a quantum conformal
field theory. As a result we are looking for a map between quantum states (CFT)
and classical objects living in a phase space (bulk). A well known example of
such a map is the map between quantum states and their corresponding classical
phase space densities (see the review \cite{Hillery:1983ms} and references there
in to the original literature). A good guess then is that the map that we are
looking for should involve a ``dressed'' version of the phase space densities
 of quantum states in the CFT.   Let us pause for a
moment to discuss the idea of a phase space distribution \cite{Hillery:1983ms}. A
particle (or statistical system) in a quantum theory is described by giving its
density matrix $\rho$. The result of any measurement can be seen as an
expectation value of an appropriate operator which is given explicitly by

\begin{equation} \label{quant}
    \langle \mathcal{O} \rangle_\rho = \text{Tr} (\rho \; \mathcal{O})
\end{equation}

\noindent This is reminiscent of classical statistical mechanics where the
measurements are averages of appropriate quantities using some statistical
distribution

\begin{equation} \label{class}
    \langle \mathcal{O} \rangle_w = \int dp \, dq \; w(p,q) \; \mathcal{O}(p.q)
\end{equation}

\noindent where the integration is over the full phase space. One
can wonder at this point if it is possible to construct a density
$w(p,q)$ so that one can rewrite equation (\ref{quant}) as
equation (\ref{class}). The answer is affirmative: for every
density matrix $\rho$ there is an associated phase space
distribution $w_{\rho}$ such that for all operators $A$ the
following equality holds

\begin{equation}
     \int dp \, dq \; w_{\rho} (p,q) \; A(p,q) = \text{Tr}\left(\rho \; A(\hat{p}, \hat{q})\right)
\end{equation}

What about the uniqueness of $w_{\rho}$? Recall that in a quantum
theory we have to face the question of operator ordering. This
comes about because the operators $\hat{q}$ and their dual momenta
$\hat{p}$ don't commute with each other. This means that the
distribution $w_{\rho}$ should somehow include information about
the chosen order of $\hat{p}$ and $\hat{q}$. As a result there
does not exist a unique phase space distribution. For example, the
distribution corresponding to Weyl ordering is the Wigner
distribution, which is given by
\begin{equation}
  w(p,q) \sim \int dy \; \langle q-y|\rho|q+y \rangle \; e^{2 i p y}  \, .
\end{equation}
This distribution suffers from the fact that it is not positive
definite in general. It is also quite sensitive to the physics at
a quantum scale \cite{Hillery:1983ms,Balasubramanian:2005mg} as it
usually has fluctuations of order $\hbar$. Another drawback of
this distribution is that it is difficult to work with from a
computational standpoint. There is another commonly used
distribution which is positive definite: the Husimi distribution.
It is roughly the convolution of the Wigner distribution with a
Gaussian. This eliminates most of the fluctuations of order
$\hbar$. The price that one pays for this is that the resulting
operators must be anti-normal ordered. However, for semi-classical
states, which by definition are states for which the classical
limit is unambiguous, $w_{\rho} (p,q)$ should be independent of
the choice of ordering prescription in the classical limit as
well, so this is not actually much of a problem.

\subsection{Typical states versus coarse grained geometry} \label{typical/average}

Let us recapitulate what we have discussed so far and what the
missing steps are to achieve our goal. On the gravity side we have
geometries with certain asymptotics that in principle yield the
one point functions of the dual operators in the CFT. On top of
that we have in principle a way to quantize the reduced phase
space of solutions by using the induced symplectic form. On the
CFT side we can consider individual states or density matrices $\rho$ and find the
corresponding expectation values of operators. The only missing
ingredient is to construct the dual effective geometry.   To do so, we expect to use some sort of  ``dressed'' version of the CFT phase space densities described above.  But how do we carry out the coarse graining that is presumably needed to reach a description in a bulk classical geometry?  At this
stage we have two options: (i) we first select a typical
representative from the CFT ensemble of states and then map this
typical state directly to geometry, or (ii) we somehow average
over all the geometries dual to the states in the ensemble.

\subsubsection{Typical states/geometries}

A typical state in an ensemble is one for which the expectation
values of macroscopic observables agree to within the observable
accuracy with the average of the observable in the entire
ensemble. Obviously, this definition depends on the appropriate
notions of macroscopic observables and observable accuracy, but in
the examples we describe we will usually have a reasonably
educated guess regarding what the typical states are. Given a typical state,
we can try to map it directly to a solution of supergravity (this
may still be a formidable task), after which one still needs to
verify that the resulting geometry has no pathologies. This
approach was followed for example in \cite{Mathur:2005zp, Balasubramanian:2005mg}.

\subsubsection{Average/coarse graining}

Alternatively, we can try to average over states and geometries directly. On the
CFT side this is trivial since it essentially involves constructing a density
matrix and following the usual rules of quantum mechanics.  But the coarse
graining procedure is difficult to implement on the bulk side because gravity
is a non-linear theory.  However, in all examples that we will study, the
equations of motion of supergravity in the BPS sector will effectively be
linearized, which allows us to solve the equations in terms of harmonic
functions with sources. In addition, the space of solutions will be in
one-to-one correspondence with distributions of the sources.  This immediately
suggests a suitable coarse-graining procedure: we simply smear the harmonic
functions against the phase space density which describes the density matrix in
question. This approach was followed for example in \cite{Alday:2006nd}.
It would be
interesting to explore in more detail whether this method gives rise to the
appropriate averaging of the one-point functions, and to what extent it agrees
with the approach based on typical states that we described in the previous
paragraph.

\subsection{Entropic suppression of variances}
\label{sec:entropicsupp}

One might object any picture where the microstates of a black hole are individually realized in spacetime as extended horizon-free bound states  would inevitably lead to radically different results for probe measurements.  If this were so, then the usual black hole could not be a good effective description, and there would be a massive violation of our usual expectations from effective field theory.  Fortunately, one can show that in {\it any} scenario where the entropy of a black hole has a statistical interpretation in terms of states in a microscopic Hilbert space, the variance of finitely local observables over the Hilbert will be suppressed by a power of $e^{-S}$ \cite{Balasubramanian:2007qv}.  

To see this, consider a quantum mechanical Hilbert space of states with energy eigenvalues lying between $E$ and $E +\Delta E$ with a basis
\begin{equation}
{\cal M}_{bas} = \left\{ \  |s\rangle  \ : \ H |s \rangle = e_s |s \rangle ~~~;~~~ E \leq e_s \leq E + \Delta E \
\right\} \, .
\label{Eeigenstates}
\end{equation}
Thus this sector of the Hilbert space consists of states
\begin{equation}
{\cal M}_{sup}  =
\left\{|\psi\rangle = \sum_s c^\psi_s  |s\rangle
\right\} \, ,
\label{allstates}
\end{equation}
with $|s\rangle$ as in (\ref{Eeigenstates}) and $\sum_s |c_s|^2 =
1$.     The expectation value of the Hamiltonian $H$ in any state
in ${\cal M}_{sup}$ also lies between $E$ and $E+\Delta E$.  If entropy of
the system is $S(E)$, then the basis in (\ref{Eeigenstates}) has
dimension $e^{S(E)}$:
\begin{equation}
1+ \dim{{\cal M}_{sup}} = | {\cal M}_{bas} | = e^{S(E)}\, .
\label{numstates}
\end{equation}
Now take ${\cal O}$ to be any local operator and consider finitely local observables of the form
\begin{equation}
c_\psi = \langle \psi | {\cal O} | \psi \rangle  
\label{momentdef}
\end{equation}
We would like to measure how these observables vary over the ensemble
${\cal M}_{sup}$.   The ensemble averages of the observables (\ref{momentdef})
and their variances over the ensemble are given by
\begin{eqnarray}
\langle c\rangle_{{\cal M}_{sup}} &=& \int D\psi \, c_\psi \label{momentdef2} \\
{\rm var}[c]_{{\cal M}_{sup}}
&=& \int D\psi \, (c_\psi)^2 -
\langle c \rangle_{{\cal M}_{sup}}^2 \, .
\label{vardef2}
\end{eqnarray}
The differences between states in the ensemble of microstates in
their responses to local probes are quantified by the
standard-deviation to mean ratios
\begin{equation}
{\sigma[c]_{{\cal M}_{sup}} \over \langle c \rangle_{{\cal M}_{sup}}} =
{\sqrt{{\rm var}[c]_{{\cal M}_{sup}}} \over \langle c
\rangle_{{\cal M}_{sup}}} \, .
\label{proxydev}
\end{equation}
It was shown in \cite{Balasubramanian:2007qv} that 
\begin{equation}
{\rm var}[c]_{{\cal M}_{sup}} < {1 \over e^{S} + 1} {\rm var}[c]_{{\cal M}_{bas}}
\label{varav5}
\end{equation}
where the variance on the right hand side is computed just over a set of basis
elements while the variance on the left hand side is over the entire Hilbert
space.   This result follows because the generic state in the Hilbert space is a
random superposition of the form (\ref{allstates}), and in the computation of
correlation functions the phases in the coefficients $c_s^\psi$ lead to
cancellations.  Thus the only avenue to having a variance large enough to
distinguish microstates by defeating the $e^S$ suppression in (\ref{varav5}) is
to find probe operators that have exponentially large correlation functions.
Finitely local correlation functions in real time typically do not grow in this
way and hence, for black holes, with their enormous entropy, a semiclassical
observer will have no hope of telling microstates apart from each other.  This
is especially so because, as we will discuss in subsequent section, even the
elements of the basis of microstates (\ref{Eeigenstates}) for a black hole will
usually possess the property of typicality, namely that they will be largely
indistinguishable using coarse probes.

Thus, even if the microstates of a black hole are realized in spacetime as some
sort of horizon free bound states, finitely local observables with finite
precision, of the kind that are accessible to semiclassical observers, would
fail to distinguish between these states.   Indeed, the semiclassical observer,
having finite precision, might as well take an ensemble average of the
observables over the microstates, as this would give the same answer.    The
ensemble of microstates gives a  density matrix with entropy $S$, and will be
described in spacetime as a black hole geometry.  In this sense, the black hole
geometry will give the effective description of measurements made by
semiclassical observers.

\section{AdS$_5$xS$^5$}

We start with the best understood case, AdS$_5 \times$ S$^5$, whose
AdS/CFT dictionary is well developed. The dual CFT is $\mathcal{N}
= 4$ SU(N) super Yang-Mills where N is the number of D3 branes that
generate the geometry. Many supergravity solutions that asymptote
to AdS$_5 \times$ S$^5$ are known, including ones with 1/2, 1/4,
1/8 and 1/16 of the original supersymmetries preserved. Black
holes with a macroscopic horizon only exist either in the 1/16 BPS
case \cite{Gutowski:2004yv} or without any supersymmetry. The
latter include AdS-Schwarzschild black holes and some of their
qualitative properties can be reproduced from the dual CFT
\cite{Balasubramanian:2005mg}. However, we are going to restrict
ourselves to the 1/2-BPS case, where completely explicit
descriptions of both supergravity solutions \cite{Cvetic:1999xp,Behrndt:1998jd,Myers:2001aq,Lin:2004nb} as well as the CFT
states \cite{Balasubramanian:2001nh,Corley:2001zk,Berenstein:2004kk} are known.   Therefore, we can test the general philosophy that we have been advocating by examining the relationship between very heavy 1/2-BPS states in the CFT  (conformal dimension of $O(N^2)$)  \cite{Balasubramanian:2001nh,Corley:2001zk,Berenstein:2004kk}, smooth gravitational solutions with the same energy  \cite{Lin:2004nb}, and the 1/2-BPS extremal black hole in AdS$_{5}$ \cite{Myers:2001aq}.   Our exposition will be brief; see \cite{Balasubramanian:2005mg, Balasubramanian:2006jt, Balasubramanian:2007zt} for details and further references.

\subsection{The 1/2-BPS sector in field theory.}
\label{halfBPS}

The Hilbert space of 1/2-BPS states in $\mathcal{N}=4$ U(N) super
Yang-Mills is isomorphic to the Hilbert space of N fermions in a
harmonic oscillator potential as shown in
\cite{Corley:2001zk,Berenstein:2004kk}.   A basis for this Hilbert space can
be enumerated in terms of Young diagrams with N rows
as follows. The ground state is composed of fermions (labelled by
$i=1, ..., N$) with energies $E^g_i = [(i -1) + 1/2] \hbar$; this
is the Fermi sea of the system.\footnote{We set the frequency of the harmonic oscillator $\omega = 1$.} When we excite these fermions, the
energies become $E_i = (e_i + 1/2) \hbar$ for some positive
disjoint integers $e_i\geq i-1$. Because we are dealing with
fermion wave functions that are completely antisymmetrized we can
always order $\{e_i\}$ in a ascending order $e_1<e_2< ...<e_N$. As
a result, the numbers $r_i$ defined by
             $$ r_i = e_i -i + 1 $$
form a non-decreasing set of integers which can be encoded in a
Young diagram where $r_i$ describes the length of the $i$-th row.
It is convenient to also introduce variables $c_j$ which count the
number of columns of length $j$ \cite{Suryanarayana:2004ig}. They are related to the $r_i$ via
             $$ c_N = r_1, \;\;\;\;\; c_{N-i} =
              r_{i+1} - r_i, \;\;\;\;\; i = 1,2,...,(N-1) $$
and clearly
             $$r_{i+1} = e_{i+1} - i = c_{N-i} + ... + c_N. $$

A property of Young diagrams that will be useful later is that a
single Young diagram corresponds to a geometry with U(1) symmetry
in the bulk. This comes about because a single Young diagram is
associated to a density matrix of a pure state, i.e. of the form
             $$\rho = |\psi\rangle \langle \psi|,$$
             where
$|\psi\rangle$ has a fixed energy eigenvalue (simply given by the
total number of boxes). The density matrix therefore commutes with
the Hamiltonian which generates rotations in phase space, and
according to the table in section~\ref{sec22} the corresponding
supergravity solution should also possess this rotational
invariance.

\subsection{The typical very heavy state} 
\label{limitshape}

Large classical objects in AdS$_5$ have masses of order $N^2$, and hence 1/2-BPS
states that correspond to black hole microstates will have conformal 
dimensions of $O(N^2)$.  In the fermion language above, such states have a total
excitation energy of $O(N^2)$, or, equivalently, $O(N^2)$ boxes in the
corresponding Young diagram.   We would like to study such very heavy states in
the semiclassical limit ($\hbar \to 0$ with $\hbar N$ fixed) and understand
their relation to the 1/2-BPS black hole \cite{Myers:2001aq} and the smooth
1/2-BPS geometries \cite{Lin:2004nb}.

Highly excited states of the large N free fermion system can
reliably be discussed in a canonical ensemble in which temperature
rather than energy is held fixed\footnote{Often, it is also
interesting to study more general canonical ensembles where additional ``chemical potentials'' are included such as the angular velocity coupling to angular momentum, or electrostatic
potentials coupling to charges. We will see some examples in the
case of AdS$_3 \times$ S$^3$.   Also note the temperature here is not physical -- it is simply a Lagrange multiplier used to fix the energy of states included in this 1/2-BPS ensemble.}. Thus we write a partition function
\begin{equation}
Z =  \sum_{c_1, c_2,\dots ,c_N=1}^\infty
e^{-\tilde\beta \omega \sum_j j c_j}    = \prod_{j=1}^N {1 \over 1 - e^{-\beta j}}  \equiv
\prod_{j=1}^N {1 \over 1 - q^j} 
\end{equation}
where we dropped the zero point energy $e^{\tilde{\beta}\hbar\omega N^2/2}$ and set $\beta = \tilde{\beta} \hbar$ and $q = e^{-\beta}$.  By fixing 
\begin{equation}
\beta \sim {1\over \sqrt{\Delta}}
\end{equation}
one studies an ensemble of operators with typical conformal dimension of
$O(\Delta)$.   We will take $\Delta \sim N^2$ since we are interested in states
with energies in the range of the large extremal black hole \cite{Myers:2001aq}.
In this range the entropy of the ensemble scales as 
\begin{equation}
S \sim \sqrt{\Delta} \sim N
\end{equation}
which grows in the large $N$ limit, but not fast enough to give rise to a finite classical horizon area in the dual gravity picture.

 Using the canonical ensemble one
can easily calculate the average energy as well as the expectation
value and standard deviation of $c_i$:
\begin{equation}  
  \langle c_j \rangle = \frac{q^j}{1-q^j}~~~~;~~~~
  {\sigma(c_j) \over \langle c_j \rangle} =
\left( {1 \over \sqrt{ \langle c_j \rangle}} \right) {1 \over \sqrt{1 - q^j}} \, .
\end{equation}
 In the thermodynamic limit $N \gg 1$ where one
rescales the rows and columns of the Young diagram by factor of $\sqrt{N}$, the Young
diagram of a state in the ensemble approaches a certain  {\it limiting shape} with probability 1
\cite{vershik}. In other words, in the large N limit almost all
states/operators belonging to the canonical ensemble under study
will have associated Young diagrams that have vanishingly small
fluctuations around this limit curve. One can check this claim by evaluating the standard deviation to mean ratio above in the large $N$ limit \cite{Balasubramanian:2005mg}. The limit shape describes a
``{\it typical}'' Young diagram in this ensemble (see section
\ref{typical/average}).

To describe this limiting shape in some more detail, let us
introduce two coordinates $x$ and $y$ along the rows and columns
of the Young diagram. We adopt the convention where the origin
$(0,0)$ is the bottom left corner of the diagram, and $x$
increases going up while $y$ increases to the right. In fermion
language, $x$ labels the particle number and $y$ its excitation
above the vacuum. One then has
 \begin{equation}
 y(x) = \langle\text{\bf{y}}(x)\rangle = \sum_{i = N-x}^N \langle c_i \rangle 
 \label{limitcurve0}
 \end{equation}
In the large N limit, $x$ and $y$ can be treated as continuous
variables and the summation above becomes an integral. Since the
$c_i$ are independent random variables in the canonical ensemble,
it is straightforward to evaluate $y(x)$ explicitly, and one
obtains an equation for the limit shape of the form \cite{Balasubramanian:2005mg} 
\be
\label{limitcurve} (1-q^N)q^y + q^{N-x}=1 \, . \ee
In sum, in the semiclassical limit ($\hbar \to 0$ with $\hbar N$ fixed), nearly
all the elements of a basis of half-BPS states of ${\cal N} = 4$, $SU(N)$
Yang-Mills theory lie close to the typical state described by
(\ref{limitcurve}).   As we will see, because of the structural similarity of
all of these states to each other, a semiclassical observer will not be able to
tell them apart.

\subsection{The 1/2-BPS sector in supergravity}

All 1/2-BPS solutions in AdS$_5 \times $ S$^5$ supergravity are given by the LLM geometries \cite{Lin:2004nb}
\begin{align}
   ds^2 & = - h^{-2} (dt + V_i dx^i)^2 + h^2 (d \eta^2 + dx^i dx^i) + \eta e^G d\Omega_3^2 + \eta e^{-G} d \tilde{\Omega}_3^2  \nonumber \\
   &h^{-2} = 2 \eta \cosh G, \;\;\; \eta \partial_{\eta} V_i = \epsilon_{ij} \partial_j z, \;\;\; \eta(\partial_i V_j - \partial_j V_i) = \epsilon_{ij} \partial_{\eta} z  \nonumber \\
   &z = \frac{1}{2} \tanh G, \;\;\;\;\; z(\eta,x_1,x_2) = \frac{\eta^2}{\pi}
   \int dy_1 \; dy_2 \; \frac{\frac{1}{2} - u (0;y_1, y_2)}{[(\vec{x} - \vec{y})^2 - \eta^2]} \label{LLMgeom}
\end{align}
where $i = 1, 2$.   In these coordinates taken from \cite{Lin:2004nb}, $y_1$ and $y_2$ have dimensions of length {\it squared}.   In addition there is a self-dual 5-form field
strength that depends on the function $z$. It is clear from the
above equations that the full geometry is specified by choosing a
boundary function $u(0;y_1,y_2)$. The requirement of smoothness of
the geometry forces $u \in \{0,1\}$. So one can see the function
$u$ as defining a droplet in the $y_1, y_2$-plane whose boundary
separates the region where $u = 1$ from the region where $u = 0$.
This means smooth 1/2 BPS geometries are in one to one
correspondence with droplets on the $(y_1, y_2)$-plane.   1/2-BPS geometries in
which $0<u <1$ have null singularities and one of the goals is to understand the
relation, if any, between such geometries and the smooth geometries and the CFT
microstates discussed above.\footnote{If $u>1$ or $u<0$ the geometry develops
pathologies such as closed time-like curves, so we will exclude these
\cite{Milanesi:2005tp}.}  If $u$ takes a constant value between $0$ and $1$
within a disk in the $y_1,y_2$ plane, it can be shown that the corresponding
geometry is precisely the ``superstar'' of
\cite{Cvetic:1999xp,Behrndt:1998jd,Myers:2001aq}, a 1/2-BPS extremal black hole.

Following \cite{Lin:2004nb}, in order to match these solutions
with states in the field theory, consider geometries for which the
regions in the $(y_1,y_2)$-plane where $u = 1$ are compact.
Quantization of the flux in the geometry leads to the following
identifications
    \begin{equation} \label{flux}
       \hbar \leftrightarrow 2\pi l_p^4, \;\;\;\;\; N = \int \frac{d^2 y}{2 \pi
	   \hbar} \; u(0;y_1,y_2)
    \end{equation}
The conformal dimension\footnote{The bulk interpretation of the
conformal dimension is energy.} $\Delta$ of a given configuration
is
     \begin{equation} \label{energy}
        \Delta = \frac{1}{2} \int \frac{d^2 y}{2 \pi \hbar} \; \frac{y_1^2 +
		y_2^2}{\hbar} \; u(0;y_1,y_2) - \frac{1}{2} \left( \int \frac{d^2 y}{2
		\pi \hbar} \; u(0;y_1,y_2) \right)^2
     \end{equation}
The formulas above suggest that we should interpret $u(0;y_1,y_2)$ as a phase
space density for a harmonic oscillator where $y_1$ and $y_2$ are treated as
phase space coordinates.  Indeed, this function has a remarkably simple
interpretation \cite{Lin:2004nb} in terms of the hydrodynamic limit of the phase
space of the dual fermionic system, once we identify the $y_1,y_2$-plane with
the single particle phase space of the fermions. This has been confirmed by
directly quantizing the phase space of smooth gravitational solutions
\cite{Mandal:2005wv,Grant:2005qc,Takayama:2005yq}, a procedure which explicitly
recovers the ``fermions in a harmonic oscillator'' description of 1/2-BPS states
that is derived from the dual field theory.   This surprising result is obtained
despite the fact that the configuration space of smooth solutions formally has
structures at the string scale -- presumably the large amount of supersymmetry
is responsible for the success in reproducing the full quantum description from
the restricted quantization of only smooth solutions in gravity.  According to
the general strategy, we should therefore try to identify $u(0;y_1,y_2)$
directly with the one-particle phase space density for any density matrix in the
quantum-mechanical fermion system \cite{Balasubramanian:2005mg}.  Specifically
we identify
\begin{equation}
y_1,y_2 \leftrightarrow p,q ~~~~;~~~~ u(0,y_1,y_2) \equiv 2\pi\hbar \, w(p,q) 
\label{identif}
\end{equation}
where $p,q$ are coordinates of the fermion one-particle phase space and $w(p,q)$ is a phase space density which encodes the expectation values of operators  as described in Sec.~\ref{sec:phasespace}.

\subsection{Geometry versus field theory states}

Given (\ref{identif}) we can explore the map between heavy 1/2-BPS states and ensembles that have a well-defined classical limit and the corresponding geometries.  In the fermion representation described above, a sufficient condition for having a well-defined semi-classical limit is that
the Young diagrams approach a fixed limiting shape with
probability one in the large $N$ limit. For the canonical
ensemble, this limiting shape was given in (\ref{limitcurve}) but
for other states and ensembles different limit curves may arise in
the large $N$ limit. We will continue to denote those curves by
$y(x)$ as in Sec.~\ref{limitshape}. They will describe the effective, coarse grained geometry
corresponding to the states/ensembles.

 To extract the geometry, we can use the identification (\ref{identif}) and work out the semiclassical limit of the phase space distributions of field theory states.  However, for ensembles with limiting Young diagrams there is a shortcut.   Using the fact that the phase space distribution should be rotationally invariant, and by
matching energy $\leftrightarrow$ conformal dimension, flux
$\leftrightarrow$ rank of the gauge group (=number of fermions),
we get
\begin{equation}
    N = \int dx = \int \frac{u(0;r^2)}{2\hbar} \; dr^2, \;\;\;\;\; E = \int (x + y(x)) dx = \int \frac{r^2 \; u(0;r^2)}{4\hbar^2} \; dr^2 =
    \Delta.
\end{equation}
The above equations should not just hold at the level of integrals
but also at the level of integrands, so that
           $$ \frac{u(0;r^2)}{2\hbar} \; dr^2 = dx, \;\;\;\;\; \frac{r^2 \; u(0;r^2)}{4\hbar^2} \; dr^2 = (y(x) + x) dx. $$
Combining these yields
            $$ y(x) + x = \frac{r^2}{2 \hbar} $$
and taking derivatives with respect to $x$, we obtain the
identification \cite{Balasubramanian:2005mg}
      \begin{equation} \label{map1}
          u(0;r^2) = \frac{1}{1 + y'}.
      \end{equation}
So given a Young diagram with a limit shape $y(x)$, one can
associate to it a geometry generated by $u(0;r^2)$ according to
(\ref{map1}).

 For the limit curve of the canonical ensemble (\ref{limitcurve}) the resulting
 phase space distribution is just that of a finite temperature Fermi gas system
 which was considered in \cite{Buchel:2004mc}.  Since $u$ will lie between $0$
 and $1$ the resulting geometry will have a null singularity.    For the
 canonical ensemble of 1/2-BPS states with the limit curve (\ref{limitcurve}),
 the explicit metric is not known.  Since this would give the closest thing to a
 ``1/2-BPS black hole,'' in the sense that it would be the 1/2-BPS geometry with
 a given mass that has the most entropy, it would be interesting to know what it
 looks like.\footnote{Notice that in our conventions $y'\geq 0$ and therefore
 the associated geometry generically has null singularities with $0< u(0;r^2)
 <1$; the unphysical cases with $u<0$ or $u>1$, which give rise to naked
 time-like singularities \cite{Milanesi:2005tp}, do not appear.}  We can also
 ask what sort of ensemble gives rise to the known extremal, charged 1/2-BPS
 black hole \cite{Myers:2001aq}, the so-called superstar.  It was shown in
 \cite{Balasubramanian:2005mg} that this is an ensemble where the number of
 columns of the BPS Young diagram is held fixed in proportion to $N$ as $N_c =
 \alpha N$ (i.e. where the maximum excitation energy in the fermionic
 description is held fixed).   In this case the limiting diagram turns out to
 have a constant slope $y^\prime = \alpha$, reproducing, via (\ref{map1}) and
 (\ref{LLMgeom}), the superstar geometry.

The above results showed that the overwhelming majority of 1/2-BPS basis states
lie close to a certain ``typical'' configuration, which, in the semiclassical
limit, corresponds to a singular geometry.   What then, is the status of the
many non-singular geometries ((\ref{LLMgeom}) with $u=0,1$) that have the same
mass and charge as the ``typical'' configuration?  We expect that almost all
of these non-singular configurations will have string scale differences from the
``typical'' configuration, and after coarse-graining as appropriate to a
semi-classical probe become indistinguishable from each other and from the
singular effective description as an extremal black hole.   This should lead to
effective information loss in the semiclassical theory even though the
underlying system is unitary.  To study this further we can ask how
semiclassical probes will respond to specific geometries derived from particular
half-BPS states according to the correspondence (\ref{identif}).

 \subsection{Detecting states and semiclassical information loss} 

The 1/2-BPS fermion basis states described above are completely characterized by the individual fermion energies $\{e_1, \cdots e_N \}$.   A gauge invariant set of variables containing the same data are the moments 
\begin{equation}
M_k = \sum_{i=1}^N e_i^k = {\rm Tr}(H_N^k/\hbar^k)  ~~~;~~~ k=0,\cdots N \, ,
\label{Mmomentdef}
\end{equation}
where $H_N$ is the Hamiltonian acting on the $N$ fermion Hilbert space with the zero point energy removed.   The $M_k$ are conserved charges of the system of fermions in a harmonic potential from which the individual energies $e_i$ can be reconstructed.  The basis of states with fixed fermion excitation energies that was described above consists of eigenstates of the moment operators.   Above we showed that almost all heavy eigenstates of the $M_k$ lie near a particular ``typical'' state.  Nevertheless, an individual state of this kind can always be identified from the $N$ eigenvalues of the $M_k$.  Hence we can ask what 1/2-BPS spacetime corresponds to an individual joint eigenstate of all the $M_k$, and how an observer might measure the conserved eigenvalues of the $M_k$, thus identifying the underlying state.

A given eigenstate of the $M_k$ corresponds to a specific Young diagram which
can be translated to a corresponding geometry via the identification
(\ref{identif}) between the phase space density and the function $u$ that
sources the LLM geometries (\ref{LLMgeom}).  In \cite{Balasubramanian:2006jt} it
was shown that in the semiclassical limit the kth multipole moment of the dual
spacetime $A_k$ is proportional to $M_k$, i.e. $A_k \propto M_k$.  Thus, in the
semiclassical limit, a 1/2-BPS basis state could be identified completely in
spacetime if the angular moments of the spacetime can be measured.

First, observe that the semiclassical  $\hbar \to 0$ with $\hbar N = \alpha$ fixed translates into gravity, using $l_p^4 \leftrightarrow \hbar$, as 
\begin{equation}
\hbar N \leftrightarrow l_p^4 N \sim g_s l_s^4 N \sim L^4 = \alpha = {\rm fixed} ~~~~~;~~~~~
L  \sim l_s (g_s N)^{1/4}
\label{classlim2}
\end{equation}
where $l_s$ is the string length, $g_s$ is the string coupling, and $L$ is the length scale associated to the asymptotic ${\rm AdS}_{5} \times {\rm S}^5$ spacetime using the standard AdS/CFT dictionary.  Thus, the semiclassical limit that we have been using is the same as the standard limit in the AdS/CFT correspondence, namely  $g_s \to 0, N \to \infty$ with   $L$ fixed.  Now, following \cite{Balasubramanian:2006jt}, one way of measuring the l$^{{\rm th}}$ multipole in is to compute the 
(2l)$^{{\rm th}}$ derivative of the metric functions or any
suitable invariant constructed from them.     Consider an
apparatus of finite size $\lambda$ that makes such a measurement.
In order to compute the k$^{{\rm th}}$ derivative of a  quantity
within a region of size $\lambda$,  the apparatus will have to
make measurements at a scale $\lambda/k$.      However, a
semiclassical apparatus can only measure quantities over distances
larger than the Planck length.  Thus, the k$^{{\rm th}}$
derivative can only be measured if
\begin{equation}
{\lambda \over k} > l_p = g_s^{1/4} l_s
\label{bound1}
\end{equation}
Setting the size of the apparatus to be a fixed multiple of the AdS scale
\begin{equation}
\lambda = \gamma L \, ,
\end{equation}
this says that
\begin{equation}
k < \gamma N^{1/4}
\end{equation}
for a derivative to be semiclassically measurable.   In order to
identify the underlying quantum state we have shown that $O(N)$
multipoles must be measured.  Since $N^{1/4}/N \to 0$ as $N \to
\infty$ we see that the semiclassical observer has access to a
negligible fraction of the information needed to identify the
quantum state.  To make matters worse, in \cite{Balasubramanian:2006jt} it was also shown that in the ensemble of states with a fixed conformal dimension the standard deviation to mean ratio is
\begin{equation}
\label{stdevtomean}
\frac{\sigma(M_k)}{\langle M_k \rangle}  = \frac{k}{\sqrt{N(2k+1)}}.
\end{equation}
This vanishes in the semiclassical limit for $k$ that grow slower than $N$ because of the typicality of heavy microstates that was discussed above.  Thus, low moments are universal and thus do not differentiate states in the ensemble.

Thus, in the semiclassical limit the high moments are unmeasurable, while the low moments are universal, leading to effective loss of information.

 \subsection{Semiclassical limit, singularities and information loss}


To study the semiclassical limit, it is convenient to follow
\cite{Balasubramanian:2007zt} and place  the $N$ fermions of Sec.~\ref{halfBPS}
in coherent states rather than in number eigenstates as we have been doing.
Recall that a coherent state labelled by a parameter $\alpha \in {\mathbb C}$,
is a Gaussian wavepacket in phase space localized around $\alpha =
\frac{y_1+i\,y_2}{\sqrt{2\hbar}}$.  Coherent states  are defined by
\begin{equation}
|\alpha \rangle = e^{-|\alpha|^2/2} \sum_{n=0}^\infty \frac{\alpha^n}{\sqrt{n!}}|n\rangle \equiv \sum_{n=0}^\infty c_n(\alpha)
|n\rangle\, . \label{eq:coherent}
\end{equation}
These states are overcomplete,
and in the semiclassical limit, a complete basis of coherent states can be
thought of as inhabiting a lattice in which the unit cells have phase space area
$2\pi \hbar$ \cite{Bargmann, Perelomov, Bacry}. 

A  semiclassical observer measures the phase plane at an area scale $\delta A =
2 \pi \hbar M \gg 2 \pi \hbar$.  Equivalently, in the dual spacetime the
semiclassical observer makes measurements at scales much bigger than the Planck
length.  At this scale, the observer is only sensitive to a smooth, coarse
grained phase space distribution $ 0 \leq 2\pi \, \hbar \,  w(p,q) \equiv
u(y_1,y_2) \leq 1$ which erases many details of the precise underlying precise
microstates.  We may view the region $\delta A $ as consisting of $M = \delta A
/ 2 \pi \hbar$ lattice sites, a fraction $u_c = \hbar W_c$ of which are occupied
by coherent states. Then the entropy of the local region $\delta A$ is
\begin{equation}
S_K = \log{\binom{M}{\hbar \, W_c \, M}} \sim -M \log{(\hbar
W_c)^{\hbar W_c} (1-\hbar W_c)^{1-\hbar W_c}} = -\frac{\delta
A}{2\pi\hbar} \log{u_c^{u_c} (1-u_c)^{1-u_c}} \, .
\label{localent}
\end{equation}
The Stirling approximation used in (\ref{localent}) is valid when
$\hbar W_c$ is reasonably far from 0 and 1.   For the total
entropy this gives
\begin{eqnarray}
S & = & \int dS = \int dA \, (\frac{dS}{dA}) \label{eq:entropy} \\
\frac{dS}{dA} & = & -\frac{u_c \log{u_c} + (1-u_c)
\log{(1-u_c)}}{2\pi\hbar} \, . \label{entropyequation}
\end{eqnarray}
Thinking about $u = 2\pi \hbar w$ as the probability of occupation of a site by
a coherent state, this is simply Shannon's formula for information in a
probability distribution.\footnote{This result was arrived at independently by
Masaki Shigemori in unpublished work.}  It would be interesting to re-derive this
result directly from the Young tableaux picture where the smooth limit shape
encapsulates the classical observer's ignorance of the underlying discrete
tableau.

These facts imply that in the semiclassical limit the function
$u(y_1,y_2)$ which completely defines a classical solution should
effectively be defined on a lattice with each plaquette of area
$\mathcal{O}(\hbar) \leftrightarrow \mathcal{O}(\ell_P^4)$, and
take values of 0 or 1 in each site.\footnote{Recall that in (\ref{LLMgeom}) $y_1$ and $y_2$ have units of length {\it squared}.  This is why an $\ell_P^4$ appears here.}  Likewise
(\ref{entropyequation}) should be interpreted as an expression for
the entropy of arbitrary half-BPS asymptotically ${\rm AdS}_{5} \times
{\rm S}^5$ spacetimes. As an example it exactly reproduces the formula
for the entropy of the typical states described in  \cite{Balasubramanian:2005mg} that correspond in spacetime to the ``superstar'' geometry \cite{Myers:2001aq}.

Note that the entropy vanishes if and only if $u$ equals $0$ or
$1$ everywhere.   Following the correspondence
(\ref{identif}) such states map into geometries that are
non-singular.   We learn that semiclassical half-BPS geometries
that are smooth all have vanishing entropy; and the  presence of
singularities $0 < u < 1$ also implies that the spacetime carries
an entropy.  Thus, in this setting, entropy is a measure of
ignorance of a part of the underlying state which is captured in
classical gravity as a spacetime singularity.

\subsection{Summary}
In this section we discussed how the almost all half-BPS states of ${\cal N} =
4$ of a given conformal dimension lie close to a certain ``typical'' state.  The
semiclassical indistinguishability of these states leads to a universal
description in AdS space as an extremal black hole.  We further discussed how
the presence of a singularity in a semiclassical geometry is directly related to
a loss of information about underlying states that have the same effective
description.   Note that most states in the Hilbert space, when represented in
either fermion excitation number basis or in the coherent state basis, will be
random superpositions of the basis states.  In \cite{Balasubramanian:2007zt} a
criterion is given for determining which of these states can be effectively
described in terms of a single geometry as opposed to a wavefunction over
geometries.

\section{AdS$_3$xS$^3$} \label{sec_ads3_s3}

In this section we are going to discuss the bound states of D1 and
D5 branes in type II-B string theory compactified on\footnote{The
same story carries over to the case of K3 $\times$ S$^1$.} T$^4
\times$ S$^1$.    These are $1/2$-BPS states (preserving 8
supercharges) that describe a black hole without a classical
horizon in 5 dimensions.\footnote{Strictly speaking the extremal black hole and the associated microstates are ground states in the Ramond sector of the theory, but by spectral flow they can be represented as 1/2-BPS states in the NS sector.   Hence we will refer to them as 1/2-BPS states.}
However, we are going to work in 6
dimensions, explicitly keeping track of the S$^1$. One of the
reasons behind this decision is that in this way one gets
solutions that are asymptotically AdS$_3 \times$ S$^3 \times$ T$^4$  after
taking a suitable decoupling limit. Thus we can employ the AdS/CFT
machinery and benefit from the known properties of the dual
two-dimensional conformal field theory.

As is well-known, the 1/2-BPS states in the CFT dual to the D1-D5
system can be identified with the states at level $L_0=N_1N_5$ in
a system with $b_1+b_3$ chiral fermions and $b_0+b_2+b_4$ chiral
bosons, where $b_i=\dim H^i(M_4)$ for a compactification on  $M_4 \times S^1$. Here $N_1$ and $N_5$ are the
quantized number of D1 and D5 branes. Notice that this
identification of 1/2 BPS states with a system of free bosons and
fermions is only valid at the level of the Hilbert space, not at
the level of correlation functions. Thus, we would ideally like to
be able to find a detailed map between states/ensembles in this
auxiliary theory of free bosons and fermions and half-BPS
solutions of six-dimensional supergravity. In what follows we will
describe such a map. We will first review the known supergravity
solutions and their quantization, and then propose a map which is
again based on the notion of phase space densities. We conclude
this section by discussing various relevant examples.

\subsection{The supergravity solution and its quantization}
\label{sugrad1d5}

Starting with a fundamental string with transversal profile
$\mathbf{F}(s)\subset\mathbb R^4$ then dualizing, one gets the
following solutions
\cite{Lunin:2001fv,Lunin:2002iz,Rychkov:2005ji}, written in the
string frame\footnote{We are going to follow the conventions of
 \cite{Rychkov:2005ji}.}

\begin{align}
\label{genericsol} ds^2=\frac{1}{\sqrt{f_1
f_5}}\left[-(dt+A)^2 + (dy+B)^2 \right] &+ \sqrt{f_1 f_5}
d\mathbf{x}^2+\sqrt{f_1/f_5}d
\mathbf{z}^2 \nonumber\\
e^{2 \Phi}=\frac{f_1}{f_5},\hspace{0.2in}C = \frac{1}{f_1}\left(dt+A
\right) &\wedge \left( dy+B \right)+\mathcal{C}
\end{align}
where $y$ parametrizes a circle with coordinate radius $R$, $z^i$
are coordinates on T$^4$ with coordinate volume $V_4$, the Hodge
star $*_4$ is defined with respect to the 4 dimensional non
compact space spanned by $x^i$ and
\begin{eqnarray} \label{aux1}
dB=*_4 dA,\;\;\; d\mathcal{C}=-*_4 df_5, \;\;\; A=\frac{Q_5}{L} \int_0^L \frac{F'_i(s)
ds}{|\mathbf{x}-\mathbf{F}(s)|^2} \nonumber \\
f_5=1+\frac{Q_5}{L}\int_0^L
\frac{ds}{|\mathbf{x}-\mathbf{F}(s)|^2}, \;
f_1=1+\frac{Q_5}{L}\int_0^L \frac{|\mathbf{F}'(s)|^2
ds}{|\mathbf{x}-\mathbf{F}(s)|^2} .
\end{eqnarray}
The solutions are asymptotically $\mathbb R^{1,4} \times S^1
\times T^4$. We can take a decoupling limit which simply amounts
to erasing the $1$ from the harmonic functions. The resulting
metric will then be asymptotically AdS$_3 \times$ S$^3
\times$ T$^4$.

As mentioned above, the solutions are parametrized in terms of a
closed curve
\begin{equation}
x_i=F_i(s),\hspace{0.2in}0<s<L,~i=1,...,4.
\end{equation}
In the sequel we are going to ignore oscillations in the $T^4$
direction as well as fermionic excitations; for a further
discussion of these degrees of freedom see
\cite{Taylor:2005db,Kanitscheider:2007wq}. The D1 (D5) charge
$Q_1$ ($Q_5$) satisfy

\begin{equation} \label{cond1}
   L = \frac{2 \pi \; Q_5}{R}, \;\;\;\;\; Q_1 = \frac{Q_5}{L} \int_0^L |\text{\bf{F}}'(s)|^2
   ds.
\end{equation}

It turns out that the space of classical solutions has
finite volume and therefore will yield a finite number of quantum
states. To see this, one first starts by expanding $\mathbf{F}$ in oscillators:
\be
\mathbf{F}(s)=\mu \sum_{k=1}^\infty
\frac{1}{\sqrt{2k}}\left(\mathbf{c}_k e^{i \frac{2\pi k
}{L}s}+\mathbf{c}_k^\dagger e^{-i \frac{2\pi k }{L}s} \right)
\ee
where $\mu=\frac{g_s}{R \sqrt{V_4}}$. Then one computes the restriction of the Poisson
bracket to the space of solutions (\ref{genericsol}) which turns out to be \cite{Donos:2005vs,Rychkov:2005ji}
\be
~[c_k^i,c_{k'}^{j \dagger}]=\delta^{ij}\delta_{kk'} .
\ee
After quantization, the relation between $Q_1$ and $Q_5$ reads
\be
\left< \int_0^L :|\mathbf{F}'(s)|^2: ds\right>=\frac{(2\pi)^2}{L}~
\mu^2 N
\ee
where
\be
N_1 = \frac{g_s}{V_4}\; Q_1, \;\;\; Q_5 = g_s N_5, \;\;\; N\equiv N_1 N_5 =\sum_{k=1}^{\infty} k \left< \mathbf{c}^\dagger_k
\mathbf{c}_k\right> .
\ee
$N_1$, $N_5$ count the number of D1 and D5 branes respectively.
The modes $c_k^i$ become the creation and annihilation modes of
four of the total of $b_0+b_2+b_4$ bosons; one can check that the
four that appear are precisely the ones associated to the
$H^{(0,0)}(M), \; H^{(2,0)}(M), \; H^{(0,2)}(M), \; H^{(2,2)}(M)$
cohomology. Finally, notice that
the number of states and hence the entropy can easily be extracted
from the known partition functions of chiral bosons and fermions.

\subsection{Geometries from coherent states}
\label{geomcoherent}

The Hilbert
space is spanned by
\begin{equation} \label{hilbert}
|\psi \rangle=\prod_{i=1}^4 \prod_{k=1}^\infty(c^{i \dagger}
_{k})^{N_{k_i}}|0\rangle,\hspace{0.3in}\sum k N_{k_i}=N
\end{equation}
Given a state, or more generically a density matrix in the CFT
\begin{equation} \label{density}
\rho=\sum_{ij} c_{ij} | \psi_i \rangle \langle \psi_j |
\end{equation}
we wish to associate to it a density on phase space. The phase
space is given by classical curves which we will parametrize as
(note that $d$ and $\bar{d}$ are now complex numbers, not
operators)
\begin{equation} \label{classcurv}
\mathbf{F}(s)=\mu \sum_{k=1}^\infty
\frac{1}{\sqrt{2k}}\left(\mathbf{d}_k e^{i \frac{2\pi k
}{L}s}+\bar{\mathbf{d}}_k e^{-i \frac{2\pi k }{L}s} \right)
\end{equation}
and which obey the classical constraint (\ref{cond1}).

We now propose to associate to a density matrix  of the form
(\ref{density}) a phase space density (compare to the general
discussion in section \ref{quant}) of the form \cite{Alday:2006nd}
\begin{equation} \label{proposal}
f(d,\bar{d})= \sum_i \frac{\langle 0|
e^{\mathbf{d}_k\mathbf{c}_k}|\psi_i \rangle \langle \psi_i |
e^{\mathbf{\bar d}_k\mathbf{c}_k^\dagger}|0 \rangle }{\langle 0|
e^{\mathbf{d}_k\mathbf{c}_k} e^{\mathbf{\bar
d}_k\mathbf{c}_k^\dagger}|0 \rangle } .
\end{equation}

The distribution corresponding to a generic state $|\psi \rangle=
\prod_{k=1}^\infty(c^{i \dagger}_{k})^{N_{k_i}}|0\rangle$ can be
easily computed
\begin{equation}
\label{genericwigner} f(d,\bar d)=\prod_{k,i} (d_k^i
\bar{d}_k^i)^{N_{k_i}}e^{-d_k^i \bar{d}_k^i}.
\end{equation}

Notice that our phase space density (\ref{proposal}), as written,
is a function on a somewhat larger phase space as $d,\bar{d}$ do
not have to obey (\ref{cond1}). To cure this discrepancy we are
going to include an extra factor $\exp(-\beta \hat{N})$ in the
calculations, where we choose $\beta$ such that the expectation
value of $\hat{N}$ is precisely $N$. This is just like passing
from a microcanonical ensemble to a canonical one, and for many
purposes this is probably a very good approximation. For a
thorough discussion of this point see \cite{Alday:2006nd}.

To further motivate (\ref{proposal}) we notice that it associates
to a coherent state a density which is a Gaussian centered around
a classical curve, in perfect agreement with the usual philosophy
that coherent states are the most classical states. It is then
also clear that given a classical curve (\ref{classcurv}) we wish
to associate to it the density matrix
\be \label{classdens}
\rho = P_N e^{\mathbf{d}_k\mathbf{c}_k}|0 \rangle \langle 0 |
e^{\mathbf{\bar d}_k\mathbf{c}_k^\dagger} P_N
\ee
where $P_N$ is the projector onto the actual Hilbert space of
states of energy $N$ as defined in (\ref{hilbert}). Because of
this projector, the phase space density associated to a classical
curve is not exactly a Gaussian centered around the classical
curve but there are some corrections due to the finite $N$
projections. Obviously, these corrections will vanish in the
$N\rightarrow \infty$ limit.

The density (\ref{proposal}) has the property that for any
function $g(d,\bar{d})$
\be \label{prop}
\int \int_{d,\bar{d}}
f(d,\bar{d}) g(d,\bar{d}) = \sum_i \langle \psi_i |
:g(c,c^{\dagger}):_A | \psi_i \rangle
\ee
where $:g(c,c^{\dagger}):_A$ is the anti-normal ordered operator
associated to $g(c,c^{\dagger})$, and $\int_{d,\bar{d}}$ is an
integral over all variables $d_i$.  Since the theory behaves like
a $1+1$ dimensional field theory the natural thing to do is to
calculate expectation values of normal ordered operators in order
to avoid infinite normal ordering contributions. Besides,
everything we do is limited by the fact that our analysis is in
classical gravity and therefore can at best be valid up to quantum
corrections. As a result a further modification to our proposal
will be to redefine $g(d, \bar{d})$ by subtracting the anti-normal
ordering effects.

Since the harmonic functions appearing in (\ref{aux1}) can be
arbitrarily superposed, we finally propose to associate to
(\ref{density}) the geometry
\begin{eqnarray} \label{finalprop}
f_5 & = & 1+\frac{Q_5}{L} {\cal N}\int_0^L \int_{d,\bar d} \frac{
f(d,\bar{d})
ds}{|\mathbf{x}-\mathbf{F}(s)|^2} \nonumber \\
f_1 & = & 1+\frac{Q_5}{L}{\cal N}\int_0^L \int_{d,\bar d} \frac{
f(d,\bar{d}) |\mathbf{F}'(s)|^2
ds}{|\mathbf{x}-\mathbf{F}(s)|^2} \nonumber \\
A^i & = & \frac{Q_5}{L}{\cal N}\int_0^L \int_{d,\bar d} \frac{
f(d,\bar{d})\mathbf{F}'_i(s) ds}{|\mathbf{x}-\mathbf{F}(s)|^2}
\end{eqnarray}
with the normalization constant
\begin{equation}
{\cal N}^{-1}=\int_{d, \bar d} f(d,\bar d)
\end{equation}

In \cite{Lunin:2002iz} it was shown that the geometries
corresponding to a classical curve are regular provided
$|\mathbf{F}'(s)|$ is different from 0 and the curve is not self
intersecting. In our setup we sum over continuous families of
curves with some weighing factor which can introduce
singularities. We expect these singularities to be rather mild,
certainly for semiclassical density matrices, and in addition in
various examples the averages will turn out to be completely
smooth anyway (see section \ref{eg}). Another point worth
mentioning is that the average will no longer solve the vacuum
type IIB equations of motion, instead a small source will appear
on the right hand side of the equations. Since these sources are
subleading in the $1/N$ expansion and vanish in the classical
limit, they are in a regime where classical gravity is not valid
and they may well be cancelled by higher order contributions to
the equations of motion. To have an idea about these sources let
us study the circular profile.

We consider the following profile
\begin{equation}
\label{circprofile} F^1(s)=a \cos{\frac{2\pi
k}{L}s},\hspace{0.2in}F^2(s)=a \sin{\frac{2\pi
k}{L}s},\hspace{0.2in}F^3(s)=F^4(s)=0
\end{equation}
which describes a circular curve winding $k$ times around the
origin in the $12$-plane. In order to simplify our discussion, we
focus on the simplest harmonic function $f_5$.
In order to evaluate the various integrals it will be convenient
to Fourier transform the $x$-dependence. Using
\begin{equation}\label{fourier}
\frac{1}{|\mathbf{x}|^2}=\frac{1}{4\pi^2}\int d^4 u
\frac{e^{i \mathbf{u}.\mathbf{x}}}{|\mathbf{u}|^2}
\end{equation}

Classically $\Box f_5$ is a delta
function with a source at the location of the classical curve, to
be precise
\begin{equation}
\Box f_5  = -4 \pi Q_5
\delta(x_1^2+x_2^2-a^2)\delta(x_3)\delta(x_4) \label{supp}.
\end{equation}

Now in the quantum theory, we associate to the classical circular curve (\ref{circprofile}) the density matrix
(\ref{classdens}) and subsequently the phase space density
(\ref{proposal}). Working this out we find out that
\begin{equation}
\label{qcircprofile} f(d,\bar{d})=((d_k^1+id_k^2)(\bar{d}_k^1-i
\bar{d}_k^2))^{N/k}e^{-\sum_{l,i} d_l^i \bar{d}_l^i}.
\end{equation}
We have ignored the delta function coming from the projection here and expect
(\ref{qcircprofile}) to be valid for large values of $N/k$. It is
therefore better thought of as a semiclassical profile rather than
the full quantum profile.

According to (\ref{finalprop}) the harmonic function $f_5$ is now
given by
\begin{equation}
f_5=1+\frac{Q_5}{4 \pi^2}{\cal N} \int_0^L ds \int_{d,\bar{d}}
f(d,\bar{d}) \int d^4u \frac{1}{|\mathbf{u}|^2}e^{i
\mathbf{u}.(\mathbf{x}-\mathbf{F}(s)) + \sum_l \frac{u^2
\mu^2}{2l}}
\end{equation}
where we have used (\ref{fourier}) and the constant $\sum_l
\frac{u^2 \mu^2}{2l}$ appears due to the fact that we want to
compute a normal ordered quantity instead of an anti-normal
ordered one. The function $\mathbf{F}(s)$ depends on an infinite
set of complex oscillators $d_{\tilde{l}}^i$. It can be easily
seen that the contribution for the oscillators different from
$d_k^1$ and $d_k^2$ cancels exactly against the normal ordering
constant $u^2\mu^2/2l$ mentioned above.

So $\Box f_5$ for this case reads
\begin{eqnarray}
\Box f_5 =-4 \pi Q_5 \delta(x_3) \delta(x_4) A(x_1,x_2)\\
A(x_1,x_2)=\int_0^\infty d\rho \rho J_0(\sqrt{x_1^2+x_2^2} \rho)
L_{N/k} \left(\frac{a^2 \rho^2}{4N/k}\right)
\end{eqnarray}
Until here we have not used any approximation. Using the identity
         $$ L_N (x) = \frac{e^x}{N!} \; \int_0^{\infty} e^{-t} t^N J_0 (2\sqrt{tx}) dt $$

and approximating $\exp(\frac{a^2
\rho^2}{4N/k}) \approx 1$ one obtains
\begin{equation}
A(x_1,x_2)=\frac{e^{-N/k~r^2/a^2} \left(N/k~r^2/a^2
\right)^{N/k}}{(N/k-1)! a^2}
\end{equation}
with $r^2=x_1^2+x_2^2$. In the limit $N/k \rightarrow \infty$
$A(x_1,x_2)$ approaches $\frac{\delta(r^2/a^2-1)}{a^2}$ and the
classical and quantum results agree. For large $N/k$ $A(x_1,x_2)$
is  approximately a Gaussian around $r^2 \approx a^2$ and width
$1/\sqrt{N/k}$, indeed, using Stirling's formula
\begin{equation}
A(x_1,x_2)\approx \frac{\sqrt{N/k}}{\sqrt{2\pi}}
e^{-N/k(r^2/a^2-1)}(r^2/a^2)^{N/k}
\end{equation}
So the geometry corresponds to a solution of the equations
of motion in presence of smeared sources. The width of the smeared
source goes to zero in the limit $N/k\rightarrow \infty$, as
expected.

\subsection{Ensembles} \label{eg}

In the following we consider the geometry of some ensembles of interest.

\subsubsection{$M = 0$ BTZ}
\label{sec:btz}

The
corresponding ensemble is characterized by the following
density matrix \footnote{We are going to ignore the $i$-index in
some equations where it does not play any role. We hope that this
will not create any confusion.}
\begin{equation}
    \rho = \sum_{N_k,\tilde{N}_k} \frac{|N_k \rangle
    \langle N_k| e^{-\beta \hat{N}} |\tilde{N}_k\rangle \langle\tilde{N}_k|}{\text{\bf Tr} e^{-\beta \hat{N}}}
\end{equation}
where $|N_k\rangle$ is a generic state labelled by collective
indices $N_k$
$$|N_k\rangle=\prod_k \frac{1}{\sqrt{N_k!}}(c_k^\dagger)^{N_k}|0
\rangle$$
and we have chosen a normalization so that $\langle N_k
| \tilde{N}_k \rangle=\delta_{N_k,\tilde{N}_k}$. The value of the
potential $\beta$ has to be adjusted such that $\langle \hat{N}
\rangle =N$. It is clear that
\begin{equation}
     \rho = \prod_n \rho_k,\hspace{0.3in}\rho_k=(1-e^{-k \beta})
     \sum_{n=0}^{\infty} e^{-n k \beta} |k,n \rangle
     \langle k,n |
\end{equation}
with $|k,n\rangle=\frac{1}{\sqrt{n!}}(c_k^\dagger)^{n}|0 \rangle$.
Then the full distribution will simply be the product
$f(d,\bar{d})=\prod_k f^{(k)}_{d_k,\bar{d}_k}$ with
\be f^{(k)}_{d_k,\bar{d}_k}=(1-e^{-k \beta})e^{-d_k
\bar{d}_k}\sum_{n=0}^{\infty}\frac{e^{-n k \beta}}{n!}(d_k
\bar{d}_k)^n=(1-e^{-k \beta})\exp{(-(1-e^{-k \beta})d_k
\bar{d}_k)} .
\ee
The needed harmonic functions \ref{aux1} are deduced from the
following generating function
\begin{equation}
    f_v = \frac{Q_5}{4 \pi^2
L}{\cal N} \int d^4u \int_0^L dr \int_{d,\bar{d}}
f(d,\bar{d})\frac{e^{\sum_k \left( \frac{|\mathbf{u}|^2
\mu^2}{2k}-\frac{2 \pi^2 k \mu^2|\mathbf{v}|^2}{L^2} \right) }e^{i
\mathbf{u}.(\mathbf{x}-\mathbf{F}(r)) + i\mathbf{v}.\mathbf{F}'(r)
}}{|\mathbf{u}|^2}
\end{equation}
which gives
\begin{align}
    f_5 = Q_5 \; \frac{1 - e^{- \frac{3 \beta}{\pi^2 \mu^2} x^2}}{x^2},& \;\;\;\;\; f_1 = Q_1 \; \frac{1 - e^{- \frac{3 \beta}{\pi^2 \mu^2} x^2}}{x^2} \nonumber\\
        A_i = 0, \;\;\;\;\; &\beta \approx \pi \; \sqrt{\frac{2}{3
        N}}.
\end{align}

A final comment is in order. The geometry obtained differs from
the {\it classical} $M=0$ BTZ black hole by an exponential piece.
Following \cite{Lunin:2002qf,Mathur:2005zp} we could put a
stretched horizon at the point where this exponential factor
becomes of order one, so that the metric deviates significantly
from the classical $M=0$ BTZ solution. Thus, using this criterion
we find for the radius of the stretched horizon\footnote{The same
value is obtained if we compute the average size of the curve in
$\mathbb R^4$, $r_0^2 \approx \langle |F|^2 \rangle$.}
\be \label{shor}
r_0 \approx \frac{\mu}{\beta^{1/2}}
\ee
with corresponding entropy proportional to $N^{3/4}$. This exceeds
the entropy of the mixed state from which the geometry was
obtained; the latter grows as $N^{1/2}$. This does not contradict
any known laws of physics, and in addition we should remember that
the notion of stretched horizon depends on the choice of observer.
It is quite likely that for a suitable choice of observer the
entropy of the stretched horizon agrees with the entropy obtained
from the dual CFT. For a further discussion of this point see
\cite{Alday:2006nd, Mathur:2007sc}.

\subsubsection{The small black ring}

In this section we consider a slightly more complicated example,
namely an ensemble consisting of a condensate of $J$ oscillators
of level $q$ plus a ensemble of effective level $N-q J$.
As argued in \cite{Bena:2004tk,Balasubramanian:2005qu,Iizuka:2005uv,Alday:2005xj} such an ensemble
should describe (in a certain region of parameter space) a small
black ring of angular momentum $J$ and dipole (or Kaluza-Klein)
charge $q$.

Using the techniques developed in the previous sections we can
compute the generating harmonic function for this case as well and
we find
\begin{equation}
f_v=Q_5 L_J \left( \frac{\mu^2}{4 q} \left[ \left( \frac{2 \pi
q}{L}v_2+i
\partial_1 \right)^2+\left( \frac{2 \pi q}{L}v_1-i
\partial_2 \right)^2 \right] \right) e^{-\frac{\mu^2 \pi^2 |v|^2}{2L^2}(N-q
J)}\frac{1-e^{-\frac{2 |\mathbf{x}|^2}{\mu^2 D}}}{|\mathbf{x}|^2}
\end{equation}
where $D=\pi \sqrt{2/3}(N-q J)^{1/2}$ so that the geometry is
purely expressed in terms of the macroscopic quantities $N,J$ and
$q$.

We would like to make contact between this geometry and the
geometry corresponding to small black rings studied in
\cite{Alday:2005xj}. As we will see, in the limit of large quantum
numbers both geometries reproduce the same one point functions.

In order to see this, first note that the exponential factor
$e^{-\frac{2 |\mathbf{x}|^2}{\mu^2 D}}$ will not contribute (as it
vanishes faster than any power at asymptotic infinity). Secondly
one has the formal expansion
\be
L_J \left( \frac{\mu^2}{4q}
\mathcal{O}\right) =J_0(\mu \sqrt{\frac{J}{q}}
\mathcal{O}^{1/2})+...
\ee
In order to estimate the validity of this approximation we can
think of ${\mathcal O}$ as being proportional to
$1/|\mathbf{x}|^2$. On the other hand $\mu \sqrt{J/q}$ can be
roughly interpreted as the radius of the black ring (see
\cite{Elvang:2004ds,Alday:2005xj}, where this parameter is called
$R$). Hence the approximation is valid for large values of $J$ at
a fixed distance compared to the radius of the ring.

Using the above approximations it is straightforward to compute
the harmonic functions
\begin{equation}
f_5 = \frac{Q_5}{r^2+\mu^2
\frac{J}{q}\cos\theta},\hspace{0.3in}f_1 = \frac{Q_1}{r^2+\mu^2
\frac{J}{q}\cos\theta}
\end{equation}
where we have used the following coordinate system
\begin{eqnarray} \label{delcoord}
  x_1 = (r^2 + a^2)^{1/2} \sin \theta \cos \varphi \; , \;\;\;\;\; x_2 = (r^2 + a^2)^{1/2} \sin \theta \sin \varphi \nonumber\\
  x_3 = r \cos \theta \cos \psi \; , \;\;\;\;\; x_4 = r \cos \theta \sin
  \psi.
\end{eqnarray}
Hence in this approximation the geometry reduces exactly to that
of the small black ring studied in \cite{Alday:2005xj}.

\subsubsection{Generic ensemble and the no-hair theorem}

In the following we consider a generic  ensemble, where
each oscillator $c_{k^{i}}$ is occupied thermally with a
temperature $\beta_{k^i}$. We further will assume that
$\beta_{k^\pm}$ for the directions $1,2$ is equal to
$\beta_{k^\pm}$ for the directions $3,4$. Restricting to, say,
directions $1,2$ we are led to consider the following distribution
\be
f(d,\bar{d})=\exp\left( -(1-e^{-\beta_{k^+}})d_k^+ \bar{d}_k^+
-(1-e^{-\beta_{k^-}})d_k^- \bar{d}_k^-\right).
\ee
Following the same steps as for the case of the small black ring we obtain
\begin{eqnarray}
f_5 & = & Q_5 \frac{1-e^{-\frac{2|\mathbf{x}|^2}{\mu^2
D}}}{|\mathbf{x}|^2}\\
f_1& =& Q_1 \left( \frac{1-e^{-\frac{2|\mathbf{x}|^2}{\mu^2
D}}}{|\mathbf{x}|^2} -\frac{J^2}{4N\mu^4
D^2}e^{-\frac{2|\mathbf{x}|^2}{\mu^2 D}} \right)
\\
A & = & \frac{\mu^2  J}{2}
\left(2\frac{e^{-\frac{2|\mathbf{x}|^2}{\mu^2 D}}}{\mu^2
D}-\frac{1-e^{-\frac{2|\mathbf{x}|^2}{\mu^2 D}}}{|\mathbf{x}|^2}
\right)(\cos^2\theta d\phi+\sin^2\theta d\psi)
\end{eqnarray}
where $(|\mathbf{x}|,\theta,\phi,\psi)$ are standard spherical
coordinates on $\mathbb R^4$.

We see that, rather surprisingly, the geometry depends only on a
few quantum numbers $N,J$ and $D$ which are given in terms of the
temperatures by
\begin{eqnarray}
N=2 \sum_k k \left( \frac{e^{-\beta_{k^+}}}{1-e^{-\beta_{k^+}}} +
\frac{e^{-\beta_{k^-}}}{1-e^{-\beta_{k^-}}} \right)\\
J=2 \sum_k \left( \frac{e^{-\beta_{k^+}}}{1-e^{-\beta_{k^+}}} -
\frac{e^{-\beta_{k^-}}}{1-e^{-\beta_{k^-}}} \right)\\
 D=2 \sum_k \frac{1}{k}
\left( \frac{e^{-\beta_{k^+}}}{1-e^{-\beta_{k^+}}} +
\frac{e^{-\beta_{k^-}}}{1-e^{-\beta_{k^-}}} \right).
\end{eqnarray}
As a result, the information carried by the geometry is much less
than that carried by the ensemble of microstates. In fact, only
$N$ and $J$ are visible at infinity while $D$ sets the size of the
``core'' of the geometry. We also find that $D$ is precisely the
expectation value of the dipole operator introduced in
\cite{Alday:2005xj}. Its presence in the density matrix is
supported by an analysis of the first law of thermodynamics
\cite{Copsey:2005se}. It is a non-conserved charge which makes its
extension to interacting theories an interesting open problem.

We interpret the above remark as a manifestation of the no-hair
theorem for black holes. The derivation in this section assumes
that the temperatures are all sufficiently large. By tuning the
temperatures, it is possible to condense one (like in the small
black ring case) or more oscillators. If this happens, we should
perform a more elaborate analysis, and we expect that the dual
geometrical description\footnote{It is not difficult to see that
the harmonic functions now will take the form of multiple Laguerre
polynomials with differential operator arguments acting on the
generating harmonic function of the $M=0$ BTZ solution.}
corresponds to concentric small black rings. In this case the
configuration will depend on more quantum numbers than just
$N,J,D$, in particular we will find solutions where the small
black rings carry arbitrary dipole charge. Thus, once we try to
put hair on the small black hole by tuning chemical potentials
appropriately, we instead find a phase transition to a
configuration of concentric small black rings, each of which still
is characterized by just a few quantum numbers.

\subsection{BTZ $M=0$ as an effective geometry}

\newcommand{\ads}[1]{{\rm AdS}_{#1}}
\def\Bracket#1{{\left\langle #1 \right\rangle}}
\def\CA{{\cal A}}
\def\CH{{\cal H}}
\def\CM{{\cal M}}
\def\CN{{\cal N}}
\def\CO{{\cal O}}
\def\CX{{\cal X}}
\def\zb{{\overline{z}}}
\def\bracket#1{{\langle #1 \rangle}}
\def\wb{{\overline{w}}}

Above we discussed how the half-BPS states of the D1-D5 CFT  and ensembles of
these states can be mapped to specific asymptotically ${\rm AdS}_3$ geometries.
In this section we will argue, following \cite{Balasubramanian:2005qu} that the
{\it typical} state in this sector responds to probes as if it were a BTZ $M=0$
black hole in spacetime.   This approach is complementary to Sec.~\ref{sec:btz}
where it was shown how BTZ arises as the effective, coarse-grained geometry of
an ensemble of D1-D5 states.   It would be interesting check more carefully if
the two-point function computed below can be somehow sensitive to corrections
from the stretched horizon described in Sec.~\ref{sec:btz}.  For concreteness we
will restrict our analysis here to $T^4$ compactifications.

The CFT dual to ${\rm AdS}_3 \times S^3 \times T^4$ is  an ${\cal N} = (4,4)$ supersymmetric sigma model whose target space is the symmetric product ${\cal M}_0=(T^4)^N/S_N$,
where $S_N$ is the permutation group of order $N$. Here we set
\begin{align}
 N&\equiv N_1 N_5.
\end{align}
More precisely, ${\cal M}_0$ is the so-called orbifold point in a family of
CFTs which are regained by turning on certain marginal deformations of
the sigma model on ${\cal M}_0$.  At the orbifold point the CFT becomes free.  The CFT has a collection of
twist fields $\sigma_n$, which cyclically permute $n \leq N$ copies of
the CFT on a single $T^4$.  One can think of these operators as creating
winding sectors of the worldsheet that wind over the different copies of
the torus.  

After spectral flow from the NS sector to the Ramond sector, all of the 1/2-BPS states discussed above become Ramond sector ground states, and we will choose to describe the system in the latter language.   A general Ramond sector ground state is constructed by multiplying together elementary bosonic and fermionic twist operators to achieve a total twist of $N = N_1 N_5$:
\begin{equation}
\begin{split}
 \sigma&= \prod_{n,\mu} (\sigma_{n}^{\mu})^{N_{n\mu}} (\tau_{n}^\mu)^{N'_{n\mu}},
 \\
 \sum_{n,\mu}n (N_{n\mu}+N'_{n\mu})&=N, \qquad
 N_{n\mu}=0,1,2,\dots,\quad N'_{n\mu}=0,1 \, .
\end{split}\label{gen_twist2}
\end{equation}
Here $\sigma_n^\mu$ and $\tau_n^\mu$, are the constituent elementary
twist operators, and $\mu = 1 \cdots 8$ labels their possible
polarizations.  This is like 8 bosonic and 8 fermionic oscillators.   Secs.~\ref{sugrad1d5},\ref{geomcoherent}, \ref{eg} above constructed states out of just 4 bosonic oscillators because a convenient subset of the states was examined which lacked oscillations in the $T^4$ and fermionic excitations.  Here we are considering the full Hilbert space of states.   The Appendix in \cite{Balasubramanian:2005qu} gives a detailed description of the construction of the twist operators and
computations using them.  For our immediate purposes, the relevant point
is that the integers
\begin{equation}
\{ N_{n\mu}, N^\prime_{n\mu} \}
\label{Rstateintegers}
\end{equation}
uniquely specify a Ramond ground  state.

When the total twist length $N=\sum_{n,\mu} n (N_{n\mu}+N'_{n\mu})$ is very
large, there are a macroscopic number ($\sim e^{2\sqrt{2}\pi\sqrt{N}}$)
of Ramond ground states.  In such a situation, most of those
$e^{2\sqrt{2}\pi\sqrt{N}}$ microstates will have a twist distribution
$\{ N_{n\mu}, N^\prime_{n\mu} \}$ that lies very close to a certain
``typical'' distribution.  In the large $N$ limit, the difference among
individual distributions is small.  Roughly, statistical mechanics says
that $\langle(\Delta N_{n\mu})^2\rangle\sim N_{n\mu}$, thus
${\langle(\Delta N_{n\mu})^2\rangle^{1/2}\over N_{n\mu}}\sim
(N_{n\mu})^{-1/2}\to 0$ as $N_{n\mu }\to\infty$.  Thus, although
correlation functions computed in individual microstates depend on the
precise form of the microstate distribution $\{N_{n\mu},N_{n\mu}'\}$,
for almost all microstates the generic responses should deviate by small
amounts from the results for the typical state.   This will be the basis for the emergence of an effective black hole description of typical Ramond ground states. 

Let us first consider the ensemble of all the Ramond ground states
\eqref{gen_twist2} with equal statistical weight. 
The canonical partition function is
\begin{align}
 Z(\beta)&
 ={\rm Tr}[e^{-\beta N}]
 =\prod_{n=1}^\infty {(1+q^n)^8\over (1-q^n)^8}
 = \left[{\vartheta_2(0|\tau)\over 2\eta(\tau)^3}\right]^4,
 \qquad
 q=e^{2\pi i \tau}=e^{-\beta}.
\end{align}
Using the modular property of the theta function,
\begin{align}
 Z(\beta)
 = \left[{\beta\over 4\pi} {\vartheta_4(0|-{1\over\tau})\over \eta(-{1\over\tau})^3}\right]^4
 \sim e^{2\pi^2/\beta} \qquad (\beta\ll 1).
\end{align}
The relation between ``energy'' $N$ and temperature $\beta$ is
\begin{align}
 N&=\Bracket{\sum_{n=1}^\infty\sum_{\mu} n(N_{n\mu} +N'_{n\mu})}
 =-{\partial \over \partial \beta}\ln Z(\beta)
 \simeq {2\pi^2\over \beta^2}.\label{N_beta}
\end{align}
Since all twist operators are independent, the average
distribution $\{N_{n\mu},N'_{n\mu}\}$ is given by the
Bose--Einstein and Fermi--Dirac distribution, respectively:
\begin{align}
 N_{n\mu}&={1\over e^{\beta n}-1},\qquad
 N'_{n\mu}={1\over e^{\beta n}+1},\qquad
 N_n=\sum_\mu (N_{n\mu}+N'_{n\mu})={8\over \sinh \beta n}.
 \label{typ_J=0}
\end{align}
For large $N$, the typical states of our ensemble have a
distribution almost identical to \eqref{typ_J=0}.  We will call
the distribution \eqref{typ_J=0} the ``representative''
distribution.

For simplicity, we will compute the 2-point  functions of non-twist
``probe'' operators $\CA$ in states created by general twist
operators.   $\CA$ can be written as a sum over copies of the CFT:
\begin{align}
 \CA&={1\over \sqrt{N}}\sum_{A=1}^N \CA_A\label{sum_CA_A2}
\end{align}
where $\CA_A$ is a {\em non-twist\/} operator that lives  in the
$A$-th copy.  For example, we can take
\begin{align}
 \CA_A=\partial X^a_A(z) \bar \partial X^b_A(\zb),
 \label{A_grvtn2}
\end{align}
which corresponds to a fluctuation of the metric in the internal
$T^4$ direction.  Although, such non-twist operators are only a
subset of the operators that correspond to spacetime excitations, we
will restrict ourselves to them because their correlation functions
are much easier to compute than those of twist operators, and
because they will be sufficient to  demonstrate that an effective
geometry emerges in the $N\to \infty$ limit.

Given a general Ramond ground state $\sigma$ (\ref{gen_twist2})  we
are interested in computing
\begin{equation}
\langle \sigma^\dagger \CA^\dagger \CA \sigma \rangle
\end{equation}
The key result, demonstrated in the Appendix of \cite{Balasubramanian:2005qu}, is that for non-twist
operators at the orbifold point in the CFT such correlation
functions decompose into independent contributions from the
constituent twists operators in (\ref{gen_twist2}).  
Using this, it is shown in \cite{Balasubramanian:2005qu} that
for a bosonic $\CA$, we
obtain
\begin{align}
 \bracket{\CA(w_1) \CA(w_2)}_{\Sigma}
 ={1\over N}\sum_{n} n N_n \sum_{k=0}^{n-1}
 {C \over \left[2 n \sin\left({w-2\pi k\over 2n}\right)\right]^{2h}
 \left[2n \sin\left({\wb-2\pi k \over 2n}\right)\right]^{2\tilde h}},
 \label{AASigma}
\end{align}
where
\begin{align}
 N_n\equiv \sum_\mu (N_{n\mu}+ N_{n\mu}') \, . \label{mnmc14Jul05}
\end{align}
Here $w =\phi - t / L$ and $\bar{w} = \phi + t/L$ with $L$ being the AdS scale are lightcone coordinates in boundary CFT.

Let us study the relative size of the contributions  to this from
terms with different $n$.  The contributions come multiplied by $n
N_{n}$, which is $8n\over \sinh\beta n$ for the typical  microstates
with $J=0$ (Eq.\ \eqref{typ_J=0}).  Because of the suppression by
the $\sinh \beta n$,  the values of $n$ that make substantial
contributions to the correlation function \eqref{AASigma} are $n
\lesssim 1/\beta\sim \sqrt{N}$.
Thus there are $O(\sqrt{N})$ twists
that make a significant contribution.   Now observe that for any
$\gamma < 1/2$,  the number of twists with $n \lesssim N^\gamma$ is
parametrically smaller than $\sqrt{N}$. Indeed, the ratio vanishes
as $N \to \infty$.  In this sense we can say that in the $N \to
\infty$ limit, (\ref{AASigma}) is dominated by twists scaling
as $n \sim \sqrt{N}$.

Next, for any $n\ge 1$, when $t \ll n$ we can approximate the sum on $k$ as
\begin{align}
 \sum_{k=0}^{n-1}
 {1 \over \left[2 n \sin\left({w-2\pi k\over 2n}\right)\right]^{2h}
 \left[2n \sin\left({\wb-2\pi k \over 2n}\right)\right]^{2\tilde h}}
 \approx
 \sum_{k=-\infty}^{\infty}
 {1 \over (w-2\pi k)^{2h} (\wb-2\pi k)^{2\tilde h}}\qquad
 (t\ll n),
\notag
\end{align}
where we assumed $h+\tilde h={\rm even}$. Putting the above statements together, we arrive at the following conclusion: for sufficiently early times
\begin{align}
 t\ll t_c=\CO(\sqrt{N}),
\end{align}
the correlation function \eqref{AASigma} can be approximated by
\begin{align}
 \bracket{\CA(w_1) \CA(w_2)}_{\Sigma}
 &\approx {1\over N}\sum_{n} n N_n
 \sum_{k=-\infty}^{\infty}
 {C \over (w-2\pi k)^{2h} (\wb-2\pi k)^{2\tilde h}}\notag\\
 &=
 \sum_{k=-\infty}^{\infty}
 {C \over (w-2\pi k)^{2h} (\wb-2\pi k)^{2\tilde h}}.\label{eff_cor_J=0bos}
\end{align}
 This  turns out to be 
{\it precisely} the correlation function computed in  the
 $M=0$ BTZ black hole background \cite{Balasubramanian:2005qu}.
 Therefore, in the orbifold CFT approximation, the
emergent effective geometry of the D1-D5 system is the $M=0$ BTZ
black hole geometry. The description in terms of this effective
geometry is valid until $t\sim t_c \sim O(S)$.   Here $e^S$ is the statistical degeneracy of the Ramond ground states which goes to infinity in the semiclassical limit where $N\to\infty$.

Notice that in (\ref{eff_cor_J=0bos}) the sum  over the twists $n$
factors out.   Thus, for $t< t_c$ we are showing that the
correlation function is largely independent of the detailed
microscopic distribution of twists.  It is this universal response
that reproduces the physics of the $M=0$ BTZ black hole.
After $t\sim t_c$, the approximation \eqref{eff_cor_J=0bos} breaks
down, and the correlation function starts to show random-looking,
quasi-periodic behavior (see the figures in \cite{Balasubramanian:2005qu}). The form of the
correlation function in this regime will depend on the precise form
of the individual microstate, no matter how close it is to the
representative state \eqref{typ_J=0}.

Note that the $M=0$ BTZ black
hole yields a correlator which decays to zero at large times as $1/t^2$.
By contrast, the microstate correlators exhibit quasi-periodic
fluctuations around a nonzero mean value. Numerical analysis indicates
that this mean value scales as ${1 \over \sqrt{N}}$ for $h=\tilde
h=1$. For an ordinary finite size, finite temperature system, one
expects the mean value to be of order $e^{-c S}$ where $S$ is the
entropy and $c$ is of order $1$.  This behavior arises because typical
interactions can explore the entire phase space of the system.  The fact
that we observe power law rather than exponential behavior is partly a result of working in the free orbifold limit of the CFT and probing the system with only non-twist operators, so that the full space of states does not come into play.    It is also worth noting that the generic state in the D1-D5 Hilbert space is a random {\it superposition} of basis states of the form (\ref{gen_twist2}).     Then the arguments of Sec.~\ref{sec:entropicsupp} and \cite{Balasubramanian:2007qv} show that over the entire Hilbert space the variance of observables will be suppressed exponentially in the entropy.

Finally, a finite $N$ microstate correlator will exhibit exact periodicity in
time because only a finite number of frequencies appear in the
Fourier expansion.  The frequencies are $\omega_n = {n\over N},~ n=
1, 2, \ldots , N$.   Let $L(N)$ denote the least common multiple of
$(1,2, \ldots, N)$.   The correlator is then periodic with period
$\Delta t = 2\pi N L(N)$.    The large $N$ behavior of $L(N)$ is
$L(N) \sim e^N$, and therefore
\begin{align}
\Delta t \sim N e^N~.
\end{align}
Our correlators have been computed in the canonical ensemble in
which the summation over $n$ extends past $N$ up to infinity, and so
we will not see this exact periodicity.  On the other hand, due to
the exponential suppression of the distribution function $N_n$ the
deviation from exact periodicity is tiny for large $N$. As was
argued above, and as can be confirmed numerically,  one finds that
for large $N$ the large time behavior of the correlator is
unaffected if we truncate the sum over $n$ at $n_{\rm max} = c
\sqrt{N}$, for $c$ of order unity. Taking this into account, we see
that our correlators will exhibit approximate periodicity with
period
\begin{align}
\Delta t \sim e^{c \sqrt{N}} = e^{\tilde{c} S} ~,
\end{align}
where $S=2\pi \sqrt{2}\sqrt{N}$ is the entropy.  This timescale is
the so-called Poincar\'{e} recurrence time, over which generic
finite size thermal systems are expected to exhibit approximate
periodicity.

\subsection{Summary}
In this section we described how appropriately coherent BPS states and density
matrices of the D1-D5 string can be mapped to specific geometries in
asymptotically AdS$_3$ spacetimes.     While specific coherent microstates
mapped to  particular horizon-free geometries, the thermal density matrices
mapped to the BTZ M=0 black hole and the small black ring.    The techniques for
producing this mapping were in analogy to the phase methods used in the previous
section to develop the relation between half-BPS geometries of $\ads{5}$ and the
dual Yang-Mills theory.   We then showed that the {\it typical} BPS microstates
of the D1-D5 string, reacts to probes in such a way that their effective
description, on timescales that go to infinity in the semiclassical limit, is as
a BTZ black hole.  This is despite the fact that the individual coherent
microstates can be mapped as described above to horizon-free BPS geometries.
Note that the latter results were demonstrated in the orbifold limit of the
D1-D5 CFT where the theory is actually free.    This is possible because the
essence of the problem of the emergence of black-hole like behavior is not the
presence of interactions per se, but rather the enormous underlying degeneracy.

\section{AdS$_3$xS$^2$} \label{sec5}

Although the extremal $D1$-$D5$ system has proved a fertile example to
test the idea that black holes are simply effective geometries, the extremal
solution has a horizon coincident with the singularity and thus a finite
horizon area only (possibly) emerges when higher derivative corrections are included.  Thus this example
has some special features that do not generalize.   It would be desirable to be
able to study a similar scenario for a system where the total charge corresponds
to a black hole with a macroscopic horizon  (i.e. a three charge black hole).
Such black holes are 1/8 BPS solutions in the full string theory or can emerge
as 1/2 BPS solutions of $\mathcal{N}=2$ four dimensional or $\mathcal{N}=1$ five
dimensional supergravity (i.e. string or $M$-theory reduced on a Calabi-Yau).

\subsection{Solution Spaces}\label{sec_sol_spaces}

The general multi-centered BPS solutions of generic
$\mathcal{N} = 2$ supergravity theories in four dimensions were constructed in \cite{Behrndt:1998ns, Denef:2002ru, Bates:2003vx}, while  \cite{Bena:2004de} classified the full set of BPS solutions for the
special case of the five-dimensional $\mathcal{N}=2$
supergravity theory which is the truncation of the $\mathcal{N}=8$
theory (i.e. the theory is invariant under 8 instead of 32
supersymmetries). The latter require specifying a four-dimensional
base metric that is restricted to be hyperk\"ahler \cite{Gauntlett:2002nw}.  A
particularly appealing class of hyperk\"ahler manifolds are
Gibbons-Hawking or multi-Taub-NUT geometries which are
asymptotically $\mathbb{R}^3\times$S$^1$ and for which we have
explicit metrics. Moreover, it has been shown that the five
dimensional solutions constructed using a Gibbons-Hawking base
manifold \cite{Bena:2005va} correspond to the four dimensional
ones via the 4d/5d connection \cite{Cheng:2006yq, Gaiotto:2005gf,  Kraus:2005gh} making them an interesting class of solutions to study \cite{Balasubramanian:2006gi}.

The five dimensional solutions, although relatively complicated, are determined
entirely in terms of $2b_2 + 2$ harmonic functions where $b_2$ is the second
Betti number of the compactification Calabi-Yau, $X$,
\begin{align}\label{harmonics}
\sH^0&=\sum_a\frac{p^0_a}{|\sx-\sx_a|} + h^0\,,\quad\ \,
\sH^A=\sum_a\frac{p^A_a}{|\sx-\sx_a|} + h^A\,,\quad\ \\
\sH_A&=\sum_a\frac{q_A^a}{\,|\sx-\sx_a|} + h_A\,,\quad\ \,
\sH_0=\sum_a\frac{q_0^a}{|\sx-\sx_a|} + h_0 \,.\nonumber
\end{align}
Here the coordinate vector $\sx_a$ gives the position in the spatial
$\mathbb{R}^3$ of the $a$'th center with charge
$\Gamma_a=(p^0_a,p^A_a,q^a_A,q_0^a)$ (note here $A$ runs from $1, \dots, b_2$).
The IIA interpretation of these charges is (D6,D4,D2,D0) wrapping cycles of $X$
while in M-theory the charge vector is (KK,M5,M2,P).   Note that the harmonics
have $2b_2 + 2$ constants $h = (h^0, h^A, h_A, h_0)$ that together determine the
asymptotic behaviour of the harmonics and hence the solutions.  We will also
have frequent occasion to use the notation $\Gamma = (p^0, p^A, q_A, q_0)$ to refer to
the total charge $\Gamma = \sum_a \Gamma_a$.

The position vectors have to satisfy the integrability constraints
\begin{equation} \label{integrabconstr}
    \sum_b\frac{\langle\Gamma_a,\Gamma_b\rangle}{|\sx_a-\sx_b|}= \langle h, \Gamma_a
    \rangle \,,
\end{equation}
where we define the symplectic intersection product
\begin{equation}
 \langle \Gamma_1,\Gamma_2 \rangle := - p_1^0 q^2_0 + p_1^A q^2_A - q^1_A p_2^A
 +  q^1_0 p_2^0.
\end{equation}
By summing (\ref{integrabconstr}) over $a$ we find that the constants $h$ have
to obey $\langle h,\Gamma \rangle=0$. Note that even once the charges of each
center have been fixed there is a large space of solutions that may even have
several disconnected components. In particular, the constraint
(\ref{integrabconstr}) implies that the positions of the centers are generally
restricted, defining a complicated moduli space of (generically) bound
solutions.  It will turn out that in order to yield bound states in the quantum
theory a symplectic form, given in Sec.~\ref{sec:sympform}, must be
non-degenerate on these moduli spaces of solutions.

The metric, gauge field and K\"ahler scalars of the solution are
now given in terms of the harmonics by
\begin{eqnarray}
ds^2_{5d}&=&2^{-2/3}\,Q^{-2}\left[-(\sH^0)^2(dt+\omega)^2-2L(dt
+\omega)(d\psi+\omega_0)
+\Sigma^2(d\psi+\omega_0)^2\right]\nonumber\\&&+2^{-2/3}\,Q\,d\sx^id\sx^i\,,\label{multicentersol}\\
A^A_{5d}&=&\frac{-\sH^0}{Q^{3/2}}(dt+\omega)+\frac{1}{\sH^0}\left(\sH^A-\frac{Ly^A}{Q^{3/2}}\right)(d\psi+\omega_0)+\mathcal{A}^A_d\,,\nonumber\\
Y^A&=&\frac{2^{1/3}y^A}{\sqrt{Q}}\,,\nonumber
\end{eqnarray}
where $\sx^i\in \mathbb{R}^3$ and $\psi$ is an angular coordinate with
period $4\pi$, and the functions appearing satisfy the relations
\begin{eqnarray}
 d\omega_0 & = &\star d\sH^0\,,\nonumber\\
 d\mathcal{A}_d^A & = &\star d\sH^A \,,\nonumber\\
 \star d\omega & = & \langle d\sH,\sH\rangle \,\nonumber\\
 \Sigma&=&\sqrt{\frac{Q^3-L^2}{(\sH^0)^2}}\,,\label{conditions}\\
 L&=&\sH_0(\sH^0)^2+\frac{1}{3}D_{ABC}\sH^A\sH^B\sH^C-\sH^A\sH_AH^0\,,\nonumber\\
 Q&=&(\frac{1}{3}D_{ABC}y^Ay^By^C)^{2/3}\,,\nonumber\\
 D_{ABC}y^Ay^B&=&-2\sH_C\sH^0+D_{ABC}\,\sH^A\sH^B\,.\nonumber
\end{eqnarray}
Here the Hodge star is with respect to the flat $\mathbb{R}^3$ spanned by the
coordinates $\sx^i$ and $D_{ABC}$ are the triple intersection numbers of the
chosen basis of $H^2(X)$. Note that the only equation in (\ref{conditions}) for
which there is no general solution in closed form is the last one.  In some
cases, e.g.\ when $b_2=1$, it is even possible to obtain a solution, in closed
form, to this equation.

The function $\Sigma$ appearing in the metric in (\ref{multicentersol}) is known as
the entropy function.  When evaluated at $\vec{x}_a$ it is proportional to the
entropy of a black hole carrying the charge of the center lying at $\vec{x}_a$.  This follows from the
Bekenstein-Hawking relation and the fact that $\Sigma$ would determine the area of
the horizon of a possible black hole at $\vec{x}_a$.  If this area is zero then the center at $\vec{x}_a$
does not have any macroscopic entropy and, if the associated geometry does not
suffer from large curvature in this region, then there is no reason to believe
stringy corrections will change this.    It was shown in
\cite{Berglund:2005vb,Balasubramanian:2006gi} that in classical ${\cal N} = 8$ supergravity, zero entropy smooth centers carry charges that are half-BPS.   The half-BPS nature of the charge vector follows from smoothness.  One way to see this is, taking $r$ to be the distance from a half-BPS center,  the associated harmonic functions would lead to a scaling of the form $r^{-1/2}$ of the entropy function
$\Sigma$ (as opposed to the $r^{-2}$ behavior near a black hole).  Such
half-BPS charge vectors, which we think of as ``zero-entropy bits",  are identified by vanishing of the entropy function, as well as its first and second derivatives \cite{Ferrara:1997ci}.    There will be some appropriate generalization of this condition to the general Calabi-Yau compactification, which  we could analyze  by studying the falloff of the metric near a center in the general Calabi-Yau.     While in classical ${\cal N} = 8$ supergravity the ``zero-entropy bits'' have to be 1/2-BPS, it is conceivable that less supersymmetric centers carrying zero entropy could also serve as ``atomic constituents'' of black holes, with higher derivative corrections and stringy degrees of freedom making them smooth.

A center with a half-BPS charge $\Gamma = (1, p/2, p^2/8, p^3/48)$ corresponds
to a single D6 brane wrapping the Calabi-Yau with all lower-dimensional charges
induced by abelian flux.  A configuration with a single such center can be
spectral flowed (see e.g. \cite{Bena:2005ni, deBoer:2008fk}) to a single D6
brane with no flux and hence no additional degrees of freedom in the Calabi-Yau;
thus ``integrating out'' the Calabi-Yau degrees of freedom does not generate an
entropy and the associated five dimensional solution is smooth.   As discussed
in \cite{Balasubramanian:2006gi, Bena:2006kb}, ``zero-entropy bits'' can also be
D4 and D2-branes with flux or D0 branes.   Generically they can carry a
``primitive'' half-BPS charge vector, i.e. some number of D6-branes with
fractional fluxes such that the induced D2, D4 and D0 brane charges are all
quantized, but have no common factor.   Note that if the D6-brane charge is
$N>1$, the solution is not strictly smooth -- there is relatively mild orbifold
singularity ($R^4 / Z_N$) in the five-dimensional theory.  

The properties of these solutions and the philosophy outlined in this paper
motivate a conjecture similar to the one in   \cite{Balasubramanian:2006gi}:
Every supersymmetric 4D black hole of finite area, preserving 4 supercharges,
can be split up into microstates made up of zero-entropy ``atoms''.    In the
context of ${\cal N} = 8$ supergravity a somewhat more restrictive conjecture
was stated in \cite{Balasubramanian:2006gi}, adding further that the
zero-entropy ``atoms'' would be 1/2-BPS (16 supercharges), and that the
``atoms'' were bound by mutual non-locality of their charges.

Multicenter configurations with every center constrained to be of the above form
have been studied in \cite{Bena:2005va, Bena:2006is, Berglund:2005vb,
Balasubramanian:2006gi} and numerous other works by the same authors.  Note that
the associated four dimensional solution can have singularities associated to
Kaluza-Klein reduction on a non-trivial S$^1$ fibration.  This high-lights an
important distinction: while the entropy, determined by replacing $H^A$ in the
definition of $\Sigma(H)$ with $\Gamma^A$, is a duality invariant notion, the
smoothness of the resulting supergravity solution is not (see e.g.
\cite{Bena:2008wt}).  Thus we will assume that solutions with all centers that
are ``zero-entropy bits'' (in the sense of vanishing ``microscopic'' entropy)
are candidate ``microstate geometries'' (in the sense described in Section
\ref{sec_rel_microstates}) even though they may have naked singularities in some
duality frames.

From (\ref{multicentersol}) and (\ref{conditions}) it may seem
that the solutions are singular if $\sH^0$ vanishes but this is
not the case as various terms in $Q$ and $L$ cancel any possible
divergences due to negative powers of $\sH^0$ (in fact, the BTZ
black hole can, in the decoupling limit introduced in the next
section, be mapped to such a solution with $\sH^0$ vanishing
everywhere).

An additional complication is the fact that even solutions satisfying the
constraint equation (\ref{integrabconstr}) may still suffer from various
pathologies, most notably closed time-like curves (CTCs).  For instance, the
prefactor to the $d\psi^2$ term in the metric may become negative if $\Sigma$
becomes imaginary\footnote{As described in \cite{Cheng:2006yq}  this would also
imply that the 4-dimensional metric associated with this 5-d solution (via the
4d/5d connection of \cite{Gaiotto:2005gf}) becomes imaginary as $\Sigma$ appears
directly in the former.}. Unfortunately there is no simple criterion which can
be used to determine if a given solution is pathology free.  To fill in this gap
\cite{Denef:2000nb} and \cite{Denef:2007vg} devised the {\it attractor flow
conjecture}, a putative criterion for the existence of (well-behaved) solutions which we
will describe in section \ref{sec_split_attr}.

An essential feature of these solutions is that they are stationary but not
static.  In particular they carry quantized intrinsic angular momentum
associated with the crossed electric and magnetic fields of the dyonic centers
\cite{Denef:2000nb}

\begin{equation}
    \vec{J}=\frac{1}{2}\sum_{ a<  b} \frac{\langle
    \Gamma_a,\Gamma_b\rangle\,\vec{\sx}_{ab}}{r_{ab}}\label{angularmomentum}\,.
\end{equation}

\noindent This will be important when quantizing the
solution space as it is a necessary (but not sufficient) condition for the
latter to be a proper phase space with a non-degenerate
symplectic form.  A solution space with vanishing angular momentum does not
enjoy this property and must be completed to a phase space by the addition of
velocities (see e.g. \cite{Michelson:1999dx}).

\subsection{Decoupling}

Some specific five dimensional solutions obtained from an
$\mathcal{N} =2$ truncation of $\mathcal{N} = 8$ (i.e.
compactification on $T^6$) have been studied using AdS/CFT
\cite{Bena:2004tk, Elvang:2004ds} by taking a decoupling limit of
a dualized form of the solutions.  This cannot be generalized to
an arbitrary compact CY because the duality group of the latter is
not known.  To study the full class of solutions using AdS/CFT it
is desirable to have a more general decoupling limit that can
embed a larger class of these solutions in an AdS$_3$ throat.  The
limit we will describe can be taken in a proper $\mathcal{N}=2$
theory and will yield an AdS$_3\times$S$^2$ near horizon geometry
\cite{deBoer:2008zn}.  This decoupling limit only works for
solutions whose total charge does not contain any
overall $D6$/KK-monopole charge so the relevant CFT is essentially
the MSW CFT\footnote{The ``MSW'' CFT is the theory that arises as
the low-energy description of $M5$-branes wrapping an ample
divisor in the Calabi-Yau. It is an $N=(0,4)$ superconformal field
theory and it owes its name to the three authors of
\cite{Maldacena:1997de}.}. Although the latter is not under very
good control it is nonetheless possible to determine, from the
asymptotics of a given geometry, the CFT quantum numbers of the
corresponding state. It is also possible to use general CFT
properties to determine various quantities such as the number of
states in a given charge sector.  

In \cite{deBoer:2008zn} the decoupling limit of the solutions described above is
defined by taking $\ell_5 \rightarrow 0$ ($\ell_5$ is the 5-d plank length)
while keeping fixed the mass of $M2$ branes stretched between the various
centers and wrapping the $M$-theory circle.  In doing so we also fix the volume
of the Calabi-Yau as measured in
5-d plank units and the length of the $M$-theory circle, $R$. Since the mass of such membranes is given by $m_{M2} \sim
R\Delta \mathsf{x}/ \ell_5^3$, the coordinate distances between the centers,
$\Delta \mathsf{x}$, must be rescaled as $\ell_5^3$.  Alternatively, we can see this limit
as a rescaling of the 5-d metric under which the Einstein part of the action is
invariant.

We now define new rescaled coordinates, $x^i$, and harmonic functions, $H$, as
\begin{equation}
    x^i = \ell_5^{-3}\sx^i \qquad H = \ell_5^{3/2} \sH
\end{equation}
By restricting to the region of finite $x^i$ we are essentially keeping the mass
of transverse, open membranes fixed while rescaling the original coordinates,
$\sx^i$.  One can see that, in these new variables, the structure of the
solution (in terms of the harmonics) does not change in the decoupling limit
except for an overall scaling of the metric by a factor of $\ell_5^{-2}$.  The
rescaled harmonics do take a new form, however,
\begin{align}
H^0&=\sum_a\frac{p^0_a}{|x-x_a|} \,,\quad\ \,
H^A=\sum_a\frac{p^A_a}{|x-x_a|} \,,\quad\ \\
H_A&=\sum_a\frac{q_A^a}{\,|x-x_a|} \,,\quad\ \,
H_0=\sum_a\frac{q_0^a}{|x-x_a|} - \frac{R^{3/2}}{4} \,.\nonumber
\end{align}
In particular note that all the constants have disappeared except
the $D0$-brane constant which now takes a fixed value (See section \ref{sec_split_attr} for the expressions of $h$).  Related to
this is the fact that the asymptotic value of the moduli are
forced to the attractor point, $Y^A \sim p^A$ (this corresponds to
sending the 4-d K\"ahler moduli to $J^A = \infty\, p^A$).

Recall that the coordinate locations of the centers must satisfy the
integrability constraint (\ref{integrabconstr}) and that this constraint depends
on the value of the constants in the harmonic functions.  As a consequence it is
possible that, in taking the decoupling limit, some solutions cease to exist.
For instance if we consider two $D4$-$D2$-$D0$ centers, $\Gamma_a$ and
$\Gamma_b$, then the constraint equation in the decoupling limit is
\begin{equation}\label{2cntr_decay}
    \frac{\langle \Gamma_a, \Gamma_b \rangle}{x_{ab}} = h_0 p^0 = 0
\end{equation}
Implying that $x_{ab}$, the inter-center distance, must be infinite (unless the
charges are parallel, $\langle \Gamma_a, \Gamma_b \rangle = 0$) so one of the
centers is forced out of the finite region of the rescaled coordinates, $x^i$,
as we take the limit.  We interpret this as implying that such centers cannot
sit in the same decoupled AdS$_3$ throat.

Because only the $D0$ constant survives, centers without $D6$ charge are no
longer bound together.  A related fact,  which will emerge presently, is that
the solution space of such centers does not have a non-degenerate symplectic
form (because, in the decoupling limit, intrinsic angular momentum is
proportional to $D6$ dipole-moment)  on it and hence cannot be quantized without
the addition of additional degrees of freedom.

Even if a set of charges admits a solutions that satisfies the constraint
equations in the decoupling limit the solution may develop other pathologies such as
CTCs.  To study these we may resort once more to the attractor flow conjecture
of \cite{Denef:2000nb} which is argued, in \cite{deBoer:2008fk}, to extend to the
decoupled solutions.  We will not discuss this in any depth here except to note
that the fact that the asymptotic moduli, $Y^A$, are forced to the attractor
point for the total charge implies that only attractor flow trees which can be
extended to trees starting from this value of the asymptotic moduli will survive
the decoupling limit.  Further discussion can be found in Section \ref{sec_split_attr}
and \cite{deBoer:2008fk}.

The decoupled solutions are asymptotically AdS$_3$xS$^2$ and
their asymptotic form is given by

\begin{eqnarray}
    ds^2_{\mathrm{5D}}&=&-\frac{\rho^2}{4U^2}dvd\psi+\frac{U^{-4}}{4}\biggl[-R^2(d^0)^2dv^2+\mathcal{D} d\psi^2\biggr]\nonumber\\
&&+4U^2\frac{d\rho^2}{\rho^2}+U^2\biggl(d\theta^2+\sin^2\theta
d\alpha^2 \biggr)+\mathcal{O}(\frac{1}{\rho^2})\,,\label{asympmetric}\\
A^A_{\mathrm{5D}}&=&-p^A\cos\theta d\alpha-D^{AB}q_Bd\psi+\mathcal{O}(\frac{1}{\rho^2})\,,\label{asympgauge}\\
Y^A&=&\frac{p^A}{U}+\mathcal{O}(\frac{1}{r^2})\,.
\end{eqnarray}
where we have introduced some coordinate redefinitions.  We first introduce
standard spherical coordinates, $(r, \theta, \phi)$, on the base spatial
$\mathbb{R}^3$ of the solution with the axis of the sphere (the $z$-axis)
aligned with the $D6$-dipole moment of the solution
\begin{equation}
    \vec{d}^0 := \sum_a p^0_a \vec{x}_a
\end{equation}
For brevity we introduce the notation $d^0 = |\vec{d}^0|$ and $\vec{e} =
\vec{r}/r$ so $d^0$ is the norm of the dipole moment and $\vec{e}$ is a unit vector
in the radial direction.  We then make a further coordinate redefinition
\begin{equation}
 v=t-\frac{R}{4}\psi\,,
\qquad \alpha=\phi+Rd^0\biggl(\frac{p^3}{3}\biggr)^{-1}v\,,\qquad
U^3=\frac{p^3}{6}\quad \mbox{and }\quad\mathcal{D}=\frac{p^3}{3}\biggl(D^{AB}q_Aq_B-2q_0\biggr)\,,\label{ascoords}
\end{equation}
and define a new radial coordinate $\rho$ via
\begin{equation}
\frac{\rho^2}{4U^2}=-\frac{U^{-4}}{2}R\biggl(\frac{e\cdot d^AD_{ABC}p^Bp^C}{3}
-\frac{p^Aq_Ad^0\cos\theta}{3}\biggr)+\frac{R}{U}r\,,
\end{equation}
Here $\vec{d}^A$ is the $D4$-dipole moment, defined analogously to
$\vec{d}^0$. To make a connection to the dual CFT the solution
needs to be reduced along the S$^2$ to give a theory defined
purely on AdS$_3$ with KK modes in representations of $SU(2)$.
This procedure is reviewed in some detail in \cite{Kraus:2006wn}
with particular attention to the subtleties involved in reducing
the 5-d Chern-Simons terms in the supergravity action.

The resultant metric is asymptotically AdS$_3$ and from this, and the asymptotic
form of the gauge field, we can determine the charges in dual field theory (see
\cite{deBoer:2008zn} for details).  In particular we find that

\begin{eqnarray}
L_0&=&\frac{(p^A \, q_A)^2}{2 p^3} - q_0 +\frac{p^3}{24}\,, \label{virasorocharges}\\
\tilde L_0&=&\frac{(p^A \, q_A)^2}{2 p^3} +\frac{p^3}{24}\,.\nonumber
\end{eqnarray}

\noindent and that the $SU(2)$ R-symmetry charge associated with a solution is
determined entirely in terms of its $D6$ dipole moment

\begin{equation}
J^3_0=-\frac{R^2d^0}{8}\label{SU2charge}
\end{equation}

\noindent While (\ref{virasorocharges}) gives the charges expected from general
considerations of the MSW CFT associated with the total charge $\Gamma$, the
$SU(2)$ charge, $J^3_0$, depends on a dipole moment and, as such, is absent in
the single center solution.  In fact, it is nothing more than
the intrinsic angular momentum, $J$, defined in (\ref{angularmomentum}) which,
in the decoupling limit, can be shown to be proportional to the $D6$ dipole moment.

In the next section we will see that this charge plays a distinguished role in
the quantization of the system as its presence is necessary in order to have  a
non-trivial symplectic form on the phase space.

\subsection{Split Attractors and State Counting}\label{sec_split_attr}

In \cite{Denef:2000nb} a conjecture is proposed whereby pathology-free solutions
are those with a corresponding {\em attractor flow tree} in the moduli space.
This conjecture was first posed for multicentered four-dimensional solutions so
we will introduce some four dimensional terminology here.  The four dimensional
moduli, $t^A(\vec{x}) = B^A(\vec{x}) + i J^A(\vec{x})$, are the complexified
K\"ahler moduli of the Calabi-Yau.  The relation between these moduli and their
five dimensional counterparts can be found in \cite{Gaiotto:2005gf}
\cite{deBoer:2008fk}.  To each charge vector, $\Gamma_i$, we can associate a
complex number, the central charge, as
\begin{equation}
Z(\Gamma_i; t) := \langle \Gamma_i, \Omega(t) \rangle \qquad \Omega(t):=
-\frac{e^{t}}{\sqrt{\frac{4}{3} J^3}}
\end{equation}
Note that, since $t^A$ is a two-form, $\Omega$ is a sum of even degree forms.
The phase of the central charge, $\alpha(\Gamma_i) := \textrm{arg}[Z(\Gamma_i; t)]$,
encodes the supersymmetry preserved by that charge at the given value of the
moduli.  The even form $\Omega$ is related, asymptotically, to the constants in
the harmonics (\ref{harmonics}) (which define both the 4-d and 5-d solutions) as
\begin{equation}
    h = -2\, \textrm{Im}(e^{-i\alpha(\Gamma)}\, \Omega)|_\infty
\end{equation}
For more details of the 4-d solution the reader should consult
\cite{Bates:2003vx}.

\begin{figure}[h]
    \begin{center}
    \includegraphics[scale=0.3,angle=-90]{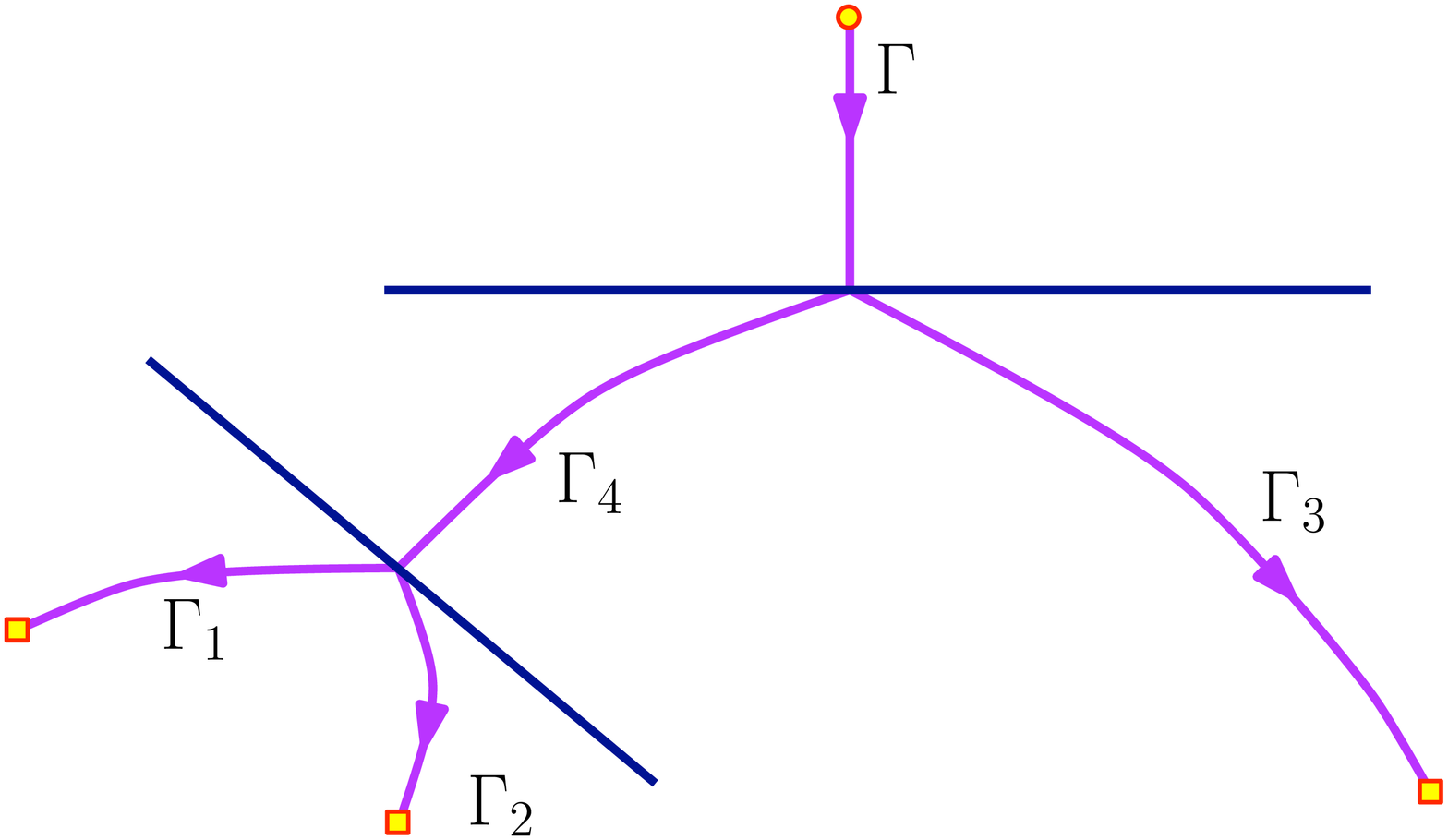}
    \caption{Sketch of a 3-center attractor flow tree from \cite{Denef:2007vg}
    \cite{deBoer:2008fk}. The dark blue lines are lines of marginal stability,
    the purple lines are single center attractor flows.  The tree starts at the
    yellow circle (the moduli at infinity) and flows towards the attractor
    points indicated by the yellow boxes.  Note here that $\Gamma_4 = \Gamma_1 +
    \Gamma_2$ and $\Gamma = \Gamma_4 + \Gamma_3$.  On the walls of marginal
    stability the moduli are such that $|Z(\Gamma; t)| = |Z(\Gamma_3;t)| +
    |Z(\Gamma_4; t)| $ (horizontal wall on top) and $|Z(\Gamma_4; t)| =
    |Z(\Gamma_1; t)| + |Z(\Gamma_2;t)|$ (diagonal wall on bottom
    left).}\label{fig_split}
\end{center}
\end{figure}

An {\em attractor flow tree} is a graph in the Calabi-Yau moduli space beginning
at the moduli at infinity, $t^A|_\infty$, and ending at the attractor points for
each center.  The edges correspond to single center flows towards the attractor
point for the sum of charges further down the tree.  Vertices can occur where
single center flows (for a charge $\Gamma = \Gamma_1 + \Gamma_2$) cross walls of
marginal stability where the central charges are all aligned ($|Z(\Gamma)| =
|Z(\Gamma_1)| + |Z(\Gamma_2)|$).  The actual flow of the moduli $t^A(\vec{x})$
for a multi-centered solution will then be a thickening of this graph (see
\cite{Denef:2000nb}, \cite{Denef:2007vg} for more details).  According to the
conjecture a given attractor flow  tree will correspond to a single connected
set of solutions to the equations (\ref{integrabconstr}), all of which will be
well-behaved.  An example of such a flow is given in figure \ref{fig_split}.

The intuition behind this proposal is based on studying the two center solution
for charges $\Gamma_1$ and $\Gamma_2$.  The constraint equations
(\ref{integrabconstr}) imply that when the moduli at infinity are moved near a
wall of marginal stability (where $Z_1$ and $Z_2$ are parallel) the centers are
forced infinitely far apart
\begin{equation}\label{2centersep}
    r_{12} =
    \frac{\langle\Gamma_1, \Gamma_2\rangle}{\langle h , \Gamma_1 \rangle } =
    \frac{\langle\Gamma_1, \Gamma_2\rangle\, |Z_1 + Z_2|}{2 \,\mathrm{Im}(\bar Z_2
    Z_1)}\biggr|_\infty
\end{equation}
In this regime the actual flows in moduli space are well
approximated by the split attractor trees since the centers are so
far apart that the moduli will assume single-center behaviour in a
large region of spacetime around each center.  Thus in this regime
the conjecture is well motivated.  Varying the moduli at infinity
continuously should not alter the BPS state count, which
corresponds to the quantization of the two center moduli space, so
unless the moduli cross a wall of marginal stability we expect
solutions smoothly connected to these to also be well defined.
Extending this logic to the general $N$ center case requires an
assumption that it is always possible to tune the moduli such that
the $N$ centers can be forced to decay into two clusters that
effectively mimic the two center case.  There is no general
argument that this should be the case but one can run the logic in
reverse, building certain large classes of solutions by bringing
in charges pairwise from infinity and this can be understood in
terms of attractor flow trees.  What is not clear is that all
solutions can be constructed in this way.  For more discussion the
reader should consult \cite{Denef:2000ar}.

Although this conjecture was initially proposed for asymptotically flat
solutions, in \cite{deBoer:2008zn} it was argued that the essential features of
the attractor flow conjecture continue to hold in the decoupling limit.

For generic charges the attractor flow conjecture also provides a
way to determine the entropy of a given solution space.  The idea is that the
entropy of a given total charge is the sum of the entropy of each possible
attractor flow tree associated with it. Thus the partition function receives
contributions from all possible trees associated with a given total charge and
specific {\em moduli at infinity}.  An immediate corollary of this is that, as
emphasized in \cite{Denef:2007vg}, the partition function depends on the
asymptotic moduli.  As the latter are varied certain attractor trees will cease
to exist; specifically, a tree ceases to contribute when the moduli at infinity
cross a wall of marginal stability (MS) for its first vertex, $\Gamma
\rightarrow \Gamma_1 + \Gamma_2$, as is evident from (\ref{2centersep}).

For two center solutions one can determine the entropy most easily
near marginal stability where the centers are infinitely far
apart.  In this regime locality suggests that the Hilbert state
contains a product of three factors\footnote{Since attractor flow
trees do not {\it have} to split at walls of marginal stability,
there will in general be other contributions to
$\mathcal{H}(\Gamma_1 + \Gamma_2; t_{ms})$ as well.}
\cite{Denef:2007vg}
\begin{equation}
    \mathcal{H}(\Gamma_1 + \Gamma_2; t_{ms}) \supset
    \mathcal{H}_{\textrm{int}}(\Gamma_1, \Gamma_2; t_{ms}) \otimes
    \mathcal{H}(\Gamma_1; t_{ms}) \otimes \mathcal{H}(\Gamma_2; t_{ms})
\end{equation}
Since the centers move infinitely far apart as $t_{ms}$ is
approached we do not expect them to interact in general. There is,
however, conserved angular momentum carried in the electromagnetic
fields sourced by the centers and this also yields a non-trivial
multiplet of quantum states.  Thus the claim is that
$\mathcal{H}_{\textrm{int}}$ is the Hilbert space of a single spin
$J$ multiplet where $J = \frac{1}{2} (|\langle \Gamma_1,
\Gamma_2\rangle| -1)$.\footnote{The unusual $-1$ in the definition
of $J$ comes from quantizing additional fermionic degrees of
freedom \cite{Denef:2002ru} \cite{deBoer:2008zn}.}
$\mathcal{H}(\Gamma_1)$ and $\mathcal{H}(\Gamma_2)$ are the
Hilbert spaces associated with BPS brane excitations in the
Calabi-Yau and their dimensions are given in terms of a suitable
entropy formula for the charges $\Gamma_1$ and $\Gamma_2$ valid at
$t_{ms}$.

Thus, if the moduli at infinity were to cross a wall of marginal stability for
the two center system above the associated Hilbert space would cease to
contribute to the entropy (or the index).  A similar analysis can be applied to
a more general multicentered configuration like that in figure \ref{fig_split}
by working iteratively down the tree and treating subtrees as though they correspond
to single centers with the combined total charge of all their nodes.  The idea
is, once more, that we can cluster charges into two clusters by tuning
the moduli and then treat the clusters effectively like individual charges.  We
can then iterate these arguments within each cluster.  This counting argument
mimics the constructive argument for building the solutions by bringing in
charges from infinity and is hence subject to the same caveats, discussed above.

Altogether the above ideas allow us to determine the entropy
associated with a particular attractor tree, which, by the split
attractor flow conjecture corresponds to a single connected
component of the solutions space.  The entropy of a tree is the
product of the angular momentum contribution from each vertex
(i.e. $|\langle \Gamma_1, \Gamma_2 \rangle|$, the dimension of
$\mathcal{H}_{\textrm{int}}$) times the entropy associated to each
node.  When we want to compare against the number of states
derived from quantizing the classical phase space, however, the
latter factor (from the nodes) will not be included as it is not
visible in the supergravity solutions.

In the following sections we will show that it is also possible, in the two and three
center cases, to quantize the solution space directly and to match the entropy so
derived with the entropy calculated using the split attractor tree.  This
provides a non-trivial check of both calculations.

Before proceeding to count the number of states associated to an attractor flow
tree we should mention an important subtlety in using the attractor flow
conjecture to classify and validate solutions.  Certain classes of charges will
admit so called {\em scaling solutions} \cite{Bena:2006kb} \cite{Denef:2007vg}
which are not amenable to study via attractor flows.  These solutions are
characterized by the fact that the constraint equations (\ref{integrabconstr})
have solutions that continue to exist at any value of the asymptotic moduli.  We
will discuss these solutions in greater detail in the next subsection but it is
important to note here that the general arguments given in this section
(such as counting of states via attractor flow trees) do not apply to scaling
solutions.

\subsection{Scaling Solutions}\label{scaling_sol}

It was observed in \cite{Denef:2002ru} that for certain choices of
charges it is possible to have points in the solution space where the coordinate
distances between the centers goes to zero.  Moreover, this occurs for any
choice of moduli so it is, in fact, a property of the charges alone.
Subsequently \cite{Bena:2006kb} noted that these ``scaling solutions'' develop
deep throats as the coordinate distance between the centers decreases, but that
the proper distance between the centers remains finite in the same limit.
Scaling solutions constructed from three centers, each of
them involving microscopic non-Abelian degrees of freedom, were emphasized in
\cite{Balasubramanian:2006gi} as likely dominant components of the underlying
state space.   The non-Abelian degrees of freedom were supposed to arise in this
picture from the open strings on the D-branes at each 1/2-BPS center, and
\cite{Denef:2007yt, Gimon:2007mha} appeared to confirm this picture by utilizing
these degrees of freedom to allow D0-branes in the solution to polarize by a
dielectric effect \cite{Myers:1999ps} into membranes.  The authors of
\cite{Denef:2007yt} argued that such microscopic configurations would dominate
the entropy of the black hole with same total charge as the solution as a whole.
The entropy of scaling solutions and their possible relation to single and
multi-centered black holes was further explored in \cite{Denef:2007vg,
deBoer:2008fk}.

Such solutions occur as follows.  We take the inter-center distances to be given
by $r_{ab} = \lambda \Gamma_{ab} + \mathcal{O}(\lambda^2)$ (fixing the order of
the $ab$ indices by requiring the leading term to always be positive).  As
$\lambda \rightarrow 0$ we can always solve (\ref{integrabconstr}) by
tweaking the $\lambda^2$ and higher terms.  The leading behaviour will be $r_{ab}
\sim \lambda \G_{ab}$ but clearly this is only possible if the $\G_{ab}$ satisfy
the triangle inequality.  Thus any three centers with intersection products $\G_{ab}$ satisfying the
triangle inequalities define a scaling solution.

We will in general refer to such solutions as {\em scaling solutions} meaning,
in particular, supergravity solutions corresponding to $\lambda \sim 0$.
The space of supergravity solutions continuously connected (by varying the
$\vec{x}_p$ continuously) to such solutions will be referred to as {\em scaling
solution spaces}.  We will, however, occasionally lapse and use the term scaling
solution to refer to the entire solution space connected to a scaling solution.
We hope the reader will be able to determine, from the context, whether a specific
supergravity solution or an entire solution space is intended.

These scaling solutions are interesting because (a) they exist for all values of
the moduli; (b) the coordinate distances between the centers go to zero; and (c)
an infinite throat forms as the scale factor in the metric blows up as
$\lambda^{-2}$.  Combining (b) and (c) we see that, although the centers naively
collapse on top of each other, the actual metric distance between them remains
finite in the $\lambda \rightarrow 0$ limit.  In this limit an infinite throat
develops looking much like the throat of a single center black hole with the
same charge as the total charge of all the centers.  Moreover, as this
configuration exists at any value of the moduli, it looks a lot more like a
single center black hole (when the latter exists) than generic non-scaling
solutions.  As a consequence of the moduli independence of these solutions it is
not clear how to understand them in the context of attractor flows; the
techniques developed in \cite{deBoer:2008zn} provide an alternative method to
quantize these solutions that applies even though the attractor tree does not.

Unlike the throat of a normal single center black hole the bottom of the scaling
throat has non-trivial structure.  If the charges, $\G_a$, are zero entropy bits
(e.g. D6's with flux) then the 5-dimensional uplifts of these solutions will
yield smooth solutions  in some duality frame  and the throat will not end in a
horizon but will be everywhere smooth, even at the bottom of the throat. Outside
the throat, however, such solutions are essentially indistinguishable from
single center black holes.  Thus such solutions have been argued to be ideal
candidate spacetime realizations of  microstates corresponding to single center
black holes \cite{Bena:2006kb, Balasubramanian:2006gi}.  In \cite{Denef:2007vg}
it was noticed that some of these configurations, when studied in the Higgs
branch of the associated quiver gauge theory, enjoy an exponential growth in the
number of states unlike their non-scaling cousins which have only polynomial
growth in the charges.

\subsection{Quantization}\label{subsec_quant}

As anticipated in \cite{Maoz:2005nk}, since the dual (0,4) CFT of
$\mathcal{N}=2$ black holes (lifted to 5-d) is less well
understood than its (4,4) cousin, in this case one can hope to
make progress by quantizing the phase space of the supergravity
solutions directly.   The quantization we will perform will be
quite general in that it will cover the original 4-d multi-center
black hole configurations \cite{Bates:2003vx}, their 5-d uplift
discussed in section \ref{sec_sol_spaces}, and the decoupled
version of the latter (which can be related to the (0,4) CFT). For
$\mathcal{N}=2$ black holes coming from Calabi-Yau
compactifications it is likely that a large portion of the entropy
will come from stringy degrees of freedom in the Calabi-Yau so it is likely to not
be possible to account for a finite fraction of the entropy using
supergravity alone. Whether this is the case or not is an
important question though the answer may be difficult to determine
as the solutions are rather complicated and even the classical
phase space is quite non-trivial.

A picture where black hole microstates are realized in spacetime as extended bound states would associate to a given total charge, $\Gamma$,  all possible decompositions of $\Gamma$ into $N$ charges, $\Gamma_a$, positioned
at different centers, such that the solutions are smooth.  The associated phase
space for each decomposition would then be the (possibly disconnected) solution
space for the charges.  Quantization of this phase space should, in principle,
yield certain microstates of the black hole and the set of (supergravity)
microstates\footnote{Actually, the full set of supergravity microstates probably
requires considering more general solutions than those with Gibbons-Hawking base
but one can at least hope that the latter set contributes a finite fraction of
the entropy.} should come from summing over all possible decompositions.  The
notion of smoothness is not necessarily (duality frame) invariant so a more
precise criterion might be that the constituent charges, $\Gamma_a$, should have
no entropy associated with them (as discussed above).

For a given decomposition into $N$ centers the phase space will be
the $2N - 2$ dimensional submanifold of $\mathbb{R}^{3N - 3}$
given by solving the constraint equations (\ref{integrabconstr})
for the positions $\vec{x}_a$.  Note that in arriving at this
counting we have subtracted the 3 center of mass degrees of
freedom; while these are present they generally decouple.  As
mentioned, this manifold may be disconnected and may posses a
rather complicated topology. Moreover, for a given total charge,
$\Gamma$, there will be many possible decompositions into
different numbers of centers implying that the total phase space
will be a disconnected sum of many manifolds of different
dimension. Determining the symplectic structure of even the lowest
dimensional solution spaces is already challenging
\cite{deBoer:2008zn}.

\subsection{Symplectic Form}
\label{sec:sympform}

In order to quantize the phase spaces described in section
\ref{subsec_quant} we will need to determine the symplectic
structure on these spaces.  This can, in principle, be derived
from the supergravity action as was done, for instance, in
\cite{Grant:2005qc}.  In this case, however, it is far more
tractable to take a different approach \cite{deBoer:2008zn}.  As
discussed in \cite{Denef:2002ru}, the four dimensional
multi-centered solutions can also be analyzed in the probe
approximation by studying the quiver quantum mechanics of D-branes
in a multicentered supergravity background.  Moreover, a
non-renormalization theorem \cite{Denef:2002ru} implies that the
terms in the quiver quantum mechanics Lagrangian linear in the
velocities do not receive corrections, either perturbatively or
non-perturbatively.  We can use this fact to calculate the
symplectic form  in the probe regime and extend it to the fully
back-reacted solution; this is because, for time-independent
solutions, the symplectic form depends only on the terms in the
action linear in the velocity.

For this approach to be consistent it is necessary that the BPS solution space,
which we interpret as a phase space, of the four and five dimensional
supergravity theories, as well as the probe theory, all match.  This follows from the fact that they are all governed by the same equation,
(\ref{integrabconstr}) \cite{Denef:2002ru}.  For instance, one can see that a
probe brane of charge $\Gamma_a$ in the background generated by a charge
$\Gamma_b$ is forced off to infinity as a wall of marginal stability is
approached \cite{Denef:2002ru} analogous to what was described below equation
(\ref{2cntr_decay}) for the corresponding supergravity solutions.

In \cite{deBoer:2008zn} the symplectic form on the solution space
is determined.  We will not review the derivation in detail but
simply note that it arises from the term coupling the probe brane
to the background gauge field, $\dot{x}^i A_i$, giving
\be
\tilde{\Omega} = \frac{1}{2} \sum_p \delta x^i_p \wedge \langle \Gamma_p ,\delta
{\mathcal A}_d^i(x_p)\rangle .
\ee
where ${\mathcal A}_d$ is the ``spatial'' part of the gauge field
defined in (\ref{multicentersol}) (this descends naturally to the
spatial part of the 4-d gauge field).  Using the definition of
${\mathcal A}_d$ we can further manipulate this expression
\cite{deBoer:2008zn} and put it in the form
\be \label{con3}
\tilde{\Omega} = \frac{1}{4} \sum_{p\neq q} \langle \Gamma_p,\Gamma_q\rangle
\frac{\epsilon_{ijk} (\delta (x_p-x_q)^i \wedge \delta
(x_p-x_q)^j) \, (x_p-x_q)^k }{|{\mathbf x}_p - {\mathbf x}_q|^3} .
\ee
This is a two form on the $(2N-2)$-dimensional solution space
which is a submanifold of $\mathbb{R}^{3N-3}$ defined by
(\ref{integrabconstr}).  Moreover, one can show that, on this
submanifold, this form is closed and, in the cases we will
investigate below, non-degenerate. Thus it endows the solution
space with the structure of a phase space.  Note that, as
anticipated, the center of mass degrees of freedom do not appear
in the symplectic form above and hence decouple in the
quantization of the system.  They will, in principle, yield an
overall pre-factor in the partition function which we will not
take into account.

Although the constraint equations (\ref{integrabconstr}) are
invariant under global SO(3) rotations these are nonetheless
(generically) degrees of freedom of the system and this is
reflected in the symplectic form.  If we contract (\ref{con3})
with the vector field that generates rotations around the 3-vector
$n^i$ (i.e. we take $\delta x_{pq}^i = \epsilon^{ijk} n^j
x_{pq}^k$) then the symplectic form reduces to
\bea
\label{redsymp} \tilde{\Omega} & \rightarrow & n^i \delta J^i
\eea
where $J^i$ are the components of the angular momentum vector defined in
(\ref{angularmomentum}).

This is nothing more than the statement that the components $J^i$
are the conjugate momenta associated to global SO(3) rotations. In
general the symplectic form on any of our phase
spaces\footnote{This does not hold for solutions spaces with
unbroken rotational symmetries, such as solution spaces containing
only collinear centers or only a single center. In these cases
some SO(3) rotations act trivially, do not correspond to genuine
degrees of freedom and do not appear in the symplectic
form.\label{footnote_degen}} will have terms like the above coming
from the global SO(3) rotations, in addition to terms depending on
other degrees of freedom.  As advertised (\ref{redsymp}) implies
that solution spaces with any $J^i = 0$ will have a degenerate
symplectic form and will therefore not constitute a proper phase
space\footnote{As mentioned in footnote \ref{footnote_degen} this
does not hold in the two center case where some SO(3) directions
decouple.  There are also potential subtleties with solution
spaces where $J=0$ at a single point but we neglect these for
now.}.

\subsection{Quantizing the Two-center Phase Space}

The inter-center distance of a two center configuration is fixed in terms of the
charges and the moduli at infinity but the axis of the centers can still be
rotated so, neglecting the center of mass degree of freedom, we are left with a
solution space that is topologically a two-sphere with diameter
\be
x_{12} = \frac{\langle h,\Gamma_1 \rangle}{\langle \Gamma_1 ,
\Gamma_2 \rangle}.
\ee
The symplectic form (\ref{con3}) is proportional to the standard
volume form on the two-sphere and is entirely of the form
(\ref{redsymp}) (note here that, as mentioned in footnote
\ref{footnote_degen}, collinearity of the solution implies that
one $U(1) \subset SO(3)$ decouples). In terms of standard
spherical coordinates it is given by
\be
\tilde{\Omega} = \frac{1}{2} \langle \Gamma_1 , \Gamma_2 \rangle \sin
\theta\, d\theta \wedge d\phi = |J|\sin \theta\, d\theta \wedge
d\phi .
\ee
We can now quantize the moduli space using the standard rules of
geometric quantization. We introduce a complex variable $z$ by
\be
z^2 = \frac{1+\cos\theta}{1-\cos\theta} e^{2i\phi}
\ee
and find that the K\"ahler potential corresponding to
$\tilde{\Omega}$ is given by
\be
K = -2|J| \log(\sin \theta) = - |J| \log\biggl(\frac{z \bar{z}}{(1 + z
\bar{z})^2}\biggr).
\ee
The holomorphic coordinate $z$ represents a section of the
line-bundle $L$ whose first Chern class equals
$\tilde{\Omega}/(2\pi)$. The Hilbert space consists of global
holomorphic sections of this line bundle and a basis of these is
given by $\psi_m(z)=z^m$. However, not all of these functions are
globally well-behaved. By examining the norm of $\psi_m$ given by
\be \label{norm1}
|\psi_m|^2 \sim \int d{\rm vol} e^{-K} |\psi_m(z)|^2
\sim \int d\cos\theta \, d\phi\,
(1+\cos\theta)^{|J| + m} (1-\cos\theta)^{|J| -m}
\ee
one finds that $\psi_m$ only has a finite norm if $m\geq - |J|$
and $m\leq |J| $. The total number of states equals $2|J|+1$. This
is in agreement with the wall-crossing formula up to a shift by 1.
It can be shown that the inclusion of fermionic degrees of freedom
will get rid of this unwanted shift \cite{deBoer:2008zn}.

The integrand in (\ref{norm1}) is a useful quantity as it is also the phase
space density associated to the state $\psi_m$.  According to the logic in
\cite{Balasubramanian:2005mg,Alday:2006nd,Balasubramanian:2007zt} (reviewed in Section
\ref{sec_rel_microstates}) the right bulk description of one of the microstates
$\psi_m$ should be given by smearing the gravitational solution against the
appropriate phase space density, which here is naturally given by the integrand
in (\ref{norm1}).  Since there are only $2|J|+1$ microstates, we cannot localize
the angular momentum arbitrarily sharply on the $S^2$; rather, it will be spread out
over an area of approximately $\pi/|J|$ on the unit two-sphere. It is therefore only in the
limit of large angular momentum that we can trust the description of the
two-centered solution (with two centers at fixed positions) in supergravity.

\subsection{Quantizing the Three-center Phase Space}

The phase space of the three center case is four dimensional.  Placing one
center at the origin (fixing the translational degrees of freedom) leaves six
coordinate degrees of freedom but these are constrained by two equations.  This
leaves four degrees of freedom, of which three correspond to rotations in SO(3)
and one of which is related to the separation of the centers.

This space is most easily visualized in the decoupling limit\footnote{We will
work in the decoupling limit simply because in this limit $\vec{J}$ is directly
proportional to the $D6$ dipole moment so it is easier to visualize it.  The
general characteristics of the solution space described hold for all
incarnations of the solutions we have considered -- the 4-d solutions, their 5-d uplift and the
decoupling limit of the latter.  We will use the language of 4-d charges simply
for brevity.} for the case when one of the centers has no
$D6$ brane charge.  In that case the solution has an angular momentum vector
$J^i$ directed between the two centers with $D6$ charge and the orientation of
the direction of this vector defines an S$^2$ in the phase space.  The third
center is free to rotate around the axis defined by this vector providing an
additional $U(1)$, which we will coordinatize by an angle $\sigma$, fibred
non-trivially over the S$^2$.  Finally the angular momentum vector has a magnitude which
may be bounded from both below and above and this provides the final coordinate
in the phase space.  This construction is perhaps not the most obvious one from
a coordinate space perspective but in these coordinates the symplectic form
takes a simple and convenient form.  This can also be used when all the centers
have $D6$ charge but then the relation between these coordinates and the
locations of the centers is less straightforward.

The symplectic form in these coordinates is (see
\cite{deBoer:2008zn} for a derivation):
\be \label{jan25}
\tilde{\Omega} = j\sin\theta\, d\theta\wedge d\phi - dj\wedge
D\sigma
\ee
with $D\sigma = d\sigma - A$, $j = |\vec{J}|$, and $dA=\sin\theta\,
d\theta\wedge d\phi$, so that $A$ is a standard monopole one-form on $S^2$. The
gauge field $A$ implements the non-trivial fibration of $\sigma$ over the S$^2$.
A convenient choice for $A$ is $A=-\cos\theta\,d\phi$ so that finally the
symplectic form can be written as a manifestly closed two-form
\be \label{fsym}
\tilde{\Omega} = - d (j\cos\theta)\wedge d\phi - dj \wedge
d\sigma.
\ee
Let us consider the shape of the solution space spanned by coordinates $[\theta,
\phi, j, \sigma]$.  $\theta$ and $\phi$ are standard spherical coordinates with
the latter defining a $U(1)$ that degenerates at $\theta=0,\pi$.  As has already
been mentioned the angular momentum usually spans some range $j \in [j_-, j_+]$
though there are cases where, for fixed charges, the angular momentum spans two
separate ranges resulting in a solution space with two disconnected components
(which must be quantized separately).  On the boundaries of these regions, at
$j=j_-$ or $j=j_+$, the centers are collinear so the coordinate $\sigma$
degenerates.   Thus the phase space is a symplectic manifold with two $U(1)$
actions (corresponding to rotations in $\sigma$ and $\phi$) which degenerate at
special points.  In fact the manifold is a toric K\"ahler manifold and can thus
be quantized using the technology of \cite{abreu-2000}\cite{Guillemin}.

This is done in detail in \cite{deBoer:2008zn} and here we will
report only the results.  Using the technology of geometric
quantization we can determine the basis of states spanning the
Hilbert space (defined as the space of normalizable holomorphic
sections of an appropriate line bundle over the phase space).  It
turns out that the number of such states is given by
\be \label{ndof}
{\cal N} = (j_+ - j_- +1)(j_+ + j_- + 1).
\ee
In fact this is not quite correct as we have neglected fermionic
degrees of freedom in defining our phase space.  It is possible to
include these degrees of freedom and quantize the resulting system
\cite{deBoer:2008zn} and this slightly changes the set of states.
The final result becomes
\be\label{3cnt_states}
{\cal N} = (j_+ - j_-)(j_+ + j_-)
\ee
This is the number we now wish to compare to the entropy determined by the wall
crossing formula.

Let us consider the attractor flow tree depicted in figure
\ref{fig_split}.  For the given charges, $\Gamma_1$, $\Gamma_2$,
and $\Gamma_3$, there are, in fact, many different possible trees
but, in terms of determining the relevant number of states, the
only thing that matters is the branching order.  In figure
\ref{fig_split} the first branching is into charges $\Gamma_3$ and
$\Gamma_4=\Gamma_1 + \Gamma_2$ so the degeneracy associated with
this split is $|\langle \Gamma_4, \Gamma_3\rangle|$ and the
degeneracy of the second split is $|\langle \Gamma_1, \Gamma_2
\rangle|$ giving a total number of states
\begin{equation}\label{tree_states}
    \mathcal{N}_{\textrm{tree}} = |\Gamma_{12}|\, |(\Gamma_{13} +
    \Gamma_{23})|
\end{equation}
where we have adopted an abbreviated notation, $\Gamma_{ij} = \langle \Gamma_i,
\Gamma_j \rangle$.

To compare this with the number of states arising from geometric
quantization of the solution space, (\ref{3cnt_states}), we need
to determine $j_+$ and $j_-$. Recall that the latter correspond to
two different collinear arrangements of the centers and, in a
connected solution space, there can be only two such
configurations \cite{deBoer:2008zn}.  To relate this to a given
attractor flow tree we will {\em assume part of the attractor flow
conjecture}, namely that we can tune the moduli to force the
centers into two clusters as dictated by the tree.  For the
configuration in figure \ref{fig_split}, for instance, this
implies we can move the moduli at infinity close to the first wall
of marginal stability (the horizon dark blue line) which will
force $\Gamma_3$ very far apart from $\Gamma_1$ and $\Gamma_2$. In
this regime it is clear that the only collinear configurations are
$\Gamma_1$-$\Gamma_2$-$\Gamma_3$ and
$\Gamma_2$-$\Gamma_1$-$\Gamma_3$; it is not possible to have
$\Gamma_3$ in between the other two charges.  Since $j_+$ and
$j_-$ always correspond to collinear configurations they must, up
to signs, each be one of
\begin{align}
    j_1 &= \frac{1}{2}(\Gamma_{12} + \Gamma_{13} + \Gamma_{23}) \\
    j_2 &= \frac{1}{2}(-\Gamma_{12} + \Gamma_{13} + \Gamma_{23})
\end{align}
$j_+$ will correspond to the larger of $j_1$ and $j_2$ and $j_-$
to the smaller but, from the form of (\ref{3cnt_states}), we see
that this will only effect $\mathcal{N}$ by an overall sign (which
we are not tracking carefully in any case).  Thus
\begin{equation}
    \mathcal{N} = \pm (j_1 - j_2)(j_1 + j_2) = \pm \Gamma_{12}\, (\Gamma_{13} +
    \Gamma_{23})
\end{equation}
which nicely matches (\ref{tree_states}).

Of course to obtain this matching we have had to assume the attractor flow
conjecture itself (in part) so it does not serve as an entirely independent
verification.  Demonstrating this matching more carefully would help validate
both methods of state counting \cite{deBoer:2008zn}.

\subsection{More than Three Centers?}

The symplectic form (\ref{con3}), when non-degenerate, defines a phase space
structure on the solution space for an arbitrary number of centers.  Analysing
the solution space for a generic set of charges however is quite difficult as
the constraint equations (\ref{integrabconstr}) imbue this space with a
complicated geometric structure.  In the two and three center case we were able
to do this because the space had a toric structure.  Fortunately there is a much
larger class of charges that also enjoy this property; namely any configuration
with two generic charges, $\Gamma_1$, $\Gamma_2$, interacting with any number,
$N$, of mutually BPS particles $\Gamma_i$.

From the constraint equation, (\ref{integrabconstr}),
\begin{align}
    \frac{\langle \Gamma_i, \Gamma_1 \rangle }{x_{1i}} + \frac{\langle \Gamma_i, \Gamma_2
    \rangle }{x_{2i}} =
    \alpha \\
    \frac{\langle \Gamma_1, \Gamma_2 \rangle }{x_{12}} + \sum_i \frac{\langle
    \Gamma_1, \Gamma_i \rangle }{x_{1i}} = \beta
\end{align}
it is clear that the $U(1)$'s around the $\vec{x}_{12}$ axis do not appear in the
constraint equations (i.e. the separation, $x_{ij}$, between mutually BPS
centers decouples) so each new coordinate $x_i$ comes with an additional
$U(1)$ isometry.

Two interesting examples in this class were studied in \cite{deBoer:2008zn}.  The
first case is obtained by setting all the $\Gamma_i = \Gamma_2$ (but leaving
$\Gamma_1 \neq \Gamma_2$).  This corresponds to having a single center
$\Gamma_1$ surrounded by a gas of particles of charge $\Gamma_2$ which do not
interact with each other and sit at a fixed distance $\langle \Gamma_2,
\Gamma_1\rangle / \alpha$ from the $\Gamma_1$.  From (\ref{integrabconstr}) the
solution space can be determined to be a product of S$^2$'s corresponding to the
position of each of the $\Gamma_i$'s on a sphere centered at $\vec{x}_1$.  This
``halo'' configuration is interesting and has occurred before in the
literature \cite{Denef:2002ru} \cite{Denef:2007vg} because it corresponds to a
system with a non-primitive charge $(N+1) \Gamma_2$.  This is, in fact, a two
charge system rather than an $N+2$ charge system since $N+1$ centers have
parallel charges.  In this case \cite{deBoer:2008zn} was able to use the
technology of geometric quantization of toric manifolds to compute the
degeneracy in this setting and match it to that computed using attractor trees
\cite{Denef:2000nb} once more providing a nice consistency check between the two
techniques.

A further generalization of this corresponds to setting $\Gamma_{1,2} = (\pm 1,
p/2, \pm p^2/8, p^3/48)$ with the + and - corresponding to center 1 and 2
respectively and setting $\Gamma_i = (0, 0, 0, -q_i)$ (with $p, q_i > 0$ and
$N = \sum_i q_i$).  This system is of physical interest because the total charge
corresponds to a D4-D0 black hole if we take $N > p^3/24$.  Moreover, note that
the charges we have selected satisfy $\Sigma(\Gamma_a) = 0$ so these geometries
have no entropy associated with them and are candidate microstate geometries for
the D4-D0 black hole.

Letting $I = p^3/6$ the regime $N < I/4$ is referred to as the polar regime
where single-centered D4-D0 black holes do not exist.  The regime $N > I/2$, on
the other hand, correspond to the regime in which the total charges, $\Gamma_1$,
$\Gamma_2$, $\Gamma_0 = \sum_i \Gamma_i$, are {\em scaling} in the sense defined
in Section \ref{scaling_sol}.  Recall that in the scaling regime the charges
can collapse to a single point in the solution space and form an infinitely deep
throat that strongly resembles a black hole to an outside observer.  Hence, for
$N > I/2$, it is the scaling solutions that are the most likely candidate
microstate geometries.

This system was studied in \cite{Denef:2007yt} \cite{Gaiotto:2004ij}  where it
was argued that D0's should yield the leading contribution to the black hole
entropy after they expand into elliptic D2's via the Myers effect.  It is thus
interesting to see how many states are captured by these smooth supergravity
solutions.  This computation was done in \cite{deBoer:2008zn} and the entropy of
these configurations was shown to have a leading behaviour of $N^{2/3}$ in the
regime $N < I/2$.  The entropy of a black hole in this regime, $I/4 < N < I/2$,
is of the order $\sqrt{NI} \sim N$ so clearly these configurations cannot
account for all the states.

It would be interesting to compute the entropy associated to these
configurations in the regime $N > I/2$ and compare this with the entropy of a
black hole in this regime.

\subsection{Large Scale Quantum Effects: Scaling Solutions}

Although it has long been understood how to account for the number of black hole
microstates in string theory \cite{Strominger:1996sh} this has generally been
done in a dual field theory making it difficult to address some fundamental
questions in black hole quantum mechanics such as information loss via Hawking
radiation. For some microscopic black holes (such as those discussed in Section
\ref{sec_ads3_s3}) the ability to dualize to an F1-P system has allowed for a
more detailed analysis of the structure of the microstates.  For these black
holes it has been argued \cite{Lunin:2002qf} that the average microstate is a
highly quantum superposition of states with the corresponding spacetime a wildly
fluctuating ``fuzzball''.  The very interesting part of this claim is that these
fluctuations extend over a region of spacetime circumscribed by the putative
black hole horizon.  The ``metrics''\footnote{Most of the relevant states are
stringy states so the term metric is not really appropriate.  A more precise
statement would be expectation values of a profile of the string in the F1-P
system.  See e.g.  \cite{Lunin:2002qf,Mathur:2005zp} for more details.}
corresponding to the states in the superposition are all very different within
the region which would be enclosed by a horizon in the naive black hole solution
but they settle down very quickly to the same metric outside the horizon.  Thus
the remarkable claim of \cite{Lunin:2002qf} is that the generic state in the
black hole ensemble has quantum fluctuations over a large region of spacetime
reaching all the way to the black hole horizon.

Unfortunately the black hole discussed in \cite{Lunin:2002qf} is microscopic and
has no horizon in supergravity (without higher derivative corrections); it would
thus be very desirable to be able to demonstrate this type of behaviour in a
system with a macroscopic black hole.  In \cite{deBoer:2008zn} an attempt was
made to do exactly this.  Scaling multicenter solutions can classically form
arbitrarily deep throats that become infinitely deep in the strict $\lambda
\rightarrow 0$ limit where the coordinate separation of the centers vanishes.
We expect, however, that quantum effects will prohibit us from localizing the
centers arbitrarily close together and will thus cap off the throat.  What is
remarkable about this is that the symplectic form, and hence the quantum
exclusion principle, is not renormalized as we increase $g_s$ (to interpolate
between quiver quantum mechanics and gravity) so, even though gravitational
effects increase the distance between the centers as the throat forms, the phase
space volume stays very small.  Thus gravitational back-reaction essentially
blows up these quantum effects to a macroscopic scale.  This is important not only because it
is reminiscent of the large scale quantum fluctuations of the D1-D5 black hole
but also because a smooth geometry with an infinite throat would be hard to
understand in the context of AdS/CFT.  Many solution spaces with a scaling point
persist and continue to exhibit scaling behaviour even after we take a
decoupling limit making all the solutions asymptotically AdS$_3$xS$^2$.  This is
problematic as general arguments suggest that an infinitely deep throat in a
smooth geometry that is asymptotically AdS would imply a continuous spectrum in
the CFT \cite{Bena:2007qc}.  Thus it is comforting that the analysis of \cite{deBoer:2008zn}
reveals the infinite throat to be an artifact of the classical limit.   Indeed, this is precisely the kind of phenomenon that was suggested in \cite{Mathur:2007sc}.

Before discussing this phenomena in more detail let us note some caveats.  The
states defined by quantizing the scaling solutions spaces are not necessarily
generic black hole microstates (in fact, it is probable that such states require
including additional stringy degrees of freedom in the phase space) so they may
not reflect the behaviour of the actual black hole ensemble.  Also, the
symplectic form was computed in the gauge theory and extended to gravity via a
supersymmetric non-renormalization theory; it would be more insightful to have a
direct supergravity computation of the symplectic form.  These caveats
notwithstanding it is remarkable that these solutions exhibit quantum structure
on a large scale even though they are smooth with everywhere small curvature.
We will return to this point presently.

What is actually determined in \cite{deBoer:2008zn} is the maximal depth of an
effective throat generated by trying to localize a state as much as possible in
the small $\lambda$ region of the phase space (recall $\lambda\rightarrow 0$ is
the scaling point where the centers coincide in coordinate space and an
infinitely deep throat forms in the geometry).  Specifically, a three center
solution similar to the one described in the previous section with a pure fluxed
D6-$\antiDp{6}$ pair and a single D0 with charge $-N$ is considered in its
lowest angular momentum eigenstate and the expectation value of the harmonic
$H_0$ and the D6-D0 separation is computed.  The latter is shown to be of order
$\epsilon \sim N/I \ge 1/2$ implying that the centers cannot be localized
arbitrarily close to each other so an infinite throat never forms.  Rather a cap
is expected to form at a scale set by the D0-D6 distance.  A wave equation analysis
for a scalar field in a simplified asymptotically AdS background with a capped throat of order $\epsilon$ reveals that the corresponding mass gap in the CFT
goes as $\epsilon/c$ where $c=6I$ is the central charge of the CFT.  The
expected result, from comparison with the D1-D5 system (see e.g.
\cite{Mathur:2005zp}), is a mass gap of order $1/c$ which is indeed what is
found in this case as $\epsilon$ is bounded from below by 1/2 (since $N \ge I/2$
for the solution to be scaling).

While the computation above is heuristic in many ways it yields two very
important qualitative lessons.  The first is that quantization of these solution
spaces as phase spaces resolves several classical pathologies such as infinitely
deep throats and also clarifies the issue of bound states (see
\cite{deBoer:2008zn} for a discussion of this).   More importantly, however, it
demonstrates that classical solutions may be invalid even though they do not
suffer from large curvature scales or singularities.  This is an important point
so let us explore it further.

In this particular system the phase space structure of the supergravity theory
can be related to that of quiver quantum mechanics by a non-renormalization
theorem.  In the latter the scaling solutions (at weak coupling) are analogous
to electron-monopole bound states.  Heisenberg uncertainty implies the minimum
inter-center distance is of order $x_{ij} \sim \hbar$.  Moreover because the
solution space is a {\em phase space} rather than a configuration space the
coordinates are conjugate to other coordinates rather than velocities so it is not
possible to localize all coordinate directions with arbitrary precision by
constructing delta-function states\footnote{Even if this were not the case delta
function localized states have a large spread in momentum and would thus destroy
any bound state.}.  Thus this quantity will have a
large variance so $\delta x_{ij}/x_{ij} \sim 1$ for very small $x_{ij}$.  At
weak coupling this is nothing more than the standard uncertainty principle and
is not particularly surprising.

What is surprising is that this behaviour persists even once gravity becomes
strong and the centers backreact stretching the infinitesimal coordinate
distance between them to a macroscopic metric distance.  Moreover, in this
regime the depth of the throat is extremely sensitive to the precise value of
$x_{ij}$ (see \cite{Bena:2007qc} for a numerical example) thus the large
relative value of $\delta x_{ij}$ translates into wildly varying depths for the
associated throat.  The associated expectation values for any component of the
metric have an extremely large variance $\delta g/g$ and so cannot possibly
correspond to good semi-classical states.  It is somewhat unusual to have
classical configurations that cannot be well approximated by semi-classical
states (i.e. those with low variance) but here this can be seen to follow from the
very small phase space volume this class of classical solutions occupy \cite{
Mathur:2007sc}.

\subsection{Summary}
The solutions described in this section constitute a sort of extended classical
foam with complicated topological structure.    The proposal of
\cite{Lunin:2001jy} translated into this context is that the classical finite area
supersymmetric black holes of string theory are simply effective descriptions of
complex, extended, horizon-free underlying bound states, of which the above
classical solutions would be simple (probably non-generic) examples.  This  is
in the same spirit as our discussion of  BPS extremal black holes in $\ads{5}$
and $\ads{3} \times S^3$ in previous sections of these notes.   A piece of
evidence in favor of this idea \cite{Denef:2002ru, Balasubramanian:2006gi}: one
can show that in the $g_s \to 0$ limit the centers in these solutions flow
together to form a single center bound state of D-branes with the charges of a
black hole.   Indeed, it is precisely these bound states that were originally
counted by Strominger and Vafa to explain the entropy of supersymmetric black
holes \cite{Strominger:1996sh}.  Turning this around, one might suggest a
picture where {\it every} black hole microstates begins life at $g_s = 0$ as a
ground state of an intersecting D-brane system, and that as the coupling is
increased to attain a finite Newton constant the bound state increases in
transverse size forming a sort of ``stringy spacetime foam''.   Further evidence
come from the results of \cite{deBoer:2008zn} that show that upon quantizing the
subset of these states that can be realized in supergravity, the class of
solutions that have a potential relation to black hole microstates will have
quantum fluctuations at spatial distances far larger than the string scale.
There are two missing pieces in the reasoning: (1) One would want to show that
these considerations apply to the generic microstate, since it is not clear that
the bound states described here are enough to account for the black hole
entropy,  (2) There is a need to reconcile the observations of an observer
falling into a black hole horizon with the sort of picture described here.
Nevertheless, perhaps an understanding involving extended underlying bound
states will eventually shed light on an enduring puzzle: why should entropy be
proportional to horizon area, rather than having some other functional
dependence on the parameters of a black hole?

\section{Conclusions}


One obvious limitation of our discussion has been the restriction
to extremal supersymmetric black holes. Although these are the
most tractable, eventually we would like to deal with
non-supersymmetric, realistic black holes that Hawking radiate. In
interesting recent work a connection between non-supersymmetric
black holes and interacting fluid dynamics was found
\cite{Bhattacharyya:2007jc,Bhattacharyya:2007vs} which suggests
that the type of dual descriptions in terms of free gases of
particles that we used in the supersymmetric cases may not be
sufficient. This does not mean that such black holes
cannot be studied using the approaches discussed in this review, but it does
indicate that the nature of microstates as well as the
extrapolation from weak to strong coupling is much more
complicated. For some recent progress in obtaining Hawking
radiation from a microstate point of view, see e.g.
\cite{Chowdhury:2007jx, Chowdhury:2008uj}. See also \cite{Balasubramanian:2005mg}
for a naive and qualitative description of the microstates of an
AdS-Schwarzschild in the weakly coupled gauge theory.

A key feature of the idea that black holes are effective descriptions of underlying extended bound states is that these bound states should roughly have an extent that agrees with that of the
black hole. In particular, they should grow as the string coupling
is increased in the same way as the black hole horizon grows, a
property emphasized and explicitly shown for the three-charge
supertube in \cite{Bena:2004wt}.   In \cite{Balasubramanian:2006gi} is was shown that a large class of multi-centered bound solutions with the same asymptotic charges as black holes shows precisely this sort of growth with the string coupling.   Another useful hint comes from the fluctuations at large proper distances in solutions of the kind needed to given effective black hole behavior (Sec. 5 and \cite{deBoer:2008zn}). It would be interesting to show whether the spacetime realization of the generic black hole microstate can grow in this way too, especially since such a growth would have important consequences for resolving the information paradox \cite{Mathur:2008wi}.

For large black holes, it is unlikely that the underlying microstates can be described in supergravity alone.  To see this, recall that in AdS/CFT the states that are dual to single particle supergravity modes in
the bulk are BPS states and their descendants. Therefore, one
might expect that supergravity solutions only contain information
about products of BPS operators and their descendants which are
the duals of general multiparticle supergraviton states. Now in
general, the phase in which AdS contains a thermal gas of
supergravitons is separated from the phase in which the AdS space
contains a black hole via a phase transition. This seems to
indicate that supergravitons alone do not have enough degrees of
freedom to account for the black hole entropy. For example, for
AdS${}_3\times$S${}^3$ one can show explicitly that this is the
case \cite{deBoer:1998us}.   Thus if it does turn out that all black holes are understood as the effective descriptions of extended bound states, the latter will likely involve many stringy degrees of freedom.

Along similar lines, it would be worthwhile exploring in more detail to
what extent the multi-centered solutions we considered in
section~\ref{sec5} can account for a finite fraction of
the entropy of the corresponding macroscopic black hole. They are
not the most general 5d supersymmetric solutions, as we took
special hyperk\"ahler manifolds as our base, and in addition we
ignored Calabi-Yau excitations and stringy excitations. Even if
they do not contribute a finite fraction of the entropy, one may
wonder whether one may at least find some typical black hole
microstates in this class. This question is very difficult to
answer, because it is not clear what set of macroscopic
observables we should precisely include in our definition of
typicality. Finally, we should be careful to not view all smooth
multi-centered solutions as associated to a single black hole.
Many are more properly thought of as being associated to multicentered
black holes. One may expect that only the solutions which are
described by a single centered attractor flow (this will include
in particular many scaling solutions) are honest microstates of a
single black hole \cite{deBoer:2008zn} . Obviously, much more work is required in order
to apply the philosophy outlined in this review in greater
detail to macroscopic supersymmetric black holes.  
Another open problem is to provide a more detailed map
between black objects in the bulk and ensembles in the boundary
CFT. What distinguishes semiclassical ensembles from non-classical
ones? What possible first laws of thermodynamics can these black
object have? How many chemical potentials do we typically need to
include in their dual description? Is there a natural way to
describe multicentered bound states in the dual field theory? We
have only begun to see some glimpses of answers to these questions
in the examples we have described.

Among the many possible generalizations, we would like to mention
the following results. For the AdS${}_5\times$S${}^5$ case,
significant progress has been made in identifying the space of 1/4
BPS geometries \cite{Lunin:2008tf}, and perhaps an extension of
the state-geometry map to the 1/4 BPS case is now feasible. There
was also recent progress in generalizing the notion of typical
states from $N=4$ SYM theory to more general four-dimensional
SCFT's \cite{Balasubramanian:2007hu}.

A key question of principle is what the infalling observer is supposed to see
in the picture outlined in these notes.  If the picture is correct, one might
imagine that  there is a tension with the usual scenario of an infalling
observer unaffected at the horizon of a large black hole.  However,   it might
be that there is some kind of ``correspondence principle'' such that classical
infalling observers still effectively see the well-known black hole interior
geometry until a short distance from the singularity.  One might imagine this
happening via a net cancellation of the many local interactions of a large object
with a diffuse quantum background, leading to an effective description in the
usual geometric terms.

\section{Literature Survey}\label{sec_lit}

Since the literature in this subject is voluminous, we will not attempt a comprehensive review but will rather merely point the reader in the direction of many relevant works and attempt to give a sense of the status of various branches of the field.  In so doing we will no doubt miss some important
developments but hope that references to these are contained in the works cited below.

\subsection{The D1-D5(-P) and Related Systems}

\begin{itemize}

\item[i-]{\bf{The D1-D5 system}}\

This configuration arises in type-IIB string compactified on $\mathcal{M}
\times$ S$^1$, where $\mathcal{M}$ is a 4 dimensional Ricci-flat manifold. The
supergravity solution describes N$_5$ D5-branes wrapping the full compact space
and N$_1$ D1-branes wrapping the S$^1$; the most general metric corresponding to
this configuration was constructed in \cite{Lunin:2001fv} (also see \cite{Cvetic:1998xh, Balasubramanian:2000rt, Maldacena:2000dr}). Originally there were
neither internal nor fermionic excitations.  In \cite{Lunin:2002iz}, the
internal excitations were included, and \cite{Taylor:2005db} included the
fermionic excitations. The near horizon geometry of this system is AdS$_3
\times$ S$^3 \times \mathcal{M}$, so the dual theory is a CFT$_2$. It was in the
context of this system that the fuzzball proposal was first proposed
\cite{Lunin:2001jy}.

A nice, but slightly out-dated, review of the D1-D5 system and the fuzzball
proposal is contained in \cite{Mathur:2005zp}.  Significant progress has
occurred since this review: the supergravity phase space has been directly
quantized \cite{Donos:2005vs,Rychkov:2005ji} (see also \cite{Bak:2004rj}), and
the idea that black holes are simply effective geometries has been studied in
detail in the context of this system \cite{Balasubramanian:2005qu,
Mathur:2005ai, Giusto:2005ag, Alday:2006nd, Kanitscheider:2006zf, Mathur:2007sc,
Kanitscheider:2007wq}.   A more up-to-date review, with an emphsis on AdS/CFT
tests of the proposal, can be found in \cite{Skenderis:2008qn}.

Supporting evidence
for the stretched horizon idea was given in \cite{Mathur:2005ai,Mathur:2007sc}.
In \cite{Mathur:2005ai} it was argued that quantum gravity effects become
important at scales larger than the Planck/string scales. In
\cite{Mathur:2007sc} a sub-ensemble of the original thermal ensemble was
considered. It was shown that, even in this case, the area of stretched horizon
matches the microscopic entropy of the subsystem to leading order. This
continues to hold even after the inclusion of 1-loop string corrections. In
\cite{Balasubramanian:2005qu,Alday:2006nd,Kanitscheider:2006zf,Kanitscheider:2007wq}
various quantities in the CFT and supergravity were compared.

There are also studies of microstates using a 1/4-BPS probe approach, see e.g
\cite{Mandal:2007ug, Raju:2007uj, Bak:2004rj, Bak:2004kz}.

\item[ii-]{\bf{Beyond the D1-D5 system}}\

Further generalizations of the D1-D5 system followed.   They can be collected
under two general themes: adding more charges and breaking supersymmetry.  In
the first case the aim was to add some charges (momentum most of the time) to
reduce the amount of preserved supersymmetry without breaking all the
supersymmetry. An incomplete list of references in this direction includes
\cite{Mathur:2003hj, Lunin:2004uu, Giusto:2004id, Giusto:2004ip, Giusto:2004kj,
Saxena:2004ap, Giusto:2004xm, Ford:2006yb}.  The D1-D5-P system is also covered
in the review \cite{Skenderis:2008qn}.

In \cite{Mathur:2003hj} a first attempt to construct dual geometries for D1-D5-P
microstates was undertaken; here the momentum was added as a small perturbation
of the D1-D5 system. The first success was achieved in \cite{Lunin:2004uu} which
was followed by other works \cite{Giusto:2004id, Giusto:2004ip, Ford:2006yb}. In
\cite{Giusto:2004xm} some arguments were given to support the claim that higher
order correction to these 3-charge geometries would not generate a
horizon or a singularity if they were not present at tree level.

A class of 5-dimensional smooth solutions, closely related to the D1-D5-P
system, have also been focus of an extended research program
\cite{Bena:2004tk,Bena:2004de,Berglund:2005vb,Bena:2005va,Bena:2006is,Bena:2006kb,Balasubramanian:2006gi,Bena:2007kg,Denef:2007yt,Gimon:2007mha,Bena:2007qc,deBoer:2008fk,Bena:2008wt,Raeymaekers:2008gk,deBoer:2008zn}.
These solutions are discussed in section \ref{sec5} of this review.  A
recent review of some of this work is \cite{Warner:2008ma}.

On the other hand, less work has been done on non-BPS solutions
\cite{Jejjala:2005yu, Cardoso:2005gj, Gimon:2007ps, Giusto:2007tt,
Chowdhury:2007jx}.  In \cite{Jejjala:2005yu} smooth non-supersymmetric
geometries were constructed.  They were then proven to be classically unstable
in \cite{Cardoso:2005gj}. This classical instability was shown to give rise to
Hawking radiation in \cite{Chowdhury:2007jx}. More details appeared in \cite{Chowdhury:2008bd}. Other properties of these
solutions were studied in \cite{Gimon:2007ps}. Another set of non-supersymmetric
solutions in 4 dimensions appeared in \cite{Giusto:2007tt}. Another interesting study was
the tunneling of a collapse of a shell to fuzz-ball geometries \cite{Mathur:2008kg} which is needed for a possible fuzz-ball like proposal
for more realistic four dimensional black holes.

\item[iii-]{\bf{Bound versus unbound systems}}\

To avoid overcounting the states responsible for a black hole's entropy one needs to know
that the solution one is dealing with describes a bound system.  For the D1-D5
system, this can be achieved by adding D3 charge. This system has been studied
in \cite{Lunin:2003gw} in the context of the fuzzball proposal.

Other works in this direction are \cite{Giusto:2004kj, Giusto:2005ag,
Bena:2008wt}. In \cite{Giusto:2004kj} known D1-D5 (D1-D5-P) geometries were
rewritten in a {\it{fiber $\times$ base}} form in order to gain some insight
into their structure. In \cite{Giusto:2005ag} another route was followed. Using
the D1-D5 system as a prototype, this paper studied the dynamical behavior of
such systems and put forth a conjecture to distinguish bound systems.  In
\cite{deBoer:2008fk} three charge systems were studied in an AdS$_3\times$S$^2$
decoupling limit and a criterion for bound configurations was proposed in terms of
split attractor trees \cite{Denef:2000nb}.  In \cite{deBoer:2008zn} some of these
solutions were quantized allowing a direct analysis of whether the relevant state
is bound or not.  In \cite{Bena:2008wt} another criterion was given using
spectral flow in AdS$_3 \times$ S$^3$.

\end{itemize}

\subsection{Asymptotically AdS$_5 \times$ S$^5$ and Related Solutions}

In AdS$_5\times$S$^5$ one can, in principle, consider 1/2, 1/4, 1/8 and 1/16
excitations of the full string theory.  The only known example of a
supersymmetric black hole  in AdS$_5$ with finite area is the Gutowski-Reall black hole
\cite{Gutowski:2004ez, Gutowski:2004yv} and this solution is only 1/16 BPS.
There are 1/2 BPS black solutions \cite{Behrndt:1998ns, Behrndt:1998jd,
Cvetic:1999xp, Myers:2001aq}, known as superstars, but these have a no horizon in
two-derivative supergravity.  Nonetheless, significant effort has gone into
studying the possible smooth, asymptotically AdS$_5$ geometries with varying
amounts of supersymmetry.  For the 1/2 BPS case  the
solutions were completely classified in \cite{Lin:2004nb}.

\begin{itemize}
\item[i-] {\bf{The 1/2 BPS Case}}\

Various generalizations of the LLM \cite{Lin:2004nb} geometries have been
considered.  The notion of typical states in an ensemble was explored in
\cite{Balasubramanian:2005mg}, while in \cite{Balasubramanian:2007zt} a
``metric'' operator was defined in the CFT and used to establish a criterion to
determine which states will have a well-defined classical dual geometry. Another
direction of research, closely related to topics discussed in this review,
involved quantization of the original LLM geometries using phase space
techniques \cite{Grant:2005qc, Maoz:2005nk}.

Thermal properties and Wilson loops in the CFT dual to these geometries were
studied in \cite{Okuda:2007kh, Liu:2007ff} while in \cite{D'Hoker:2007xz,
D'Hoker:2007xy, Gomis:2007fi} 1/2 BPS geometries corresponding to defects in the
CFT were considered.  In \cite{D'Errico:2007jm} the relation between phase space
densities in the fermion formulation of the theory and generalized Young
tableaux is studied. In \cite{Sato:2007zu} the on-shell action for the LLM
geometries was derived.

In \cite{Ogawa:2008kz} where different methods were used to calculate
the entropy of a ``black hole'' resulting from coarse graining over LLM geometries. There was a perfect agreement between entropies calculated by CFT and gravity coarse grainings .

There is, in fact, a large volume of literature on the 1/2 BPS case and our short
survey is by no means intended to be comprehensive.  For more references the
reader should consult the works cited above.

\item[ii-] {\bf{Less SUSY}}\

Backgrounds which preserve only 1/4, 1/8 and 1/16 supersymmetry have also been
considered and explicit supergravity solutions have been constructed.  For
instance smooth 1/4 BPS solutions were constructed in \cite{Gauntlett:2004zh,
Donos:2006iy, Donos:2006ms, Chong:2004ce} and an LLM-like prescription to derive
them from droplets on the plane is related to constraints on brane webs in
\cite{Lunin:2007mj, Lunin:2008tf}.

Probe solutions (giant gravitons) preserving 1/8 of the supersymmetry were
studied in \cite{Mikhailov:2000ya, Mandal:2006tk, Biswas:2006tj, Kim:2005mw,
Berenstein:2005aa} and the back-reaction of such probes was worked out in
\cite{Gava:2006pu,Gava:2007qs}.  The 1/8 BPS sector of the dual CFT was explored
in \cite{Lucietti:2008cv}.

As mentioned above, the 1/16 BPS sector is distinguished by having black holes with macroscopic
horizons \cite{Gutowski:2004yv, Gutowski:2004ez, Chong:2005da, Chong:2005hr,
Kunduri:2006ek}.  The 1/16 sector of the CFT has also been studied; operators
potentially related to the black hole have been identified \cite{Berkooz:2006wc}
and the entropy has been (qualitatively) reproduced \cite{Kinney:2005ej}.  Giant
gravitons preserving 1/16 of the supersymmetry were found in \cite{Kim:2006he}.

Attempts to treat the full set of 1/2, 1/4, and 1/8 solutions in a common
framework can be found in \cite{Liu:2007rv, Chen:2007du}.  Other related work
includes \cite{Hoppe:2007tv, Dutta:2007ws, Gava:2007kr}.  Once more, this list
is only intended to serve as an introduction to the literature and no doubt has
failed to include many important works.

\item[iii-] {\bf{No Supersymmetry}}\

Another direction of investigation involves breaking all supersymmetries.  Some
research in this direction includes \cite{Balasubramanian:2007bs, Liu:2007xj, Chen:2007gh}.
\end{itemize}

The considerations described above for the D1-D5 system and the 1/2-BPS states of AdS$_5$ were applied to other backgrounds (an M-theory solution) in \cite{Shieh:2007xn}.
In \cite{Balasubramanian:2007qv},  general considerations are applied to
study the differences between correlators of an operator in a typical state and a
thermal state.  In this paper it is also shown that for a system with an entropy $S$, the variance in finitely local correlation functions over the entire Hilbert space will by suppressed by $e^{-S}$.  Because of this, regardless of the detailed origin of black hole entropy, if there is {\it any} statistical interpretation of the underlying degeneracy semiclassical observers will have difficulty telling apart the microstates.

\section*{Acknowledgements}

JdB would like to thank the organizers and participants of the RTN
Winter School on Strings, Supergravity and Gauge Theories at CERN for
the opportunity to present these lectures in a stimulating environment.  VB thanks the organizers and participants of the PIMS Summer Schools in String Theory where some of this material was presented.  VB and JdB  also thank the organizers of the Sowers workshop, and in particular Mark Sowers, for the opportunity to present some of this work at Virginia Tech.   

The work of VB was supported in part by DOE grant DE-FG02-95ER40893 and in part
by a Helen and Martin Chooljian membership at the Institute for Advanced Study,
Princeton.  The work of JdB, SES and IM is supported financially by the
Foundation of Fundamental Research on Matter (FOM).   We are also grateful to
Bartek Czech, Frederik Denef, Eric Gimon, Veronika Hubeny, Per Kraus, Klaus
Larjo, Tommy Levi, Don Marolf, Mukund Rangamani,  Masaki Shigemori,  Joan Simon
and Dieter Van den Bleeken for collaborations on some of the work described
here.

The authors are very grateful to Eric Gimon for providing useful feedback on 
these notes and to Samir Mathur for some insightful discussions.


\bibliographystyle{jhep}
\bibliography{refslist-proc}

\end{document}